\documentclass[aps,prb,twocolumn,superscriptaddress,floatfix,showpacs,groupedaddress]{revtex4-1}
\usepackage{graphicx}
\usepackage{amsfonts}
\usepackage{amssymb}
\usepackage{amsmath}
\usepackage{txfonts}
\usepackage{lipsum}
\DeclareMathOperator{\Tr}{Tr}

\DeclareMathOperator{\Imag}{Im}
\DeclareMathOperator{\Arg}{Arg}

\DeclareMathOperator{\pv}{p.v.}

\begin{document}
\title{Friedel oscillations at the surfaces of rhombohedral $N$-layer graphene}
\author{C. Dutreix}
\author{M. I. Katsnelson}
\affiliation{Radboud University, Institute for Molecules and Materials, Heyendaalseweg 135, 6525AJ Nijmegen, The Netherlands}


\begin{abstract}
The low-energy physics of rhombohedral $N$-layer graphene mainly arises on the external layers, where most of the $\pi$ electrons are located. Their Bloch band structure defines a two-band semimetal; the dispersion relation scales as $\pm q^{N}$ with the momentum norm $q$ in the vicinity of two non-equivalent valleys. In this paper, we address the problem of elastic scattering through a localized impurity located either on the surface of the material or within the bulk, and focus on the quantum interferences it induces on the two external layers. It is apprehended in the framework of a $T$-matrix approach, both numerically and analytically, regardless of the impurity magnitude, which enables the description of realistic scatters. In rhombohedral multilayer graphene, the impurity induces Friedel oscillations that always decay as $1/r$. As a result, monolayer graphene is the only material of the rhombohedral class that exhibits $1/r^{2}$-decaying Friedel oscillations. The interference patterns are subsequently analyzed in momentum space. This analysis enables a clear distinction between monolayer graphene and multilayer graphene. It also shows that the interference pattern reveals the whole Bloch band structure, and highlights the number of layers stacked in the material, as well as the $\pi$-quantized Berry phases that characterize the existence of nodal points in the semimetallic spectrum. Experimentally, these features may be probed from scanning tunneling microscopy, when imaging the local density of states at the surfaces of suspended rhombohedral $N$-layer graphene.
\end{abstract}

\maketitle

\section*{Introduction}
\label{Introduction}
In the 1950s, Friedel addressed the problem of the response of conduction electrons to a localized charge in metals [\onlinecite{friedel1952xiv}]. Whereas negative point charges would tend to exponentially screen a positively charged impurity over the Debye length, Friedel reported algebraically decaying oscillations in the charge density, due to a nesting wave vector given by twice the Fermi momentum $k_{F}$. These long-range modulations, referred to as Friedel oscillations, subsequently found a counterpart in terms of magnetic interactions [\onlinecite{PhysRev.96.99},\onlinecite{vonsovsky1974magnetism}]. Even in the case of non-interacting electrons, the elastic scattering through a localized impurity also yields Friedel oscillations in the local density of states (LDOS). For a two-dimensional non-relativistic electron gas, these $2k_{F}$-wave-vector oscillations were found to decay as $1/r$ with the distance to the impurity [\onlinecite{adhikari1986quantum}], which was observed at the Cu(111) surface by scanning tunneling microscopy (STM) [\onlinecite{crommie1993imaging}]. This technique, earlier developed by Binnig and Rohrer [\onlinecite{RevModPhys.59.615}], relies on the measurement of the tunneling differential conductance between a sharp tip and a metallic surface and, therefore, images the LDOS of the surface with atomic-scale resolution. Later, it was realized that the Fourier analysis of STM data could be useful too, as it provides information about the Fermi contours involved in the elastic scattering [\onlinecite{Sprunger21031997},\onlinecite{PhysRevB.57.R6858}]. For the two-dimensional non-relativistic electron gas, the LDOS Fourier transform reveals a $2k_{F}$-radius ring. Since the scattering is elastic and the dispersion relation is parabolic, the scattering wave vectors are obtained by joining two states of the circular Fermi contour. Consequently, the $2k_{F}$-radius ring means that backscattering is the more efficient process. In this sense, STM can be thought of as a technique to probe the iso-energy contours of the dispersion relation in two-dimensional electronic systems.

Soon after the isolation of graphene [\onlinecite{Novoselov22102004}], it was found out that STM could also provide features of the quasiparticles associated to the Fermi contours. Being a one-atom-thick layer of graphite, graphene is a two dimensional material and offers a natural playground for STM [\onlinecite{PhysRevB.75.125425, PhysRevLett.103.206804, PhysRevB.81.085413, PhysRevLett.104.096804, 2011NatPh...7...43B}]. Its low-energy Bloch band-structure is based on a dispersion relation made of two non-equivalent Dirac cones and, thus, describes relativistic electrons [\onlinecite{PhysRev.71.622}]. A remarkable property for such fermions is that backscattering is totally suppressed. For example it implies that incoming electrons with a normal incidence tunnel as holes through a barrier potential with a transmission probability of 1. This phenomenon, known as Klein tunneling with reference to Klein's paradox in relativistic quantum mechanics [\onlinecite{klein1929reflexion}], plays of course an important role in issues such as localization and transport [\onlinecite{ando1998berry, Katsnelson:2006qf, Cheianov02032007, young2009quantum, 1402-4896-2012-T146-014010, Reijnders2013155, katsnelson2012graphene}]. The suppression of backscattering in graphene is also responsible for the unconventional decay of the Friedel oscillations induced by a localized impurity in the LDOS [\onlinecite{ PhysRevLett.97.226801},\onlinecite{PhysRevLett.100.076601}]. Indeed, when the elastic scattering occurs within a single Dirac cone, these oscillations exhibit a $1/r^{2}$ decay with the distance to the impurity and their Fourier transform consists of a $2k_{F}$-radius disk [\onlinecite{PhysRevLett.100.076601}]. The observation of this unconventional signature has demonstrated the ability of STM to probe the relativistic nature of Dirac electrons [\onlinecite{PhysRevLett.101.206802},\onlinecite{Simon:2009eu}]. This is the reason why this feature has also been used in the context of three-dimensional topological insulators, in order to prove the presence of surface states which exist as Dirac fermions  in these systems [\onlinecite{xia2009observation},\onlinecite{PhysRevLett.103.266803}].

In this paper, we address the problem of elastic scattering through a localized impurity in rhombohedral $N$-layer graphene. Whereas the issue of the dc conductivity in the presence of such scatterers has been reported recently [\onlinecite{PhysRevB.89.165308},\onlinecite{PhysRevB.91.195104}], we rather focus on the pattern of the quantum interferences induced in the LDOS, namely the Friedel oscillations and their Fourier transform.
Similarly to monolayer graphene, the low-energy Bloch band structure of rhombohedral $N$-layer graphene reveals a two-band semimetal with two non-equivalent valleys [\onlinecite{haering1958band},\onlinecite{PhysRevB.73.245426}]. The valence and conduction bands refer to electronic orbitals of two sublattices located on the external surfaces. As a result, their LDOS can be imaged by STM when suspending the material [\onlinecite{C2NR30162H}]. If a localized impurity is introduced on a given layer, then Friedel oscillations are induced on both outer surfaces. It turns out that, from a $T$-matrix approximation, these oscillations always exhibit a conventional $1/r$ decay. Therefore, monolayer graphene is the single one material of the rhombohedral class that shows $1/r^{2}$-decaying Friedel oscillations. Then, we show that the LDOS Fourier transform  reveals the whole Bloch band structure, as well as the $\pi$-quantized Berry phases that characterize the nodal points in the semimetallic spectrum, which also highlights the number of layers stacked in the material.

The first section reminds the reader of the Bloch band structures of graphene and rhombohedral multilayer graphene. The Friedel oscillations induced in the LDOS by a localized impurity are discussed in the second section in the framework of a T-matrix approach. The third section is devoted to their Fourier analysis.

\section{Rhombohedral $N$-layer graphene}
\label{Rhombohedral multilayer graphene}

\subsection{Some reminders about graphene}
\begin{figure}[t]
\centering
$\begin{array}{cc}
\includegraphics[trim = 0mm 0mm 0mm 0mm, clip, width=3.cm]{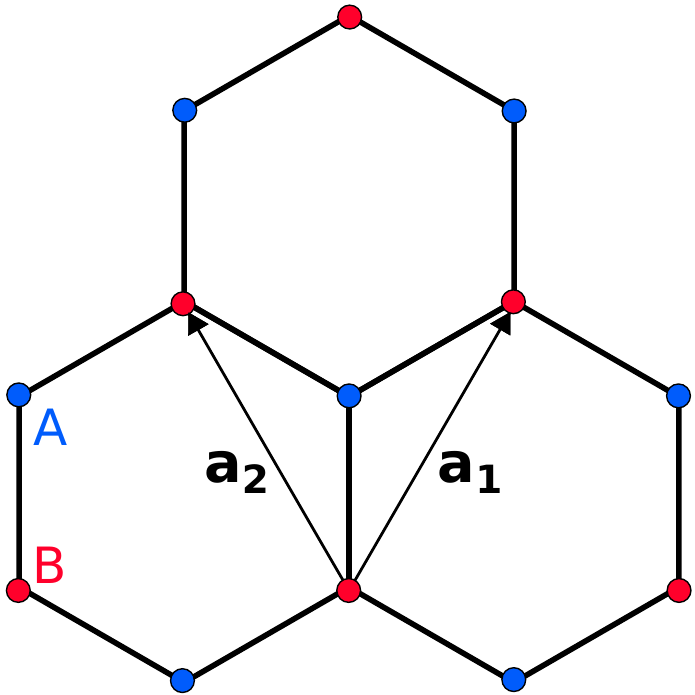} &
\includegraphics[trim = 20mm 0mm 15mm 0mm, clip, width=5.0cm]{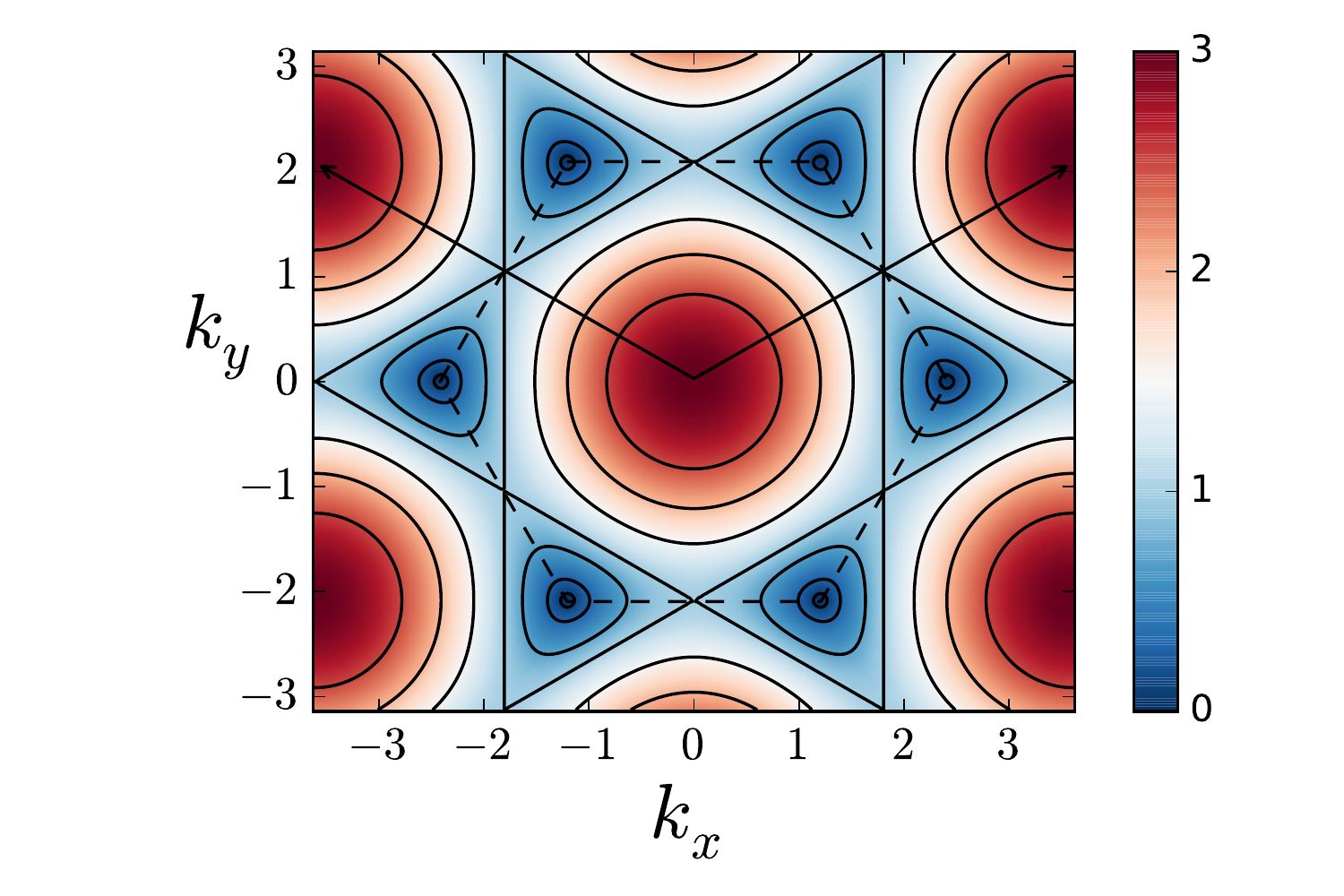}
\end{array}$
\caption{\small (Color online) Honeycomb lattice of graphene (left) and iso-energy contours of its electronic dispersion relation (right). The hexagon made of dashed lines outlines the first Brillouin zone and the two arrows depict the vectors ${\bf b_{1}}$ and ${\bf b_{2}}$ that span the reciprocal lattice. The energy is given in units of the hopping amplitude $t$ .}
\label{Honeycomb Lattice}
\end{figure}

Graphene consists of a superposition of two triangular Bravais lattices. They are denoted A and B according to Fig. \ref{Honeycomb Lattice}. Every atom A (respectively B) has three nearest neighbors B (respectively A), whose relative positions are given by the three unit vectors
\begin{align}
{\bf d_{1}}=\Big( \frac{\sqrt{3}}{2},~\frac{1}{2} \Big)~,~
{\bf d_{2}}=\Big( -\frac{\sqrt{3}}{2},~\frac{1}{2} \Big)~,~
{\bf d_{3}}=\Big( 0,~ -1 \Big)~.
\end{align}
The definitions above implicitly mean that the lattice constant a$_{0}\simeq1.42~{\AA}$ has been chosen as the unit of length. The vectors that span the two triangular Bravais lattices are chosen to be
\begin{align}
{\bf a_{1}}=\Big( \frac{\sqrt{3}}{2},~\frac{3}{2} \Big)~,~~~~~
{\bf a_{2}}=\Big( -\frac{\sqrt{3}}{2},~\frac{3}{2} \Big)~.
\end{align}
The triangular geometry leads to hexagonal Brillouin zones in momentum space, whose basis vectors ${\bf b_{1}}$ and ${\bf b_{2}}$ are defined from the following scalar product
\begin{align}
{\bf a_{i}}\cdot {\bf b_{j}}=2\pi~\delta_{ij}~.
\end{align}

Although the atoms A and B are not equivalent from the viewpoint of the lattice structure, both are carbon atoms. So each atom has a single free valence electron which refers to a $\pi$ electronic orbital. Within a spinless nearest-neighbor tight-binding approximation, the electronic Bloch band-structure of graphene relies on the following Hamiltonian matrix 
\begin{align}\label{Hk graphene}
H({\bf k}) = 
\left( \begin{array}{cc} 
0 & f({\bf k}) \\
f^{*}({\bf k}) & 0
\end{array} \right)~,
\end {align}
where $\hbar =1$ and $f({\bf k})=t\sum_{j=1..3}e^{i{\bf k}\cdot{\bf d_{j}}}$ is a momentum space representation of intersublattice hopping processes for electrons with wave vector ${\bf k}$. The parameter $t$, which is about -3eV in graphene [\onlinecite{PhysRevB.66.035412}], denotes the hopping amplitude.

The low-energy physics of isotropic graphene arises at the corners of the Brillouin zones [\onlinecite{PhysRev.71.622}]. Each corner is defined in a unique way by
\begin{align}\label{Korner Vector}
{\bf K^{\xi}_{mn}} = \xi\frac{{\bf b_{1}}-{\bf b_{2}}}{3}+m{\bf b_{1}}+n{\bf b_{2}} ~,
\end{align}
where $m$ and $n$ are integers and $\xi=\pm1$. Two corners which are labeled by opposite values of $\xi$ cannot be connected to one another by a linear combination of ${\bf b_{1}}$ and ${\bf b_{2}}$, and are said to be non-equivalent. In their neighborhood, it is found that
\begin{align}
f({\bf K^{\xi}_{mn}}+{\bf q}) &\simeq -\xi v_{F}qe^{i\theta^{\xi}_{mn}({\bf q})} ~,
\end{align}
where $q=|{\bf q}|\ll |{\bf K^{\xi}_{00}}|$. The Fermi velocity is $v_{F}=3t/2$ and the phase is defined as
\begin{align}\label{Phase Definition}
\theta^{\xi}_{mn}({\bf q}) = {\bf K_{mn}^{\xi}}\cdot{\bf d_{3}} + \theta^{\xi}({\bf q}) ~,
\end{align}
with $\theta^{\xi}({\bf q})=\xi\theta_{\bf q}$ and $\theta_{\bf q}$ the polar angle of the wave vector ${\bf q}$ with respect to the direction ${\bf a_{1}}-{\bf a_{2}}$.

The dispersion relation, which is depicted in Fig. \ref{Honeycomb Lattice}, relies on two non-equivalent Dirac cones 
\begin{align}
\epsilon_{\pm}({\bf K^{\xi}_{mn}}+{\bf q}) \simeq \pm v_{F} q ~.
\end{align}
It becomes degenerate right at the corners of the Brillouin zone. The presence of these nodal points at zero energy defines a semimetallic phase. Due to the energy landscape around the momenta ${\bf K^{\xi}_{mn}}$, their vicinity is also referred to as a valley. As far as we will be concerned thereafter, the coincidence between these valleys and the corners of the Brillouin zones is not crucial. For example, anisotropies in the hopping parameters are likely to make the Dirac cones move away from the corners [\onlinecite{PhysRevB.74.033413},\onlinecite{PhysRevLett.100.236405}]. Then it may happen that, for a threshold value of the anisotropy, non-equivalent cones merge at a time-reversal invariant point, before opening an energy gap in the spectrum. Nevertheless, we disregard such a merging transition, as it is not achievable in graphene [\onlinecite{PhysRevB.76.064120}]. For more simplicity, we also restrict the discussion to the isotropic case, so that the Dirac cones are indeed centered at the Brillouin zone corners. What is crucial, however, is that the non-equivalent cones always come in pairs at opposite momenta. This is a consequence of the time-reversal symmetry which implies $H({\bf k})=H^{*}(-{\bf k})$ in the spinless description. Finally, note that, even if the two non-equivalent cones are related to one another by this symmetry, it is not sufficient to guarantee their existence. The latter turns out to be guaranteed if the crystal additionally has the inversion symmetry, i.e. $\sigma_{1}H({\bf k})\sigma_{1}=H(-{\bf k})$, which prevents any mass term in the Hamiltonian matrix (\ref{Hk graphene}). Under those two symmetries, the semimetallic phase is then protected [\onlinecite{PhysRevB.75.155424}].

Regarding the massless Dirac electrons, they are described by the following Bloch eigenstates
\begin{align}
|\Psi_{\pm}({\bf K^{\xi}_{mn}}+{\bf q}) \rangle \simeq \frac{1}{\sqrt{2}}
\left( \begin{array}{c} 
1 \\
\mp \xi e^{-i{\bf K_{mn}^{\xi}}\cdot{\bf d_{3}}}e^{-i\theta^{\xi}({\bf q})}
\end{array} \right) ~.
\end{align}
Their spinor structure is the one of a two-level system but, more interestingly, it is momentum dependent. This momentum dependence simply appears as a phase factor between the two sublattice components of the Bloch spinor and is based on the phase $\theta^{\xi}({\bf q})$. Therefore, the low-energy band structure, which consists of the set of the dispersion relation and the Bloch eigenstates, is entirely characterized by $|{\bf q}|$ and $\theta_{\bf q}$, namely the norm and the orientation of the wave vector ${\bf q}$.

This plays an important role in issues such as elastic scattering. In graphene, it crucially relies on the absence of backscattering [\onlinecite{ando1998berry,Katsnelson:2006qf, Cheianov02032007}]. This peculiar feature of the massless Dirac electrons comes from the following relation
\begin{align}\label{Phase symmetry relation}
\theta^{\xi}({-{\bf q}}) - \theta^{\xi}({\bf q}) = \xi \pi ~,
\end{align}
which requires the incoming and backscattered electrons to have orthogonal wave functions and interfere destructively. In the framework of a Born approximation, the backscattering probability within the valley ${\bf K^{\xi}_{mn}}$ is indeed proportional to
\begin{align}\label{Born backscattering}
|\langle \Psi_{\pm}({\bf K^{\xi}_{mn}}-{\bf q}) | \Psi_{\pm}({\bf K^{\xi}_{mn}}+{\bf q}) \rangle|^{2}=0 ~,
\end{align}
regardless of the momentum dependence of the potential [\onlinecite{Allain:2011sf}]. Note that this suppression of backscattering is based on a single Dirac cone description, which is sufficient when the scattering potentials are smooth enough compared to the atomic length-scale.

The case of localized scatterers turns out to be more informative about the band structure. Since they allow elastic scattering in the whole momentum space, the scattering can occur within a single valley, as well as between two distant and eventually non-equivalent valleys.
The localized impurity induces long-range Friedel oscillations in the LDOS, which can be Fourier analyzed and compared to STM experiments [\onlinecite{PhysRevB.86.045444}]. In the case of monolayer graphene, both sublattices belong to the same surface, so that the Fourier transform of STM data involves the contributions of the two sublattices together.
\begin{itemize}
\item For scattering processes which take place within a single valley, Eq. (\ref{Phase symmetry relation}) is responsible for a reduction of the Friedel oscillations to a $1/r^{2}$ decay [\onlinecite{PhysRevLett.97.226801}], instead of their usual $1/r$ decay in a two dimensional non-relativistic electron gas [\onlinecite{adhikari1986quantum}]. This yields a clear signature (see central disk in Fig. \ref{FT STM graphene}) in momentum space [\onlinecite{PhysRevLett.100.076601},\onlinecite{PhysRevB.78.014201}], which has been confirmed by STM experiments [\onlinecite{PhysRevLett.101.206802},\onlinecite{Simon:2009eu}]. Importantly, this confirmation proved that STM can probe the relativistic nature of charge carriers.
\begin{figure}[t]
\includegraphics[trim = 45mm 0mm 10mm 0mm, clip, width=6cm]{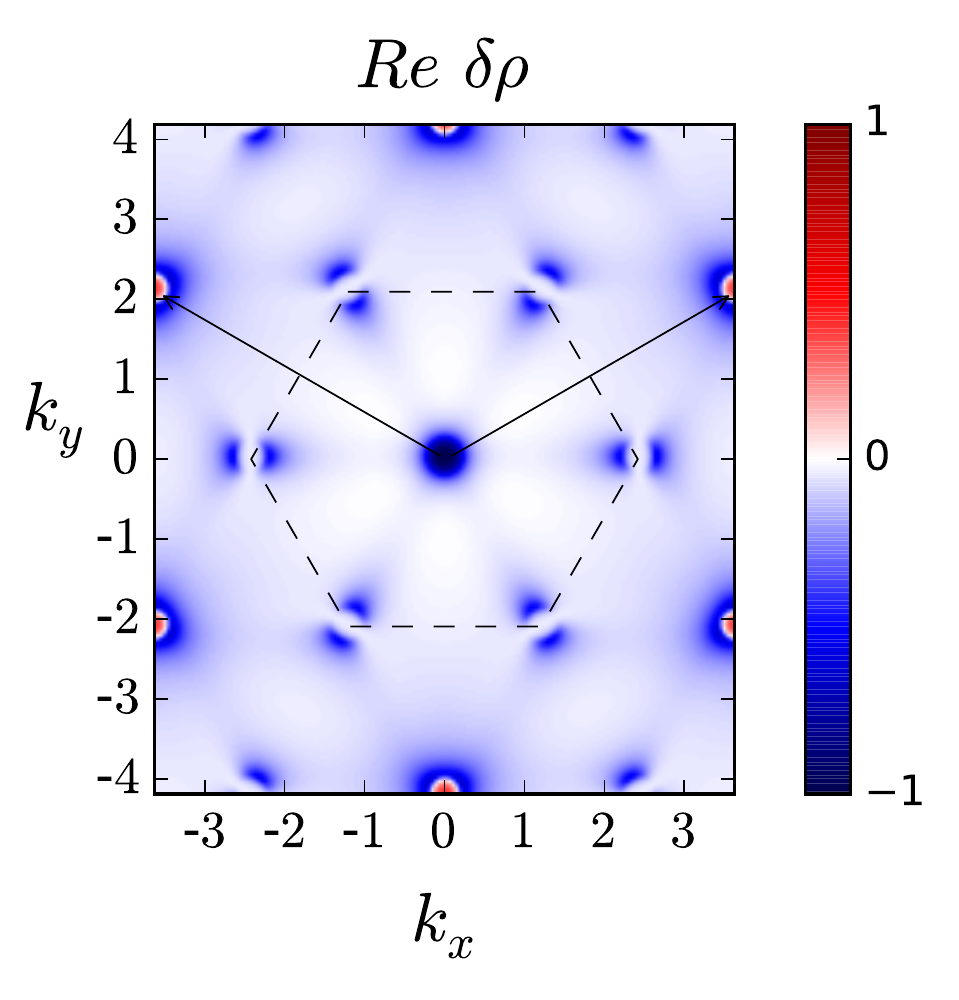}
\caption{\small (Color online) Momentum space representation of $\delta \rho$, the LDOS modulation induced by a localized impurity of magnitude $t$. Only the real part of the LDOS Fourier transform is depicted. The energy is $\omega=0.150t$. The two vectors that span the reciprocal space are depicted by the black arrows. The LDOS modulations which can be connected to the origin by a linear combination of these two vectors refer to scattering processes that take place between equivalent valleys. The LDOS modulations at the corners of the dashed-line-made hexagon arise from scattering processes that take place between two nearest non-equivalent valleys.}
\label{FT STM graphene}
\end{figure} 
\item For scattering processes which take place between two non-equivalent valleys, the momentum space signature depicts a ring with broken circular symmetry (see hexagon corners in Fig. \ref{FT STM graphene}), as reported in [\onlinecite{PhysRevB.86.045444}] and [\onlinecite{PhysRevB.78.014201}]. In the last section, we shall show that, if one could measure the LDOS Fourier transform of the pristine sublattice independently of the one with impurity, then one would directly access the phase $\theta^{\xi}({\bf q})$ of the Bloch spinors, additionally to the iso-energy contours. In other words, it would be possible to know the full Bloch band-structure from the localized impurity scattering at low energy. We will discuss this possibility in the case of rhombohedral multilayer graphene. In this semimetal, the low-energy physics only involves two sublattices which are located at the external layers. Thus their LDOS Fourier transforms could be imaged independently of one another from STM.
\end{itemize}

On top of that, the phase $\theta^{\xi}({\bf q})$ is involved in another feature of the band structure, which is generally known as Berry phase [\onlinecite{Pancharatnam, 1984, shapere1989geometric}]. This is a gauge-invariant geometrical phase picked up by wave functions along an adiabatic cycle. It has been discussed in various fields, such as polarization [\onlinecite{PhysRevB.47.1651},\onlinecite{RevModPhys.66.899}], orbital magnetization [\onlinecite{PhysRevLett.95.137205},\onlinecite{PhysRevLett.112.026402}] and symmetry-protected topological order [\onlinecite{PhysRev.56.317, PhysRevLett.49.405, PhysRevLett.95.146802}], whose topological classification of gapped single-particle Hamiltonians relies on a quantized Berry phase [\onlinecite{schnyder2008classification}]. The importance of the Berry phase in the semiclassical dynamics of Bloch electrons under electromagnetic fields has been reviewed in [\onlinecite{RevModPhys.82.1959}]. This notion plays an important role in graphene [\onlinecite{katsnelson2012graphene}]. Indeed, each nodal point supports a topological characterization which is given by a $\pi$-quantized Berry phase defined as follows
\begin{align}\label{Dirac point Berry phase}
\gamma_{\xi} &= i\oint_{{\cal{C}}^{\xi}_{mn}}d{\bf q} \cdot \langle \Psi_{\pm}({\bf K^{\xi}_{mn}}+{\bf q}) | \nabla_{\bf q} | \Psi_{\pm}({\bf K^{\xi}_{mn}}+{\bf q}) \rangle \notag \\
&= \frac{1}{2} \oint_{{\cal{C}}^{\xi}_{mn}}d{\bf q}\cdot \nabla_{\bf q} \theta^{\xi}({\bf q}) \notag \\
&= \xi \pi ~,
\end{align}
where ${\cal{C}}^{\xi}_{mn}$ is a closed path enclosing once a nodal point in the valley ${\bf K^{\xi}_{mn}}$. Thus, non-equivalent valleys, which are related to one another by the time-reversal symmetry, have opposite Berry phases.
This phase can be probed by applying a magnetic field perpendicularly to the material, which yields closed cyclotron orbits in momentum space. When they enclose a Dirac cone, they are associated to a non-vanishing Berry phase, which leads to a zero-energy Landau level [\onlinecite{katsnelson2012graphene},\onlinecite{PhysRevLett.82.2147},\onlinecite{10.1140/epjb/e2010}]. This results in the half-integer plateaus of the Hall conductivity that have been measured in experiments of anomalous quantum Hall effect in graphene [\onlinecite{novoselov2005two},\onlinecite{zhang2005experimental}]. These observations have confirmed the presence of massless Dirac electrons at low energy. Besides, Eq. (\ref{Dirac point Berry phase}) implies that the integer $\gamma_{\xi}/\pi$ is nothing but the number of times that the phase $\theta^{\xi}({\bf q})$ runs over the interval $[-\pi,+\pi]$, when ${\bf q}$ encloses once the nodal point ${\bf K^{\xi}_{mn}}$ along the path ${\mathcal C}^{\xi}_{mn}$. So it is clear that if the localized impurity scattering is a mean to access $\theta^{\xi}({\bf q})$, then it directly leads to the $\pi$-quantized Berry phase which characterizes the existence of Fermi points in the spectrum.

\subsection{Rhombohedral $N$-layer graphene}
Rhombohedral $N$-layer graphene is an ABC stacking of $N$ graphene layers [\onlinecite{haering1958band}]. Only processes occurring between nearest-neighbor layers are considered. In the framework of a tight-binding approximation, the electronic properties are then obtained from the following Hamiltonian matrix
\begin{align}\label{Hamiltonian matrix}
H({\bf k}) = 
\left( \begin{array}{ccccc} 
H_1({\bf k}) & T_{\perp} & 0 & \cdots & 0  \\
T^{\dagger}_{\perp} & H_{2}({\bf k}) & \ddots & \ddots & \vdots \\
0 & \ddots & \ddots & \ddots & 0\\
\vdots & \ddots & \ddots & \ddots & T_{\perp} \\
0 & \cdots & 0 & T^{\dagger}_{\perp} &  H_{N}({\bf k})
\end{array} \right)~.
\end{align}
The $2\times2$ matrix $H_{n}$ describes the intralayer processes. They are limited to nearest-neighbor hopping. In the sublattice basis $\{A_{n}$, $B_{n}\}$, this matrix is given by
\begin{align}
H_{n}({\bf k}) = 
\left( \begin{array}{cc} 
0 & f({\bf k}) \\
f^{*}({\bf k}) & 0
\end{array} \right)~,
\end {align}
where $f({\bf k})$ has been introduced in Eq. (\ref{Hk graphene}). Regarding the interlayer processes, they are described by
\begin{align}
T_{\perp} = 
\left( \begin{array}{cc} 
0 & 0 \\
t_{\perp} & 0
\end{array} \right)~,
\end {align}
where the hopping amplitude $t_{\perp}$ only couples sublattices B$_{\text{n}}$ and A$_{\text{n}+1}$.

\begin{figure}[t]
\includegraphics[trim = 0mm 0mm 0mm 0mm, clip, width=6cm]{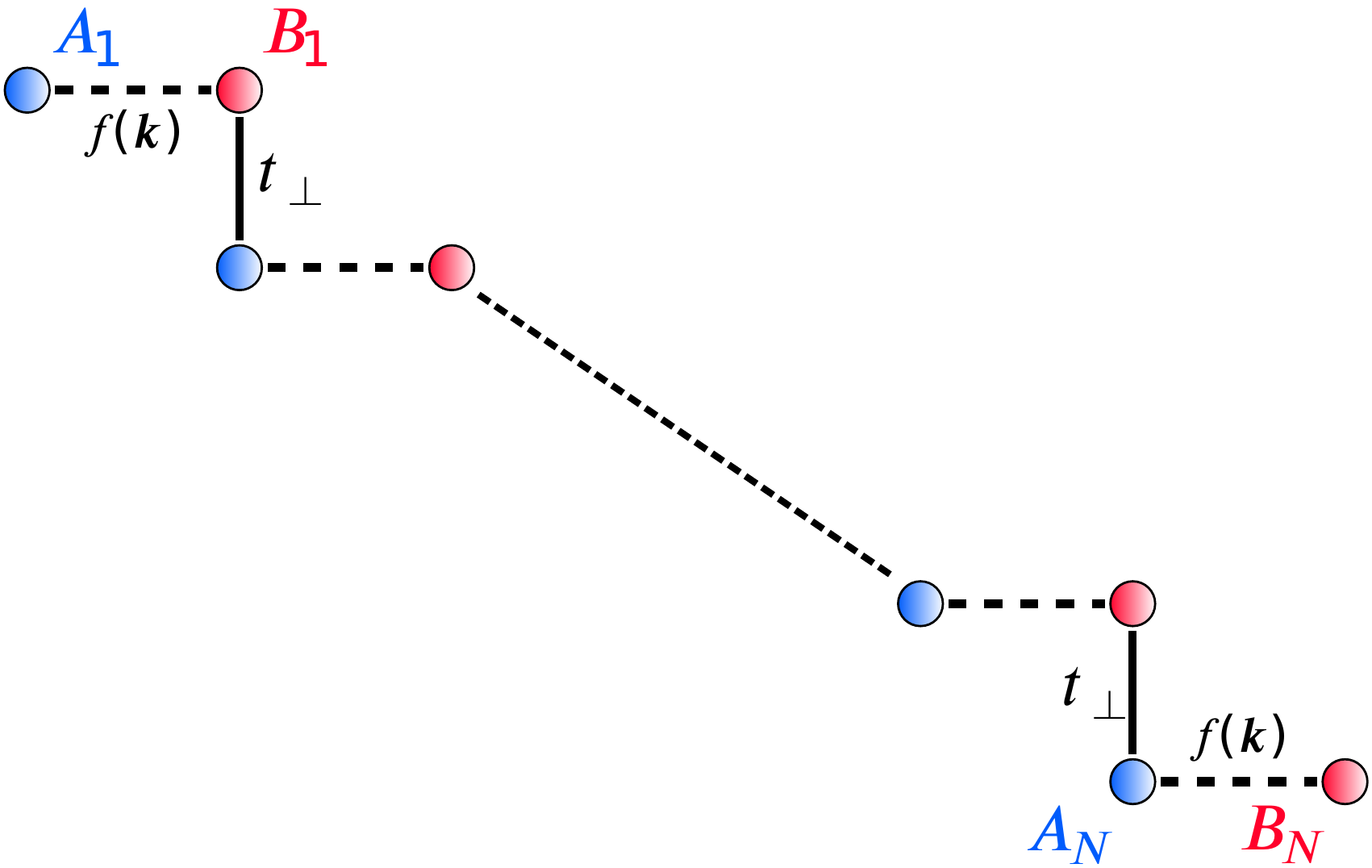}
\caption{\small (Color online) Illustration of the momentum space mapping onto a one-dimensional chain made of $N$ dimers. In this effective system, $f({\bf k})$ and $t_{\perp}$ denote the nearest-neighbor hopping amplitudes.}
\label{Dimer chain}
\end{figure}

As already mentioned in [\onlinecite{PhysRevB.73.245426}], the case of rhombohedral stacking is rather special in the sense that its low-energy physics supports a two-band description. Indeed, the Hamiltonian matrix (\ref{Hamiltonian matrix}) effectively describes a one-dimensional chain made of $N$ dimers with two nearest-neighbor hopping amplitudes, namely $f({\bf k})$ and $t_{\perp}$. This effective model is illustrated in Fig. \ref{Dimer chain}. For such a system with sublattice symmetry, Shockley realized, a long time ago before the boom of topological insulators, that there may exist low-energy edge states within the bulk energy gap [\onlinecite{PhysRev.56.317}]. He found that the existence of the surface states, previously pointed out by Tamm [\onlinecite{tamm1932possible}], depends on the ratio of the two nearest-neighbor hopping parameters. In particular, a surface state does exist at each end of the system when $|f({\bf k})|<|t_{\perp}|$. For the rhombohedral stacking that is considered here, Shockley's criterion guarantees the localization of the electrons in the vicinity of sublattices A$_{1}$ and B$_{\text{N}}$. This is all the more true that the dispersion relation of each graphene layer vanishes, i.e. $|f({\bf k})|\rightarrow 0$. In the limit $f({\bf k})=0$, there exist two zero-energy states that are strictly localized on sublattices A$_{1}$ and B$_{\text{N}}$. Not only does this suggest that the low-energy physics of the rhombohedral stacking reduces to a two-band model [\onlinecite{PhysRevB.73.245426}], but also the two sublattices involved in this model refer to the outer layers of the material. Crucially, this means that their LDOS could be imaged via STM experiments when suspending the material. For example, STM analyses of suspended monolayer and bilayer graphene have been reported in [\onlinecite{C2NR30162H}].

\subsection{Low-energy band structure}
\begin{figure}[t]
\includegraphics[trim = 0mm 0mm 0mm 0mm, clip, width=3.5cm]{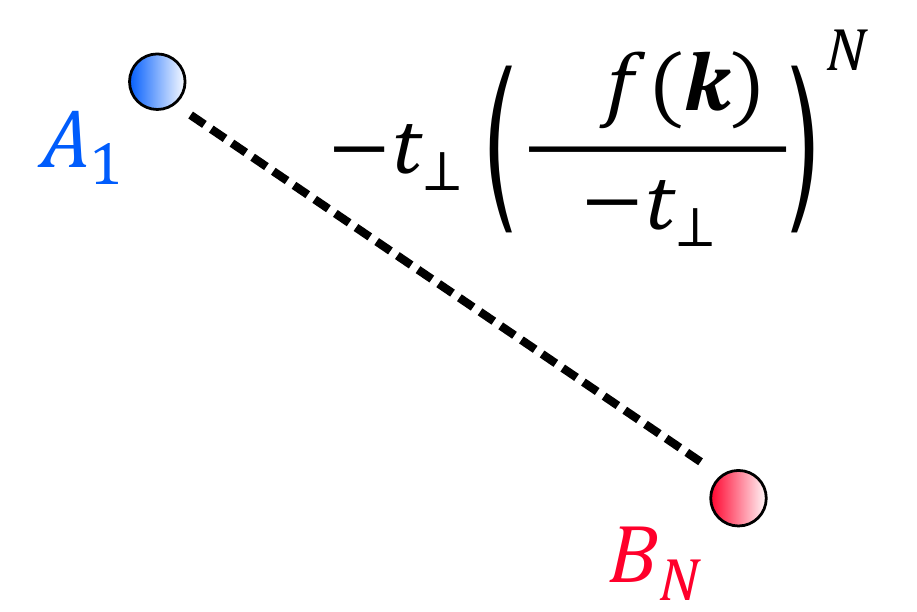}
\caption{\small (Color online) Illustration of the momentum space mapping onto a monomer with renormalized hopping amplitude in the limit $E\ll t_{\perp}$.}
\label{DimerChainTwoBands}
\end{figure}

The band structure of rhombohedral multilayer graphene, first studied in [\onlinecite{haering1958band}], can be obtained from the Hamiltonian matrix (\ref{Hamiltonian matrix}) by solving the following recursive system:
\begin{align}\label{Recursive System}
\left \{
\begin{aligned}
f({\bf k})~ B_{1} &= E~ A_{1} \\
f^{*}({\bf k})~ A_{n-1}+t_{\perp}~ A_{n} &= E~ B_{n-1} \\
t_{\perp}~ B_{n-1}+f({\bf k})~ B_{n} &= E~ A_{n} \\
f^{*}({\bf k})~ A_{N} &= E~ B_{N}
\end{aligned}
\right . ~,
\end{align}
where $E$ denotes the eigenenergy, $A_{n}$ (respectively $B_{n}$) refers to the electronic orbitals of the sublattice A$_{\text{n}}$ (respectively B$_{\text{n}}$), and the index $n$ runs from $2$ up to $N$. In the low-energy limit $E \ll t_{\perp}$, the electronic properties only involve sublattices A$_{1}$ and B$_{\text{N}}$, both located at the outer surfaces of the material. For more details, the reader may refer to Appendix \ref{Appendix Low-energy Band Structure}. The effective band structure is given by
\begin{align}\label{Effective hamiltonian}
\mathcal{H}_{N}({\bf k})= -t_{\perp}
\left( \begin{array}{cc} 
0 & \Big(-\frac{f({\bf k})}{t_{\perp}}\Big)^{N} \\
\Big(-\frac{f^{*}({\bf k})}{t_{\perp}}\Big)^{N} & 0
\end{array} \right)
\end{align}
and is illustrated in Fig. \ref{DimerChainTwoBands}.
The dispersion relation is simply given by
\begin{align}\label{Dispersion relation}
E_{\pm}({\bf k}) \simeq \pm t_{\perp} \Big| \frac{f({\bf k})}{t_{\perp}} \Big|^{N} ~.
\end{align}
So the condition $E\ll t_{\perp}$ implies that the two-band description is effective around the corners of the Brillouin zone, where
\begin{align}\label{Effective hamiltonian}
\mathcal{H}_{N}({\bf K^{\xi}_{mn}}+{\bf q})=t_{\perp}
\left( \begin{array}{cc} 
0 & -\Big(\xi \frac{v_{F}}{t_{\perp}}q~e^{i\theta^{\xi}_{mn}({\bf q})} \Big)^{N} \\
-\Big(\xi \frac{v_{F}}{t_{\perp}}q~e^{-i\theta^{\xi}_{mn}({\bf q})} \Big)^{N} & 0
\end{array} \right) ~
\end{align}
and the dispersion relation behaves as
\begin{align}\label{Dispersion relation}
E_{\pm}({\bf K^{\xi}_{mn}}+{\bf q}) \simeq \pm t_{\perp} \Big(\frac{v_{F}}{t_{\perp}}q\Big)^{N} ~.
\end{align}
The valence and conduction bands touch each other right at the Brillouin-zone corners, so that the ABC stacking is semimetallic as well. Again, the stability of these nodal points at zero energy relies on the inversion and time-reversal symmetries [\onlinecite{PhysRevB.75.155424}]. As a consequence of these two symmetries, the Bloch eigenstates can be written as
\begin{align}\label{Bloch eigenstates}
\Psi_{\pm}({\bf K^{\xi}_{mn}}+{\bf q}) \simeq \frac{1}{\sqrt{2}}
\left( \begin{array}{c} 
1 \\
\mp \xi^{N} e^{-i N{\bf K_{mn}^{\xi}}\cdot{\bf d_{3}}}e^{-iN\theta^{\xi}({\bf q})}
\end{array} \right) ~,
\end{align}
where the momentum dependence is simply encoded into the phase $\theta^{\xi}({\bf q})$ and appears as a phase factor between the two sublattice components of the Bloch spinor, similarly to the case of graphene. The definition of the Berry phase given in Eq. (\ref{Dirac point Berry phase}) implies that each valley ${\bf K^{\xi}_{mn}}$ is characterized by a quantized geometrical phase of $\xi N \pi$.

\section{Localized Impurity scattering}
\label{Localized Impurity scattering}

\subsection{$T$-matrix approach}

\begin{figure}[t]
\centering
\includegraphics[trim = 0mm 0mm 0mm 0mm, clip, width=7cm]{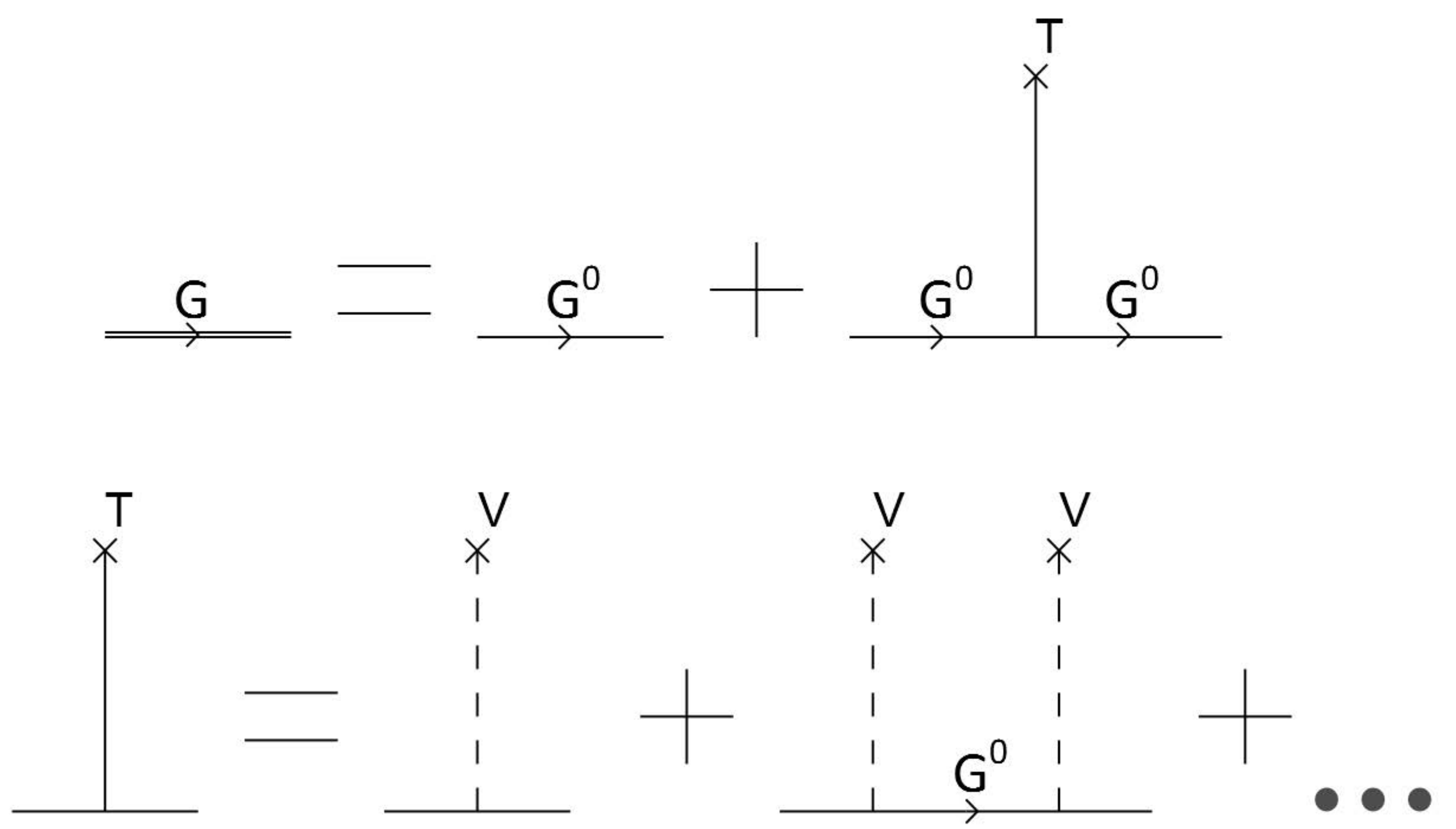}
\caption{\small Diagrammatic representation of the $T$-matrix approach.}
\label{Diagrams}
\end{figure}

When free electrons are described by a given Hamiltonian matrix $H({\bf k })$ in momentum space, their bare Green function $G^{(0)}$ can be defined in the following way:
\begin{align}\label{BareGreenFunctionDefinition}
G^{(0)}({\bf k}, \omega) = [\omega I-H({\bf k})]^{-1} ~,
\end{align}
where $I$ is the identity matrix and $\omega$ denotes the frequency. Then the electron scattering through a localized impurity, whose potential is simulated by $V~\delta({\bf r})$, can be apprehended in the framework of a $T$-matrix approach. The latter consists in a perturbative expansion of the Green function in all orders in the impurity scattering, as illustrated in Fig. \ref{Diagrams}. Because the impurity potential is localized, the infinite sum of diagrams turns out to be a geometric series and can be performed exactly. The momentum space expression of the $T$-matrix is
\begin{align}\label{T-matrix definition}
T(\omega)=\Big(1-V~ \int_{BZ}G^{(0)}({\bf k}, \omega)\Big)^{-1}~ V ~.
\end{align}
The integral runs over the whole Brillouin zone (BZ) and the $T$-matrix is momentum independent.

The perturbed Green function, which describes the system in the presence of the localized impurity, satisfies
\begin{align}\label{Perturbed Green function definition}
{\cal{G}}({\bf r_{1}}, {\bf r_{2}}, \omega) &= G^{(0)}({\bf r_{1}}-{\bf r_{2}}, \omega) \notag \\
&+ G^{(0)}({\bf r_1}, \omega)~T(\omega)~G^{(0)}(-{\bf r_2}, \omega)~.
\end{align}
The imaginary part of its trace defines the local density of states (LDOS)
\begin{align}\label{LDOS definition}
\rho({\bf r}, \omega) =-\frac{1}{\pi}~\Imag~\big[\Tr~{\cal{G}}({\bf r}, {\bf r}, \omega) \big] ~,
\end{align}
an observable quantity that reveals the pattern of the quantum interferences induced by the impurity. Its momentum space representation is given by
\begin{align}\label{Momentum Space LDOS}
\rho({\bf k},\omega) = \frac{i}{2\pi} \Tr \int_{BZ}d{\bf q}~ \Big[ {\cal{G}}({\bf k}+{\bf q}, {\bf q}, \omega)-{\cal{G}}^{*}({\bf q}, {\bf k}+{\bf q}, \omega) \Big] ~.
\end{align}
Importantly, note that the quantity we are interested in thereafter is not the LDOS itself, but rather the correction to the LDOS of the unperturbed system. It is defined from the correction to the bare Green function, namely
\begin{align}\label{GF correction}
\delta{\cal G}({\bf r_{1}}, {\bf r_{2}}, \omega) = G^{(0)}({\bf r_1}, \omega)~T(\omega)~G^{(0)}(-{\bf r_2}, \omega) ~,
\end{align}
as
\begin{align}\label{LDOS correction}
\delta \rho({\bf r}, \omega)=-\frac{1}{\pi}~\Imag~ \big[ \Tr~ \delta{\cal G}({\bf r}, {\bf r}, \omega) \big] ~.
\end{align}
Nevertheless it is still referred to as LDOS along the subsequent lines.

\begin{figure}[t]
\centering
$\begin{array}{cc}
\includegraphics[trim = 30mm 0mm 10mm 0mm, clip, width=4.2cm]{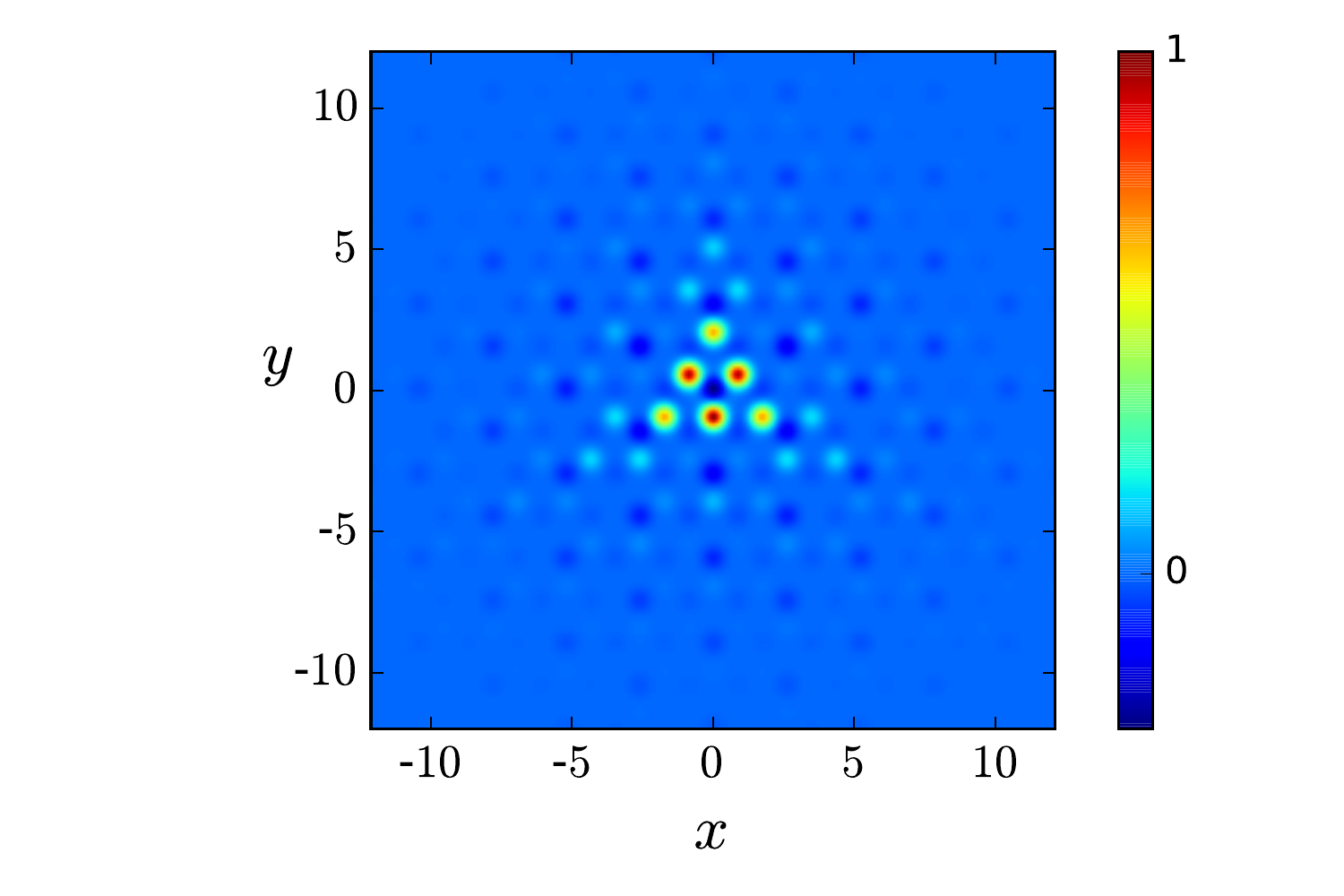}&
\includegraphics[trim = 30mm 0mm 10mm 0mm, clip, width=4.2cm]{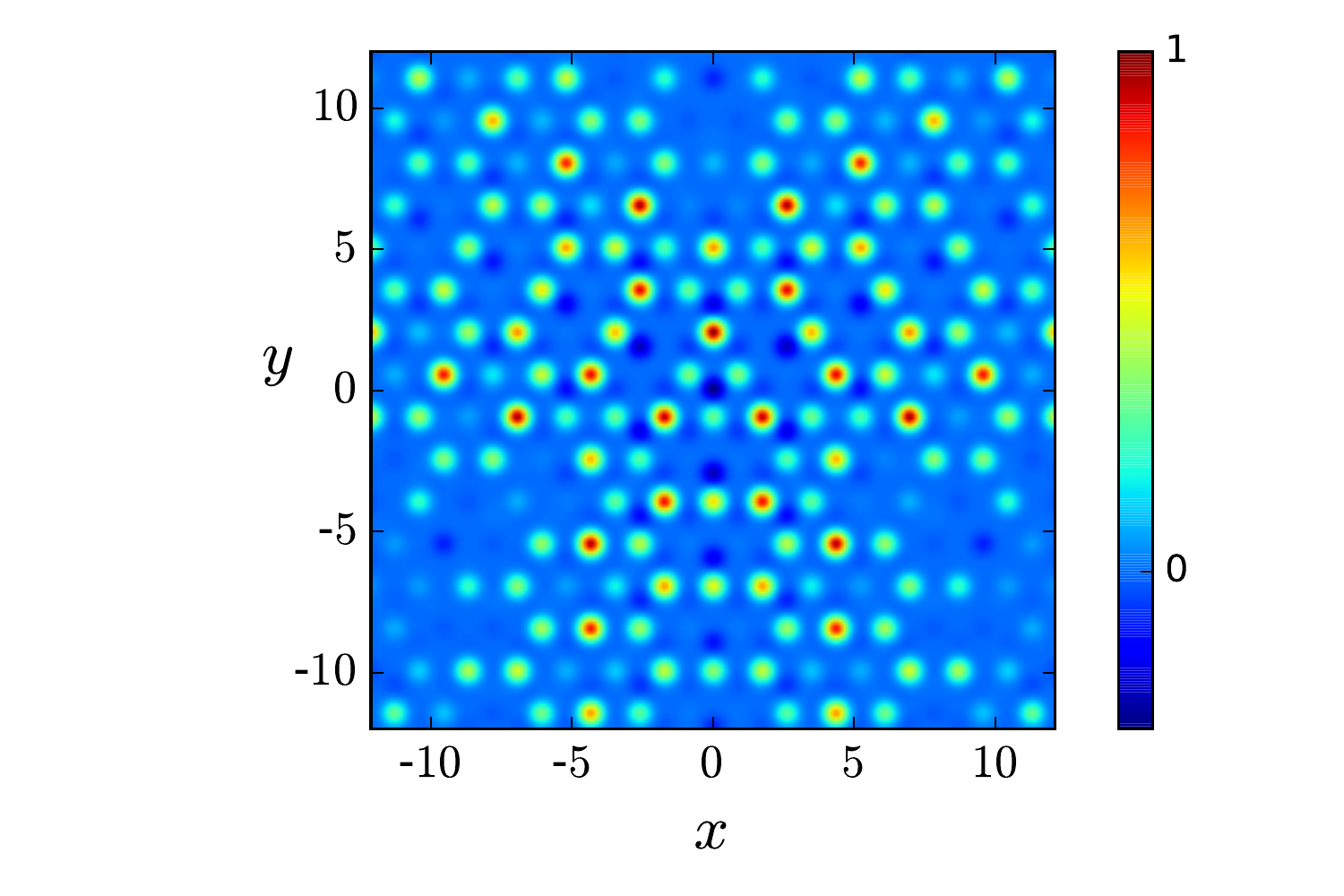}\\
\includegraphics[trim = 30mm 0mm 10mm 0mm, clip, width=4.2cm]{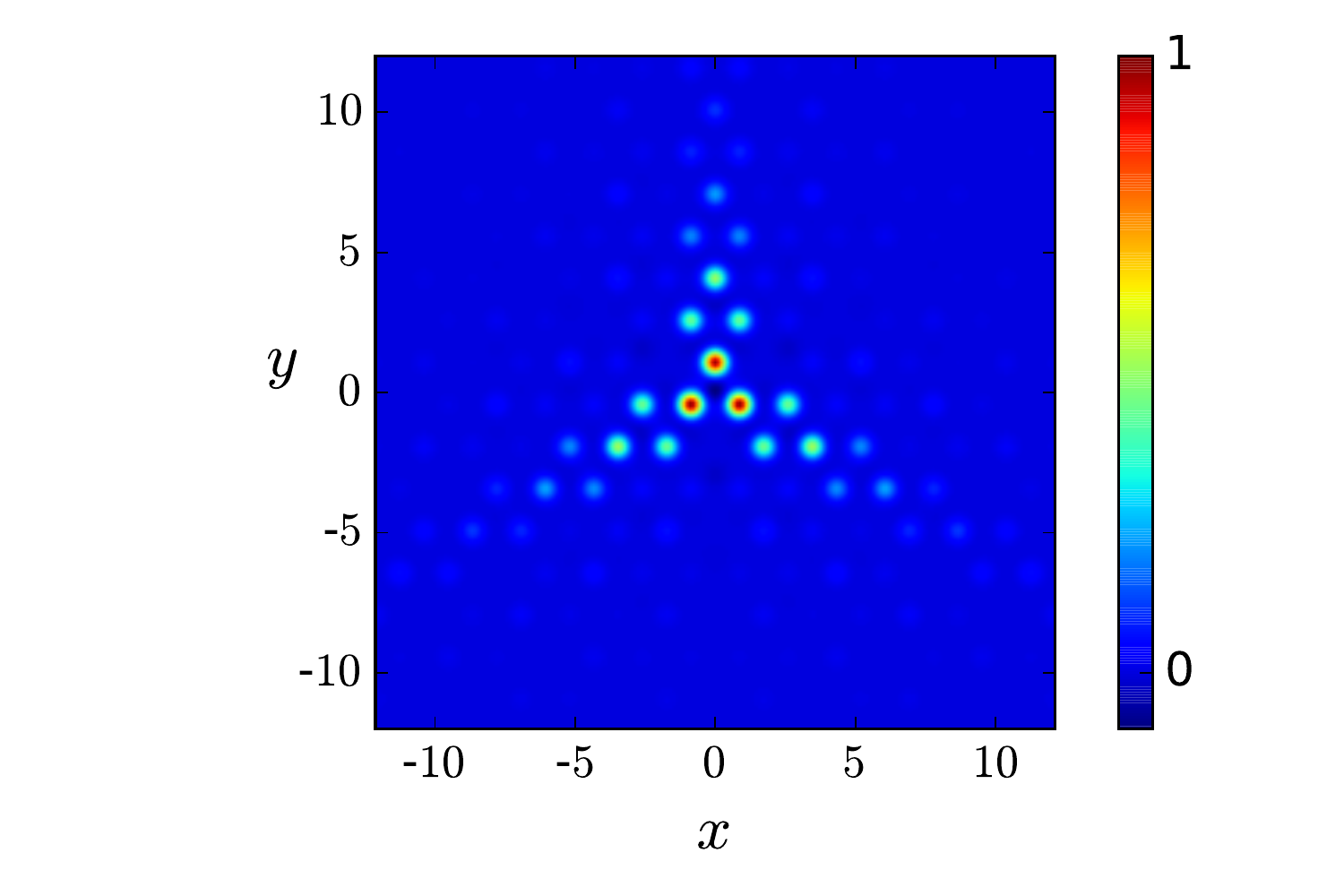}&
\includegraphics[trim = 28mm 0mm 10mm 0mm, clip, width=4.2cm]{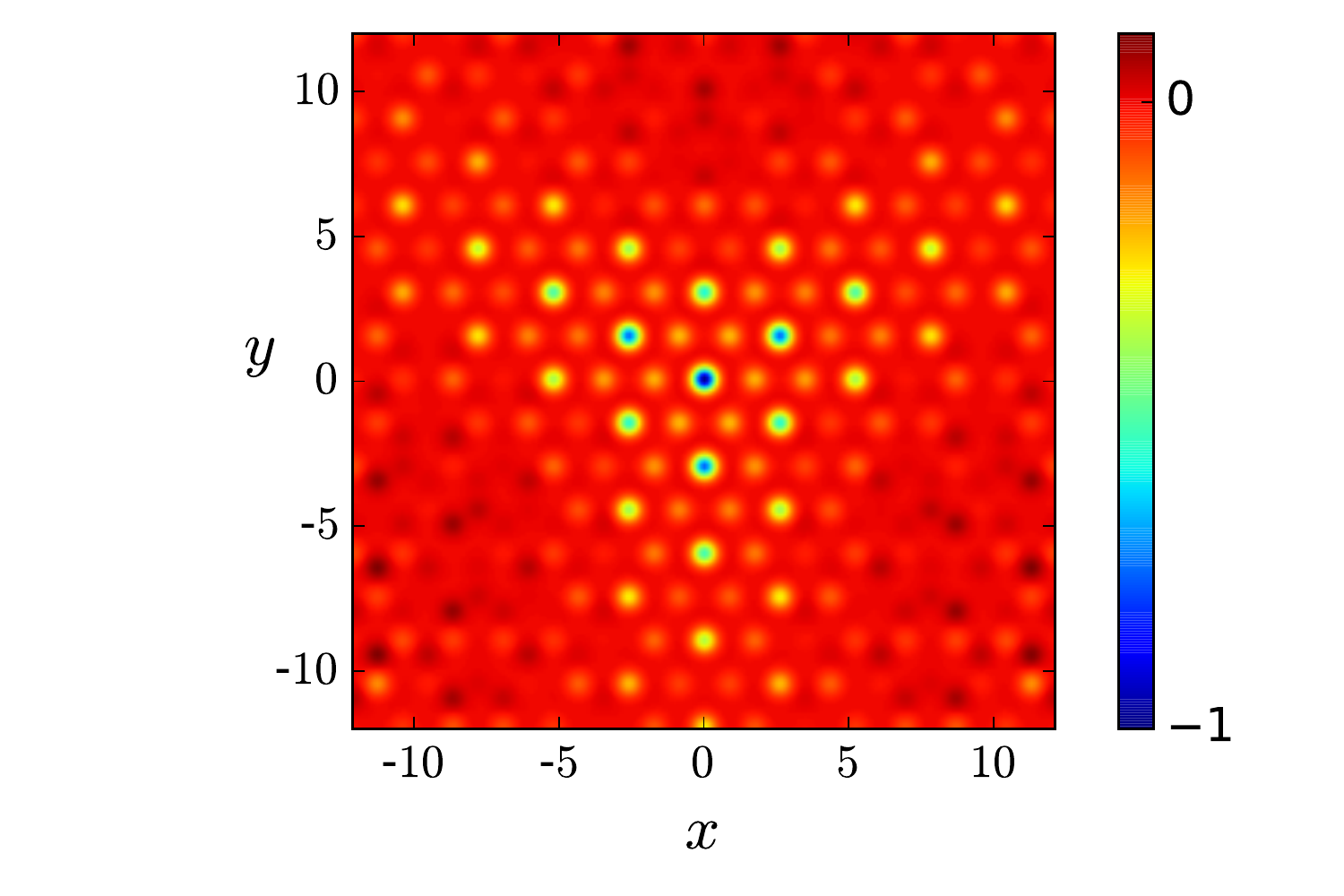}\\
\end{array}$
\caption{\small (Color online) Simulations of the LDOS modulations at the opposite surfaces of ABC trilayer graphene ($N=3$) for $\omega=0.038t$. The surface of the impurity is shown in the left-hand column. The impurity potential, whose magnitude satisfies $V_{0} \gg t$, is localized at the origin of the surface which corresponds either to sublattice A$_{1}$ (top), or to sublattice B$_{1}$ (bottom). In both cases, the LDOS modulations induced on the opposite pristine surface are depicted in the right-hand column. Distances are given in units of the lattice constant $a_{0}$.}
\label{LDOS}
\end{figure}

For instance, suppose that a localized impurity is introduced on a given surface of rhombohedral multilayer graphene, say the one that is made of sublattices A$_{1}$ and B$_{1}$. In momentum space, it is simulated by the following $N\times N$ matrix:
\begin{align}\label{MS Impurity Hamiltonian}
V = 
\left( \begin{array}{cccc} 
v & 0 & \cdots & 0  \\
0 & 0 & \ddots & \vdots \\
\vdots & \ddots & \ddots & 0 \\
0 & \cdots & 0 & 0
\end{array} \right)~,
\end{align}
where the $2\times2$ matrix $v$ describes the impurity potential in the sublattice basis $\{ A_{1}, B_{1} \}$. Numerically, it is then sufficient to compute the LDOS accordingly to the $T$-matrix formalism described above, when substituting the Hamiltonian matrix (\ref{Hamiltonian matrix}) into Eq. (\ref{BareGreenFunctionDefinition}). As an example, Fig. \ref{LDOS} illustrates the impurity-induced LDOS at the surfaces of ABC trilayer graphene, which still shows a threefold rotational symmetry. Note that it is convenient to add a small imaginary part to the energy in the definition of the bare Green functions for numerical simulations. Thus, the quasiparticles they describe have a finite lifetime. This is achieved via the substitution $G^{0}({\bf k},\omega) \mapsto G^{0}({\bf k},\omega+i\delta)$, with $\delta=0.05$ for the numerical results presented here. We also choose $t_{\perp}=0.3t$ as interlayer hopping amplitude.

Finally, note that the Fourier transform of the LDOS, which can be obtained numerically from Eq. (\ref{Momentum Space LDOS}), is not discussed for the moment. It will be the purpose of the last section.

\subsection{Low-energy description of the impurity problem}

It turns out that one can also get more insight at low energy, which means when the energy of the system is small compared to the interlayer hopping amplitude $t_{\perp}$. In this limit, the impurity problem can be tackled analytically and independently of the impurity magnitude, which can then simulate realistic localized scatters. For example, this can describe adatoms deposited onto the surface, either on sublattice A$_{1}$ or on sublattice B$_{1}$. But it can also refer to a bulk impurity, i.e. an impurity located on an intermediate layer in between the the two surfaces of the material. This may describe structural point defects like vacancies created by irradiation. Those three possible cases are discussed below.

\subsubsection{Localized impurity on sublattice A$_{1}$}

If the localized impurity is located on the sublattice A$_{1}$ and simulated by the potential $V_{0}\delta({\bf r})$, then the matrix $v$ in Eq. (\ref{MS Impurity Hamiltonian}) has the following form
\begin{align}\label{Surface Impurity Matrix A1}
v = 
\left( \begin{array}{cc} 
V_{0} & 0 \\
0 & 0 \\
\end{array} \right)~.
\end{align}
The spectrum of $H({\bf k})+V$ is obtained from the following recursive system
\begin{align}\label{Recursive System Impurity A1}
\left \{
\begin{aligned}
V_{0}~A_{1}+f({\bf k})~ B_{1} &= E~ A_{1} \\
f^{*}({\bf k})~ A_{n-1}+t_{\perp}~ A_{n} &= E~ B_{n-1} \\
t_{\perp}~ B_{n-1}+f({\bf k})~ B_{n} &= E~ A_{n} \\
f^{*}({\bf k})~ A_{N} &= E~ B_{N}
\end{aligned}
\right .~,
\end{align}
where $E$ denotes the eigenenergy, $A_{n}$ (respectively $B_{n}$) refers to the electronic orbitals of sublattice A$_{\text{n}}$ (respectively B$_{\text{n}}$), and $n$ runs from $2$ up to $N$. Following the derivation given in Appendix \ref{Appendix Low-energy Band Structure}, one finds that, in the limit $E\ll t_{\perp}$, the impurity problem still supports a two-band description given by
\begin{align}\label{Impurity Problem Description A1}
\mathcal{H}_{N}({\bf k})+
\left( \begin{array}{cc} 
V_{0} & 0 \\
0 & 0 \\
\end{array} \right) ~,
\end{align}
where $\mathcal{H}_{N}({\bf k})$ is the matrix already introduced in Eq. (\ref{Effective hamiltonian}) and written in the sublattice basis $\{ A_{1}, B_{N} \}$.

\begin{figure}[t]
$\begin{array}{ccc}
\includegraphics[trim = 4mm 0mm 0mm 0mm, clip, width=3.2cm]{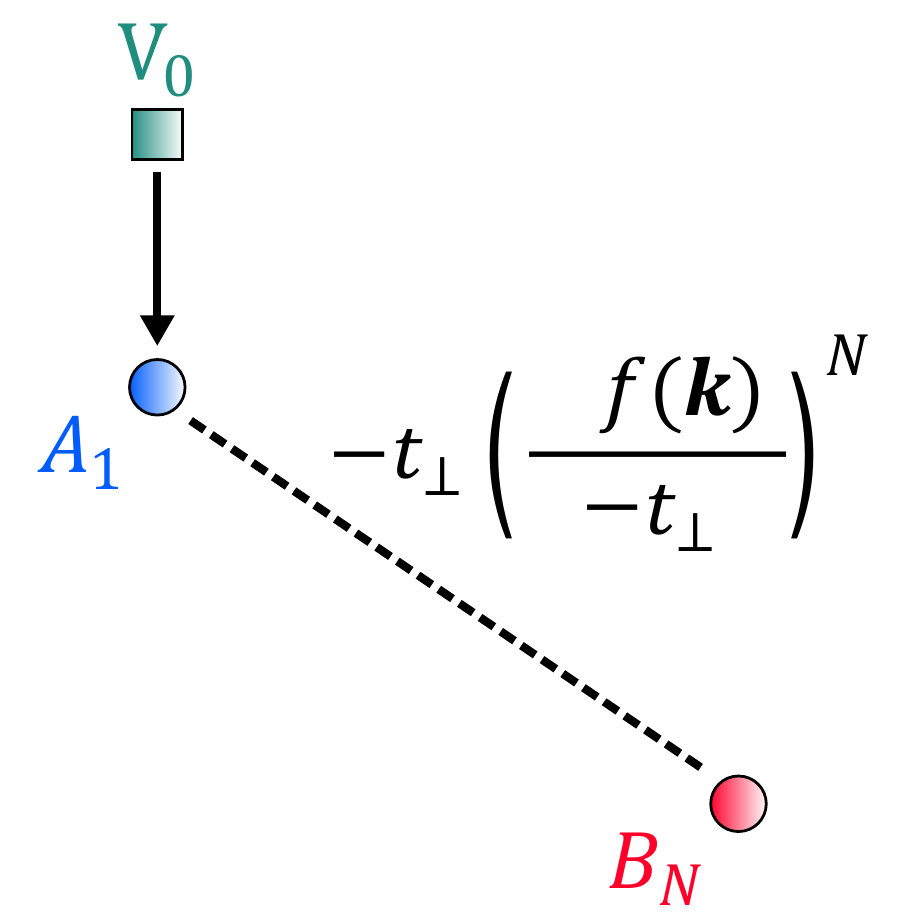}&~~~~&
\includegraphics[trim = 0mm 0mm 0mm 0mm, clip, width=4.5cm]{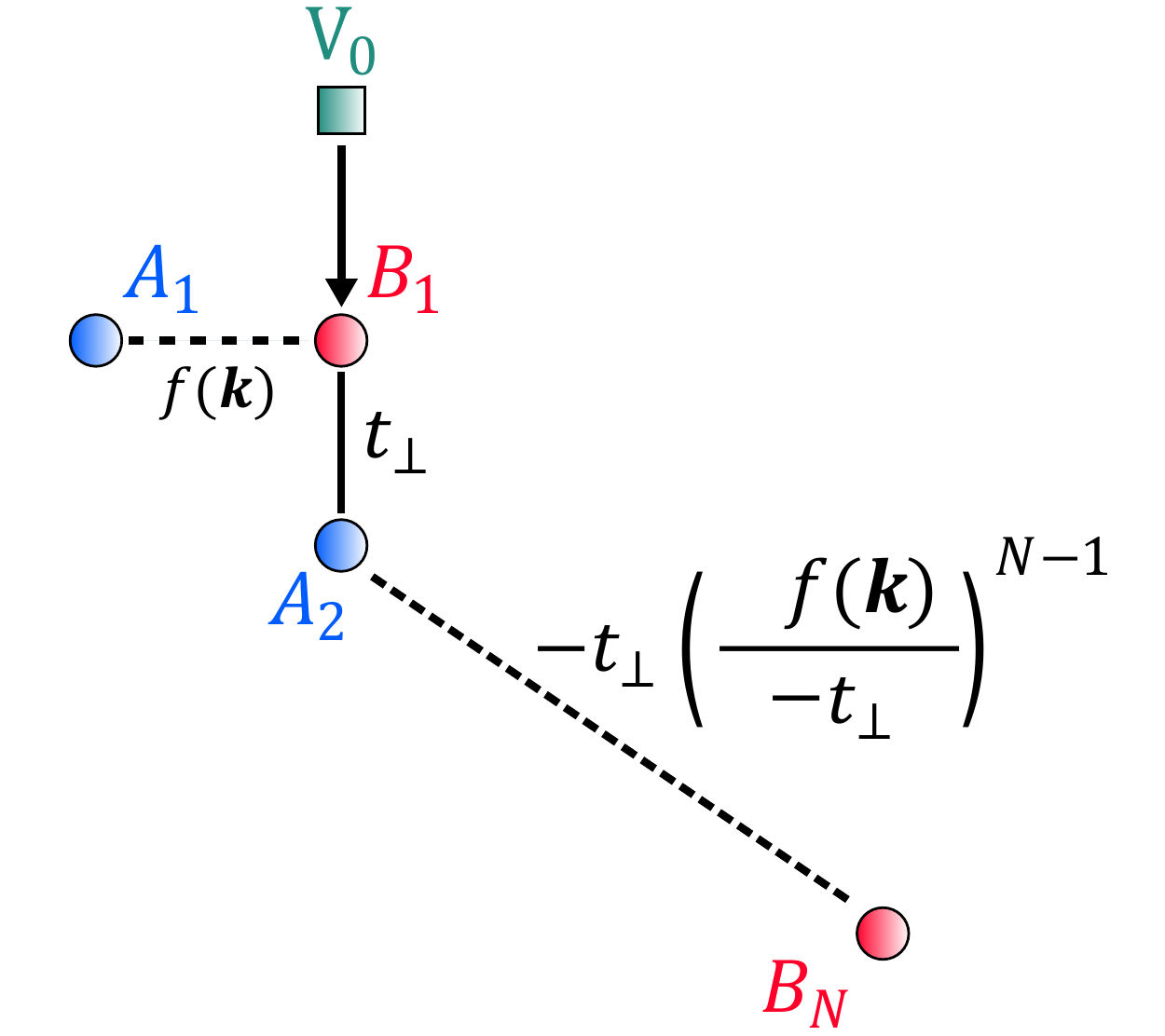}\\
\end{array}$
\caption{\small (Color online) Momentum space representation of the impurity problem in the limit $E\ll t_{\perp}$, when the scatterer, whose potential is denoted $V_{0}$,  is located on the surface sublattice A$_{1}$ (left) or B$_{1}$ (right).}
\label{Momentum Representation Impurity Problem}
\end{figure}

Importantly, this effective two-band picture of the impurity problem only involves sublattices A$_{1}$ and B$_{\text{N}}$, as illustrated in Fig. \ref{Momentum Representation Impurity Problem}. As they are both located at the outer surfaces of the material, they can be probed by STM. On top of that, the description holds regardless of the magnitude of $V_{0}$.

\subsubsection{Localized impurity on sublattice B$_{1}$}

If the localized impurity is located on sublattice B$_{1}$ and is still simulated by the potential $V_{0}\delta({\bf r})$, then
\begin{align}\label{Surface Impurity Matrix B1}
v = 
\left( \begin{array}{cc} 
0 & 0 \\
0 & V_{0} \\
\end{array} \right)~.
\end{align}
The spectrum of $H({\bf k})+V$ is obtained from
\begin{align}\label{Recursive System Impurity B1}
\left \{
\begin{aligned}
f({\bf k})~ B_{1} &= E~ A_{1} \\
f^{*}({\bf k})~ A_{1} + V_{0}~ B_{1} + t_{\perp}~ A_{2} &= E~ B_{1} \\
t_{\perp}~ B_{1}+f({\bf k})~ B_{2} &= E~ A_{2} \\
f^{*}({\bf k})~ A_{n-1}+t_{\perp}~ A_{n} &= E~ B_{n-1} \\
t_{\perp}~ B_{n-1}+f({\bf k})~ B_{n} &= E~ A_{n} \\
f^{*}({\bf k})~ A_{N} &= E~ B_{N}
\end{aligned}
\right .~,
\end{align}
where $E$ denotes the eigenenergy, $A_{n}$ (respectively $B_{n}$) refers to the electronic orbitals of sublattice A$_{\text{n}}$ (respectively B$_{\text{n}}$), and $n$ runs from $3$ up to $N$. Following a similar prescription to the one given in the appendix \ref{Appendix Low-energy Band Structure}, we find that the spectrum mainly relies on
\begin{align}
\left \{
\begin{aligned}
f({\bf k})~ B_{1} &= E~ A_{1} \\
f^{*}({\bf k})~ A_{1} + V_{0}~ B_{1} + t_{\perp}~ A_{2} &= E~ B_{1} \\
t_{\perp}~ B_{1}-t_{\perp}~ \Big( \frac{f({\bf k})}{-t_{\perp}} \Big)^{N-1}~ B_{N} &= E~ A_{2} \\
-t_{\perp}~ \Big( \frac{f^{*}({\bf k})}{-t_{\perp}} \Big)^{N-1}~ A_{2} &= E~ B_{N} \\
\end{aligned}
\right. ~
\end{align}
 in the limit $E\ll t_{\perp}$. So the low-energy physics involves four bands that refer to sublattices A$_{1}$, B$_{1}$, A$_{2}$ and B$_{\text{N}}$, as illustrated in Fig. \ref{Momentum Representation Impurity Problem}. In this sublattice basis, the impurity problem is finally described by
\begin{align}\label{Impurity Problem Description B1}
{H}({\bf k}) \simeq \tilde{\mathcal{H}}_{N}({\bf k}) + 
\left( \begin{array}{cccc} 
v & 0 \\
0 & 0 \\
\end{array} \right) ~,
\end{align}
where
\begin{align}
\tilde{\mathcal{H}}_{N}({\bf k})= t_{\perp}
\left( \begin{array}{cccc} 
0 & \frac{f({\bf k})}{t_{\perp}} & 0 & 0 \\
\frac{f^{*}({\bf k})}{t_{\perp}} & 0 & 1 & 0 \\
0 & 1 & 0 & -\Big(\frac{f({\bf k})}{-t_{\perp}} \Big)^{N-1} \\
0 & 0 & -\Big(\frac{f^{*}({\bf k})}{-t_{\perp}} \Big)^{N-1} & 0 \\
\end{array} \right)~.
\end{align}

Again, $E \ll t_{\perp}$ is the single condition required for this four-band description and no assumption at all has been made about $V_{0}$ which can then simulate realistic localized scatters. Note also that the form of the $2\times2$ matrix $v$, which has been introduced in Eq. (\ref{Surface Impurity Matrix B1}), could be more universal with four non-vanishing components. This would not affect the four-band description of the impurity problem given in Eq. (\ref{Impurity Problem Description B1}).

\subsubsection{Localized impurity in bulk}
Let us now generalize the previous cases to a localized impurity located on sublattice A$_{N_{0}}$ simulated by $V_{0}\delta({\bf r})$, which describes an impurity in the bulk of the material for $1<N_{0}<N$. Note that, choosing B$_{N_{0}}$ as the impurity sublattice would lead to equivalent results.

In the same way as for the two previous cases, one can show that the impurity problem is well characterized by
\begin{align}
\left \{
\begin{aligned}
-t_{\perp}~ \Big( \frac{f({\bf k})}{-t_{\perp}} \Big)^{N_{0}-1}~ B_{N_{0}-1} &= E~ A_{1} \\
-t_{\perp}~ \Big( \frac{f^{*}({\bf k})}{-t_{\perp}} \Big)^{N_{0}-1}~ A_{1} +  t_{\perp}~ A_{N_{0}} &= E~ B_{N_{0}-1} \\
t_{\perp}~ B_{N_{0}-1}+V_{0}A_{N_{0}}-t_{\perp}~ \Big( \frac{f({\bf k})}{-t_{\perp}} \Big)^{N-N_{0}+1}~ B_{N} &= E~ A_{N_{0}} \\
-t_{\perp}~ \Big( \frac{f^{*}({\bf k})}{-t_{\perp}} \Big)^{N-N_{0}+1}~ A_{N_{0}} &= E~ B_{N} \\
\end{aligned}
\right. ~
\end{align}
 in the limit $E\ll t_{\perp}$. So the low-energy physics relies on the following four-band Hamiltonian matrix:
\begin{align}
\frac{\tilde{\mathcal{H}}_{N}({\bf k})}{-t_{\perp}}= 
\left( \begin{array}{cccc} 
0 & \Big( \frac{f({\bf k})}{-t_{\perp}} \Big)^{N_{0}-1} & 0 & 0 \\
\Big( \frac{f^{*}({\bf k})}{-t_{\perp}} \Big)^{N_{0}-1} & 0 & -1 & 0 \\
0 & -1 & 0 & \Big( \frac{f({\bf k})}{-t_{\perp}} \Big)^{N-N_{0}+1} \\
0 & 0 & \Big( \frac{f^{*}({\bf k})}{-t_{\perp}} \Big)^{N-N_{0}+1} & 0 \\
\end{array} \right)~.
\end{align}
which is written in the sublattice basis $\{ A_{1}, B_{N_{0}-1}, A_{N_{0}}, B_{N} \}$.

\subsection{Friedel oscillations in the LDOS}

From the low-energy description introduced above, we can get the analytic expression of the Green functions at large distances, as well as the LDOS they lead to in the framework of the $T$-matrix approach. We choose $t_{\perp}$ and $v_{F}/t_{\perp}$ as new units of energy and length, respectively. So the variable ${\bf q}$, which refers to the momentum, is dimensionless and so is its conjugate variable ${\bf r}$ in real space.

\subsubsection{Localized impurity on sublattice A$_{1}$}

In the sublattice basis $\{ A_{1},B_{N} \}$, the bare Green function can be evaluated as
\begin{align}\label{Low-energy Bare GF Impurity A1}
G^{(0)}({\bf K^{\xi}_{mn}}+{\bf q}, \omega) &\simeq [\omega I-\mathcal{H}_{N}({\bf K^{\xi}_{mn}}+{\bf q})]^{-1} \\
&\simeq \frac{1}{\omega^{2}-q^{2N}} \notag \\
& \times
\left( \begin{array}{cc} 
\omega & -\Big(\xi q e^{i\theta^{\xi}_{mn}(\bf q)} \Big)^{N} \\
-\Big(\xi q e^{-i\theta^{\xi}_{mn}(\bf q)} \Big)^{N} & \omega
\end{array} \right) ~. \notag
\end{align}
The diagonal components depend on the momentum only through the dispersion relation, whereas the off-diagonal ones additionally involve the phase $N\theta^{\xi}_{mn}({\bf q})$ which characterizes the momentum dependence of the Bloch spinors in the vicinity of the valley ${\bf K^{\xi}_{mn}}$. The momentum space representation of the $T$-matrix is then given by
\begin{align}
T(\omega)=
\left( \begin{array}{cc} 
t(\omega) & 0 \\
0 & 0
\end{array} \right)~,
\end{align}
where
\begin{align}
t(\omega)=\frac{V_{0}}{1-V_{0}\int_{BZ}G^{(0)}_{A_{1}A_{1}}({\bf k}, \omega)} ~.
\end{align}
The integral runs over the whole Brillouin zone (BZ) and the T-matrix is momentum independent.

The real space representation of the bare Green function is introduced as
\begin{align}\label{Bare GF Fourier Transform Real Space}
G^{(0)}({\bf r}, \omega) &\simeq \frac{1}{n_{BZ}} \sum_{m,n,\xi} e^{i{\bf K^{\xi}_{mn}}\cdot{\bf r}}~ \int_{{\mathbb R}^{2}}\frac{d^{2}q}{(2\pi)^{2}}~ G^{(0)}({\bf K^{\xi}_{mn}}+{\bf q}, \omega)~e^{i{\bf q}\cdot{\bf r}} ~,
\end{align}
where $n_{BZ}$ is the number of unit cells in momentum space. This factor will not be mentioned anymore in the Green-function expressions for more convenience, but one has to keep in mind that $\sum_{m,n}=1$. From Eq. (\ref{GF correction}), it follows that the large-distance behaviors of $\delta{\cal G}$ diagonal components satisfy
\begin{align}\label{Bare GF Correction A1 Impurity A1}
\delta{\cal G}_{A_{1}A_{1}}({\bf r}, {\bf r}, \omega) &\simeq G^{(0)}_{A_{1}A_{1}}({\bf r}, \omega)~ t(\omega)~ G^{(0)}_{A_{1}A_{1}}(-{\bf r}, \omega) \notag \\
&\simeq - \frac{t(\omega)}{4^{2}N^{2}\omega^{2-4/N}}~\Big(H_{0}(\omega^{\frac{1}{N}}r)\Big)^{2} \\
&\times \sum_{m,n,\xi} \sum_{m',n',\xi'} e^{i{\bf \Delta K}\cdot{\bf r}} \notag
\end{align}
and
\begin{align}\label{Bare GF Correction BN Impurity A1}
\delta{\cal G}_{B_{N}B_{N}}({\bf r}, {\bf r}, \omega) &\simeq G^{(0)}_{B_{N}A_{1}}({\bf r}, \omega)~ t(\omega)~ G^{(0)}_{A_{1}B_{N}}(-{\bf r}, \omega) \notag \\
&\simeq - \frac{t(\omega)}{4^{2}N^{2}\omega^{2-4/N}}~\Big(i^{N}H_{N}(\omega^{\frac{1}{N}}r)\Big)^{2} \\
&\times \sum_{m,n,\xi} \sum_{m',n',\xi'} e^{i{\bf \Delta K}\cdot{\bf r}}~ \Big( \xi \xi'~ e^{-i\Delta \theta ({\bf r})} \Big)^{N} ~, \notag
\end{align}
where ${\bf \Delta K} = {\bf K^{\xi}_{mn}}-{\bf K^{\xi'}_{m'n'}}$ and $\Delta \theta({\bf r}) = \theta^{\xi}_{mn}({\bf r}) - \theta^{\xi'}_{m'n'}(-{\bf r})$. The reader may refer to Appendix \ref{Appendix Bare Green Functions} for more details about the derivations of those expressions. The function $H_{N}$ denotes the $N$-th order Hankel function of the first kind. From the definitions (\ref{Korner Vector}) and (\ref{Phase Definition}), it turns out that ${\bf \Delta K}$ and $\Delta \theta({\bf r})$ depend on the four indices $m$, $m'$, $n$, $n'$ only through their differences, namely $\delta m = m-m'$ and $\delta n = n-n'$. Then one can change variables in the discrete sums according to
\begin{align}
\sum_{m,n,\xi} \sum_{m',n',\xi'}~ \mapsto \sum_{\mu, \nu} \sum_{\delta m, \delta n} \sum_{\xi,\xi'} ~,
\end{align}
where $\mu = m+m'$ and $\nu=n+n'$. The two sums over $\mu$ and $\nu$ are separable from the others and simply satisfy $\sum_{\mu,\nu}=1$. Then, the scattering wave vector ${\bf \Delta K}$ is entirely fixed by the set of the indices $\delta m$, $\delta n$, $\xi$ and $\xi'$. The latter all take opposite integer values and, since ${\bf K_{-m-n}^{-\xi}}=-{\bf K_{mn}^{\xi}}$ and $\theta^{-\xi}_{-m-n}=-\theta^{\xi}_{mn}~[2\pi]$, one can rewrite Eqs. (\ref{Bare GF Correction A1 Impurity A1}) and (\ref{Bare GF Correction BN Impurity A1}) as follows
\begin{align}
\delta{\cal G}_{A_{1}A_{1}}({\bf r}, {\bf r}, \omega) &\simeq - \frac{t(\omega)}{4^{2}N^{2}\omega^{2-4/N}}~\Big(H_{0}(\omega^{\frac{1}{N}}r)\Big)^{2}~~~~~~~ \\
&\times \sum_{\delta m, \delta n}\sum_{\xi,\xi'}  \cos \big({\bf \Delta K}\cdot{\bf r} \big) \notag 
\end{align}
and
\begin{align}
\delta{\cal G}_{B_{N}B_{N}}({\bf r}, {\bf r}, \omega) &\simeq - \frac{t(\omega)}{4^{2}N^{2}\omega^{2-4/N}}~\Big(i^{N}H_{N}(\omega^{\frac{1}{N}}r)\Big)^{2} \\
&\times \sum_{\delta m, \delta n}\sum_{\xi,\xi'} \big(\xi \xi' \big)^{N} \cos \big({\bf \Delta K}\cdot{\bf r} - N\Delta \theta({\bf r}) \big) ~. \notag
\end{align}
In the above sums, every single term refers to a unique scattering process, i.e. to a unique scattering wave vector ${\bf \Delta K}$.

In the limit $\omega^{\frac{1}{N}}r\gg1$, the LDOS associated to a given wave vector ${\bf \Delta K}$ is
\begin{widetext}
\begin{align}\label{LDOS Impurity on A1}
\left \{
\begin{aligned}
\delta\rho_{A_{1}}({\bf r}, \omega) &\simeq -\frac{1}{\pi}~\Imag\Big\{~ i~ \frac{t(\omega)}{4^{2}N^{2}\omega^{2-\frac{3}{N}}}~ \frac{e^{i2\omega^{\frac{1}{N}}r}}{r}~ \cos \big({\bf \Delta K}\cdot{\bf r} \big)~ \Big[ 1 - \frac{i}{4}\frac{1}{\omega^{\frac{1}{N}}r} +... \Big]~\Big\} \\
\delta\rho_{B_{N}}({\bf r}, \omega) &\simeq -\frac{1}{\pi}~\Imag\Big\{~ i~ \frac{t(\omega)}{4^{2}N^{2}\omega^{2-\frac{3}{N}}}~ \frac{e^{i2\omega^{\frac{1}{N}}r}}{r}~ \cos \big({\bf \Delta K}\cdot{\bf r} - N\Delta \theta({\bf r}) \big)~ \Big(\xi \xi' \Big)^{N}~ \Big[ 1 +i~\Big(N^{2}-\frac{1}{4}\Big) \frac{1}{\omega^{\frac{1}{N}}r} +... \Big]~\Big\} \end{aligned}
\right . ~.
\end{align}
\end{widetext}
The asymptotic expansion of the Hankel functions is reminded in Appendix \ref{Appendix Bare Green Functions}. As a result, the localized impurity induces Friedel oscillations in the LDOS on the two opposite surfaces of the material. These long-range quantum interferences exhibit a conventional algebraic decay at large distances on both surfaces. Before evaluating these oscillations induced by an impurity located on sublattice B$_{1}$, we would like to point out the respective contributions of the dispersion relation and the Bloch eigenstates to these quantum interferences in the two-band description. 

On the one hand, it turns out that only the dispersion relation is responsible for the algebraic decay of these oscillations. This can be seen explicitly for the sublattice A$_{1}$. For this sublattice indeed, the LDOS only involves the diagonal components of the bare Green function in Eq. (\ref{Low-energy Bare GF Impurity A1}) which, themselves, only depend on the dispersion relation introduced in Eq. (\ref{Dispersion relation}). Although it is not as explicit for the sublattice B$_{\text{N}}$, one can check from Eq. (\ref{Bare GF Correction BN Impurity A1}) that this is also the dispersion relation that is responsible for the Hankel function $H_{N}$ which yields the $1/r$ decay in the LDOS. For example, a semi-Dirac dispersion relation, i.e. a dispersion that scales linearly with the momentum in one direction and quadratically in the orthogonal one, would lead to $1/\sqrt{r}$ decaying oscillations [\onlinecite{PhysRevB.87.245413}].

On the other hand, the phase $\theta^{\xi}({\bf q})$, which characterizes the momentum dependence of the Bloch spinors, is not involved in the interferences on the sublattice of the impurity, namely A$_{1}$. In the two-band description of the impurity problem, this phase manifests itself solely through the LDOS of the pristine sublattice B$_{\text{N}}$ via
\begin{align}
N\Delta \theta ({\bf r}) &= N\theta^{\xi}_{mn}({\bf r}) - N\theta^{\xi'}_{m'n'}(-{\bf r}) \notag \\
&= N{\bf \Delta K}\cdot{\bf d_{3}} + N(\xi-\xi')\theta_{\bf r} - N\xi' \pi ~,
\end{align}
where we have used the relation (\ref{Phase symmetry relation}). For scattering processes that couple two equivalent valleys ($\xi=\xi'$), this is likely to yield a reduction of the long-range oscillations to a $1/r^{2}$ algebraic decay. It can be seen from Eq. (\ref{LDOS Impurity on A1}) that, when
\begin{align}\label{D3 Criterion}
N{\bf \Delta K}\cdot{\bf d_{3}} &= N(\delta m+\delta n) \frac{2\pi}{3} \notag \\
&= 0~[2\pi] ~,
\end{align}
the $1/r$ contributions of the LDOS of the two sublattices satisfy
\begin{align}\label{Antiphase}
\delta\rho_{B_{N}}({\bf r}, \omega) = (-1)^{N}~ \delta\rho_{A_{1}}({\bf r}, \omega) ~.
\end{align}
The two LDOSs are in antiphase if there is an odd number of layers stacked in the material. Then the sum of the two contributions vanishes, and one has to consider the next leading order which is responsible for a $1/r^{2}$ decay in the LDOS. Nevertheless, summing the two contributions is not relevant experimentally, except if the two sublattices A$_{1}$ and B$_{\text{N}}$ belong to the same surface, which only happens in monolayer graphene ($N=1$) [\onlinecite{PhysRevLett.100.076601},\onlinecite{PhysRevLett.101.206802}].

\subsubsection{Localized impurity on sublattice B$_{1}$}

In the sublattice basis $\{ A_{1},B_{1},A_{2},B_{N} \}$, the bare Green function satisfies
\begin{align}\label{Low-energy Bare GF Impurity B1}
&G^{(0)}({\bf K^{\xi}_{mn}}+{\bf q}, \omega) \simeq [\omega I-\tilde{\mathcal{H}}_{N}({\bf K^{\xi}_{mn}}+{\bf q})]^{-1} \\
&\simeq \frac{1}{\omega-q^{2N}}
\left( \begin{array}{cccc} 
.... & -\xi q e^{i\theta^{\xi}_{mn}(\bf q)}~(\omega^{2}-q^{2(N-1)}) & .... & ....  \\
.... & \omega~(\omega^{2}-q^{2(N-1)}) & .... & .... \\
.... & .... & .... & .... \\
.... & -\Big( \xi q e^{-i\theta^{\xi}_{mn}(\bf q)} \Big)^{N-1}\omega & .... & ....
\end{array} \right) \notag
\end{align}
in the vicinity of any valley ${\bf K^{\xi}_{mn}}$. In the above expression, we have only mentioned explicitly the components that turn out to be useful when evaluating the LDOS at the outer surfaces of the material. The momentum space representation of the $T$-matrix is then given by
\begin{align}
T(\omega)=
\left( \begin{array}{cccc}
0& 0 & 0 & 0 \\
0& t(\omega) & 0 & 0 \\
0& 0 & 0 & 0 \\
0& 0 & 0 & 0
\end{array} \right)~,
\end{align}
where
\begin{align}
t(\omega)=\frac{V_{0}}{1-V_{0}\int_{BZ}G^{(0)}_{B_{1}B_{1}}({\bf k}, \omega)} ~,
\end{align}
so that the $T$-matrix is momentum independent. From the real-space representation of the bare Green function defined in Eq. (\ref{Bare GF Fourier Transform Real Space}), it follows that
\begin{align}
\delta{\cal G}_{A_{1}A_{1}}({\bf r}, {\bf r}, \omega) &\simeq G^{(0)}_{A_{1}B_{1}}({\bf r}, \omega)~ t(\omega)~ G^{(0)}_{B_{1}A_{1}}(-{\bf r}, \omega) \notag \\
&\simeq -\frac{t(\omega)~\omega^{\frac{2}{N}}}{4^{2}N^{2}}~ \Big( i H_{1}(\omega^{\frac{1}{N}}r) \Big)^{2} \\
&\times \sum_{m,n,\xi} \sum_{m',n',\xi'} e^{i{\bf \Delta K}\cdot{\bf r}}~\xi \xi'~ e^{i\Delta \theta ({\bf r})} ~, \notag
\end{align}
\begin{align}
\delta{\cal G}_{B_{1}B_{1}}({\bf r}, {\bf r}, \omega) &\simeq G^{(0)}_{B_{1}B_{1}}({\bf r}, \omega)~ t(\omega)~ G^{(0)}_{B_{1}B_{1}}(-{\bf r}, \omega) \\
&\simeq -\frac{t(\omega)~ \omega^{2}}{4^{2}N^{2}}~\Big( H_{0}(\omega^{\frac{1}{N}}r) \Big)^{2} \sum_{m,n,\xi} \sum_{m',n',\xi'} e^{i{\bf \Delta K}\cdot{\bf r}} ~, \notag
\end{align}
\begin{align}
\delta{\cal G}_{B_{N}B_{N}}({\bf r}, {\bf r}, \omega) &\simeq G^{(0)}_{B_{N}B_{1}}({\bf r}, \omega)~ t(\omega)~ G^{(0)}_{B_{1}B_{N}}(-{\bf r}, \omega) \notag \\
&\simeq -\frac{\omega^{\frac{2}{N}}}{4^{2}N^{2}}~ \Big( i^{N-1} H_{N-1}(\omega^{\frac{1}{N}}r) \Big)^{2} \\
&\times \sum_{m,n,\xi} \sum_{m',n',\xi'} e^{i{\bf \Delta K}\cdot{\bf r}}~ \Big( \xi \xi'~ e^{-i\Delta \theta ({\bf r})}\Big)^{N-1} ~. \notag
\end{align}
Details about the derivations of those expressions may be found in Appendix \ref{Appendix Bare Green Functions}. 

Following the prescription already detailed in the case of a localized impurity on sublattice A$_{1}$, one finds that, for a given scattering wave vector ${\bf \Delta K}$, the LDOS behaves as
\begin{widetext}
\begin{align}\label{LDOS Impurity on B1}
\left \{
\begin{aligned}
\delta\rho_{A_{1}}({\bf r}, \omega) &\simeq -\frac{1}{\pi}~\Imag\Big\{~ i~ \frac{t(\omega)}{4^{2}N^{2}\omega^{-\frac{1}{N}}}~ \frac{e^{i2\omega^{\frac{1}{N}}r}}{r}~ \cos \big({\bf \Delta K}\cdot{\bf r} + \Delta \theta({\bf r}) \big)~\xi \xi' ~ \Big[ 1 - i\frac{3}{4}\frac{1}{\omega^{\frac{1}{N}}r} +... \Big]~\Big\} \\
\delta\rho_{B_{1}}({\bf r}, \omega) &\simeq -\frac{1}{\pi}~~\Imag\Big\{~ i~ \frac{t(\omega)}{4^{2}N^{2}\omega^{\frac{1}{N}-2}}~ \frac{e^{i2\omega^{\frac{1}{N}}r}}{r}~ \cos \big({\bf \Delta K}\cdot{\bf r} \big)~ \Big[ 1 - i\frac{1}{4}\frac{1}{\omega^{\frac{1}{N}}r} +... \Big]~\Big\} \\
\delta\rho_{B_{N}}({\bf r}, \omega) &\simeq -\frac{1}{\pi}~\Imag\Big\{~ i~ \frac{t(\omega)}{4^{2}N^{2}\omega^{-\frac{1}{N}}}~ \frac{e^{i2\omega^{\frac{1}{N}}r}}{r}~ \cos \big({\bf \Delta K}\cdot{\bf r} - (N-1)\Delta \theta({\bf r}) \big)~ \Big(\xi \xi' \Big)^{N-1}~ \Big[ 1 +i~\Big((N-1)^{2}-\frac{1}{4}\Big) \frac{1}{\omega^{\frac{1}{N}}r} +... \Big]~\Big\}
\end{aligned}
\right .
\end{align}
\end{widetext}
at large distances. The localized impurity induces Friedel oscillations in the LDOS of the two external material surfaces. These quantum interferences exhibit a conventional algebraic decay on any of the three surface sublattices involved at low energy, namely A$_{1}$, B$_{1}$ and B$_{\text{N}}$.

From the experimental point of view, it is possible to image the LDOS of sublattices A$_{1}$ and B$_{1}$ together, as they belong to the same surface. For scattering processes that couple two equivalent valleys ($\xi=\xi'$), Eq. (\ref{LDOS Impurity on B1}) shows that $\delta\rho_{A_{1}}$ and $\delta\rho_{B_{1}}$ are always in antiphase when the condition (\ref{D3 Criterion}) is met, regardless of the number of layers. One could think that, \textit{a priori}, the situation is similar to the one in graphene, and that the intravalley scattering (${\bf \Delta K}=0$) reduces the Friedel oscillations to a $1/r^{2}$ power law. Nevertheless, this reduction does not arise since $\delta\rho_{A_{1}}$ and $\delta\rho_{B_{1}}$ have not the same dependence on the energy. Indeed, the $1/r$-decaying contributions lead to
\begin{align}
\delta\rho_{B_{1}}({\bf r}, \omega) = -~\omega^{2(1-\frac{1}{N})}~ \delta\rho_{A_{1}}({\bf r}, \omega) ~.
\end{align}
The LDOS modulation induced on the impurity sublattice, namely B$_{1}$, is then negligible compared to the one of sublattice A$_{1}$ at low energy ($\omega \ll 1$).

\subsubsection{Localized impurity in bulk}
In the sublattice basis $\{ A_{1}, B_{N_{0}-1}, A_{N_{0}}, B_{N} \}$, one can show in a similar manner as above that the low-energy bare Green function is well estimated by
\begin{align}\label{Low-energy Bare GF Impurity Bulk}
&G^{(0)}({\bf K^{\xi}_{mn}}+{\bf q}, \omega) \simeq \frac{1}{\omega-q^{2N}}  \\
&\times
\left( \begin{array}{cccc} 
.... & -\Big(\xi q e^{i\theta^{\xi}_{mn}(\bf q)}\Big)^{N_{0}-1}~\omega & .... & ....  \\
.... & \omega^{2} & .... & .... \\
.... & \omega~\big(\omega^{2}-q^{2(N_{0}-1)}\big) & .... & .... \\
.... & -\Big( \xi q e^{-i\theta^{\xi}_{mn}(\bf q)} \Big)^{N-N_{0}+1}~\big(\omega^{2}-q^{2(N_{0}-1)}\big) & .... & ....
\end{array} \right) \notag
\end{align}
in the vicinity of any valley ${\bf K^{\xi}_{mn}}$. In the above expression, we have only mentioned explicitly the components that turn out to be useful when evaluating the LDOS at the outer surfaces of the material. The momentum space representation of the $T$-matrix is then given by
\begin{align}
T(\omega)=
\left( \begin{array}{cccc}
0& 0 & 0 & 0 \\
0& 0 & 0 & 0 \\
0& 0 & t(\omega) & 0 \\
0& 0 & 0 & 0
\end{array} \right)
~\text{and}~ t(\omega)=\frac{V_{0}}{1-V_{0}\int_{BZ}G^{(0)}_{A_{N_{0}}A_{N_{0}}}({\bf k}, \omega)}
\end{align}
so that the $T$-matrix is momentum independent. If one follows the same procedure as the one used in the case of the previous four-band description that describes an impurity on sublattice B$_{1}$, one shows that the dominant oscillations are given by
\begin{widetext}
\begin{align}\label{LDOS Impurity in bulk}
\left \{
\begin{aligned}
\delta\rho_{A_{1}}({\bf r}, \omega) &\simeq -\frac{1}{\pi}~\Imag\Big\{~ i~ \frac{t(\omega)}{4^{2}N^{2}\omega^{2-\frac{2N_{0}+1}{N}}}~ \frac{e^{i2\omega^{\frac{1}{N}}r}}{r}~ \cos \big({\bf \Delta K}\cdot{\bf r} + (N_{0}-1)\Delta \theta({\bf r}) \big)~\Big(\xi \xi' \Big)^{N_{0}-1} \Big\} \\
\delta\rho_{B_{N}}({\bf r}, \omega) &\simeq -\frac{1}{\pi}~\Imag\Big\{~ i~ \frac{t(\omega)}{4^{2}N^{2}\omega^{2-\frac{2N_{0}+1}{N}}}~ \frac{e^{i2\omega^{\frac{1}{N}}r}}{r}~ \cos \big({\bf \Delta K}\cdot{\bf r} - (N-N_{0}+1)\Delta \theta({\bf r}) \big)~ \Big(\xi \xi' \Big)^{N-N_{0}+1}\Big\}
\end{aligned}
\right .
\end{align}
\end{widetext}
for low energies ($\omega\ll1$) and large distances. Once again, sublattices A$_{1}$ and B$_{\text{N}}$ belong to opposite surfaces in the case of rhombohedral multilayer graphene ($N\geq2$), so that the two signals $\delta\rho_{A_{1}}$ and $\delta\rho_{B_{N}}$ should be observed independently from one another in STM, thus avoiding any reduction of the Friedel oscillations to a $1/r^{2}$ decay.

\begin{figure*}[t]
\centering
$\begin{array}{cccc}
\includegraphics[trim = 47mm 0mm 10mm 0mm, clip, width=4.3cm]{kDOSrealw150t100tp30N01.pdf}&
\includegraphics[trim = 47mm 0mm 10mm 0mm, clip, width=4.3cm]{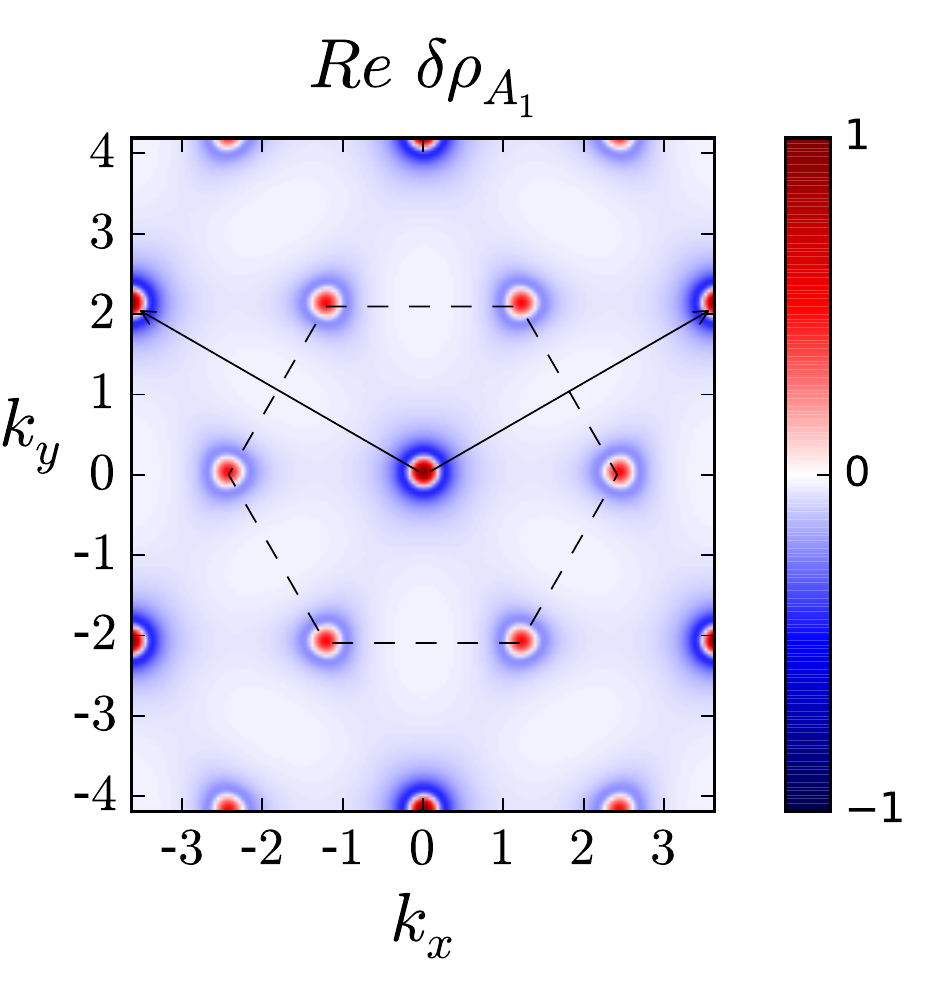}&
\includegraphics[trim = 47mm 0mm 10mm 0mm, clip, width=4.3cm]{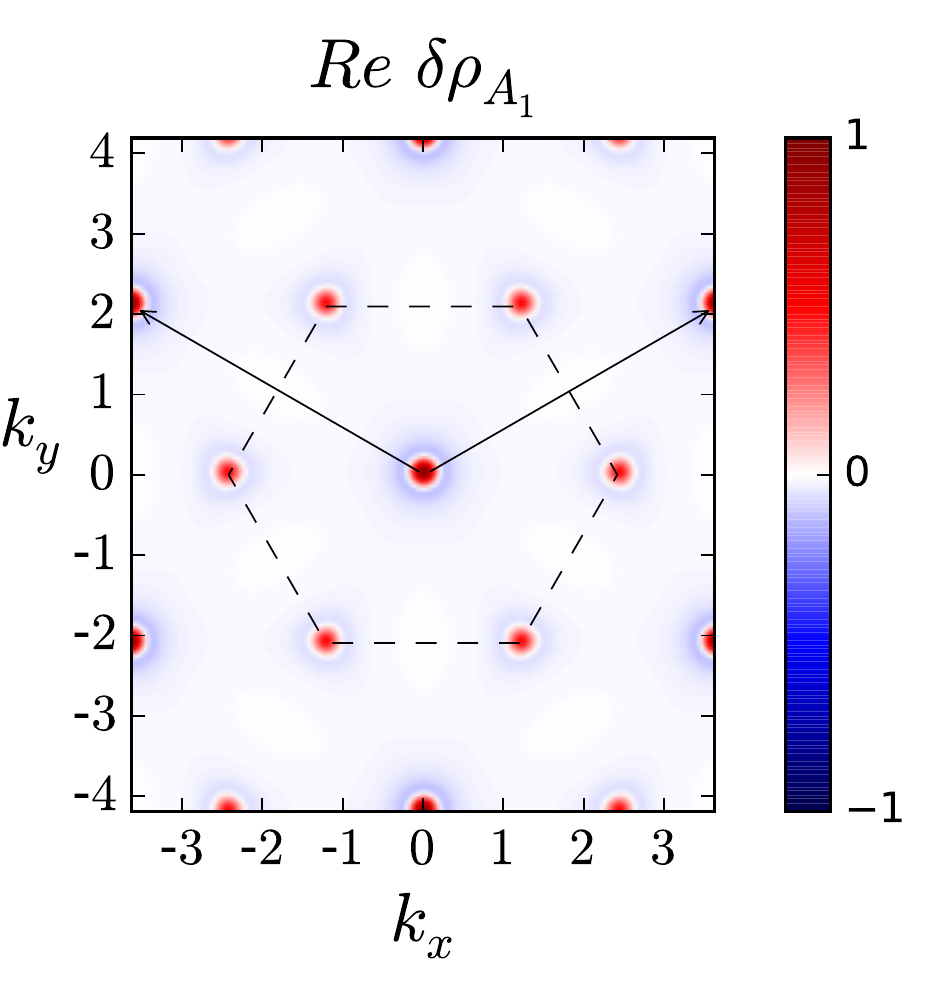}&
\includegraphics[trim = 47mm 0mm 10mm 0mm, clip, width=4.3cm]{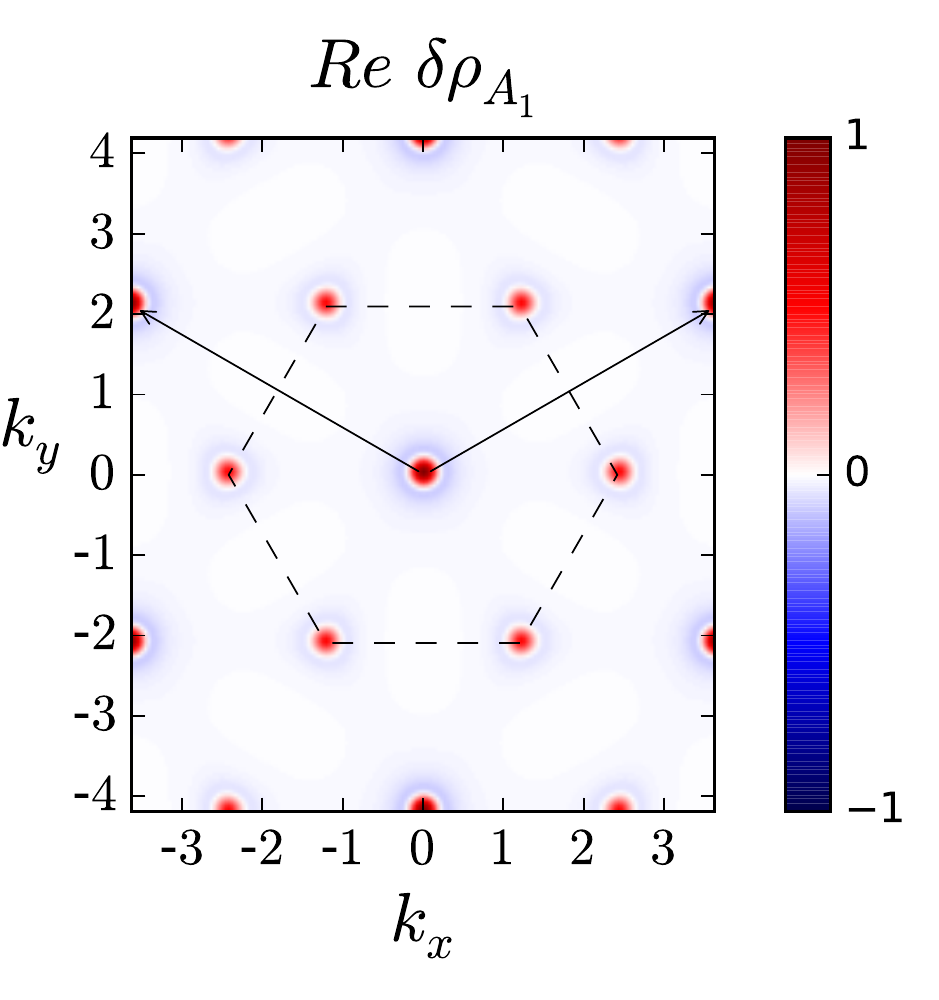}\\
~&~&~&~\\
\includegraphics[trim = 47mm 0mm 10mm 0mm, clip, width=4.3cm]{kDOSrealw150t100tp30N01.pdf}&
\includegraphics[trim = 47mm 0mm 10mm 0mm, clip, width=4.3cm]{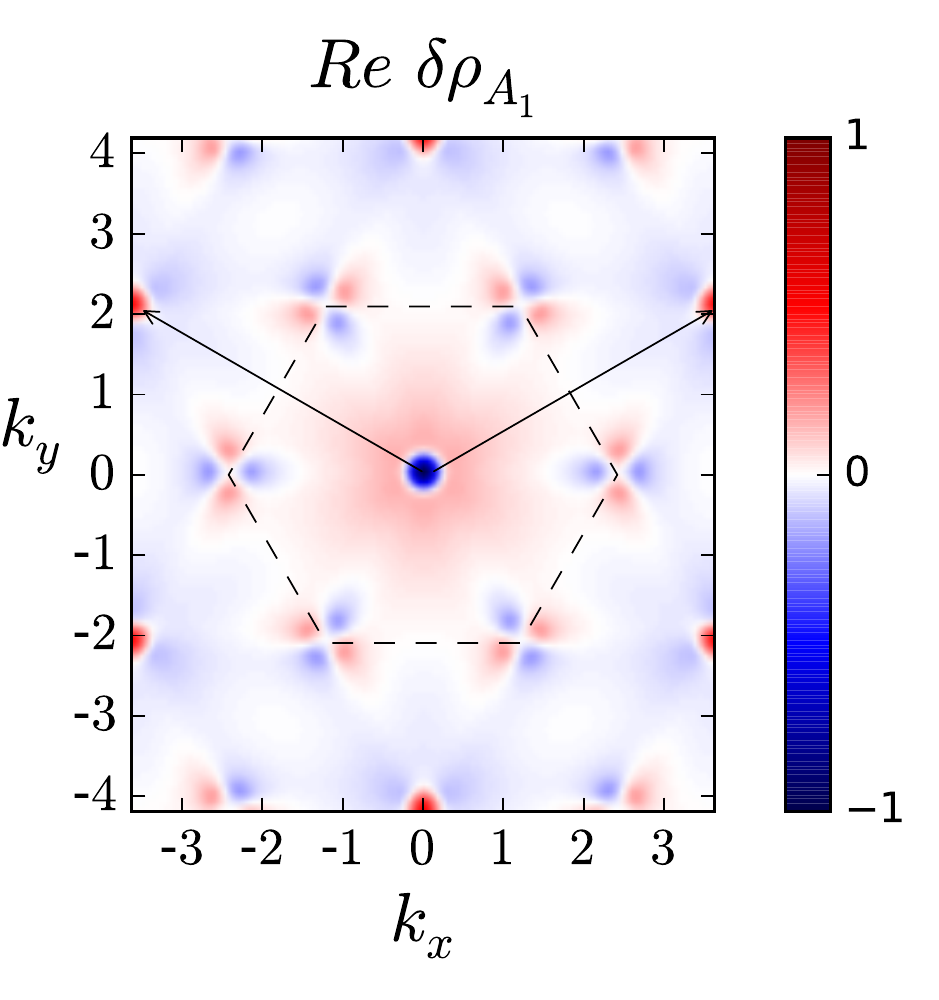}&
\includegraphics[trim = 47mm 0mm 10mm 0mm, clip, width=4.3cm]{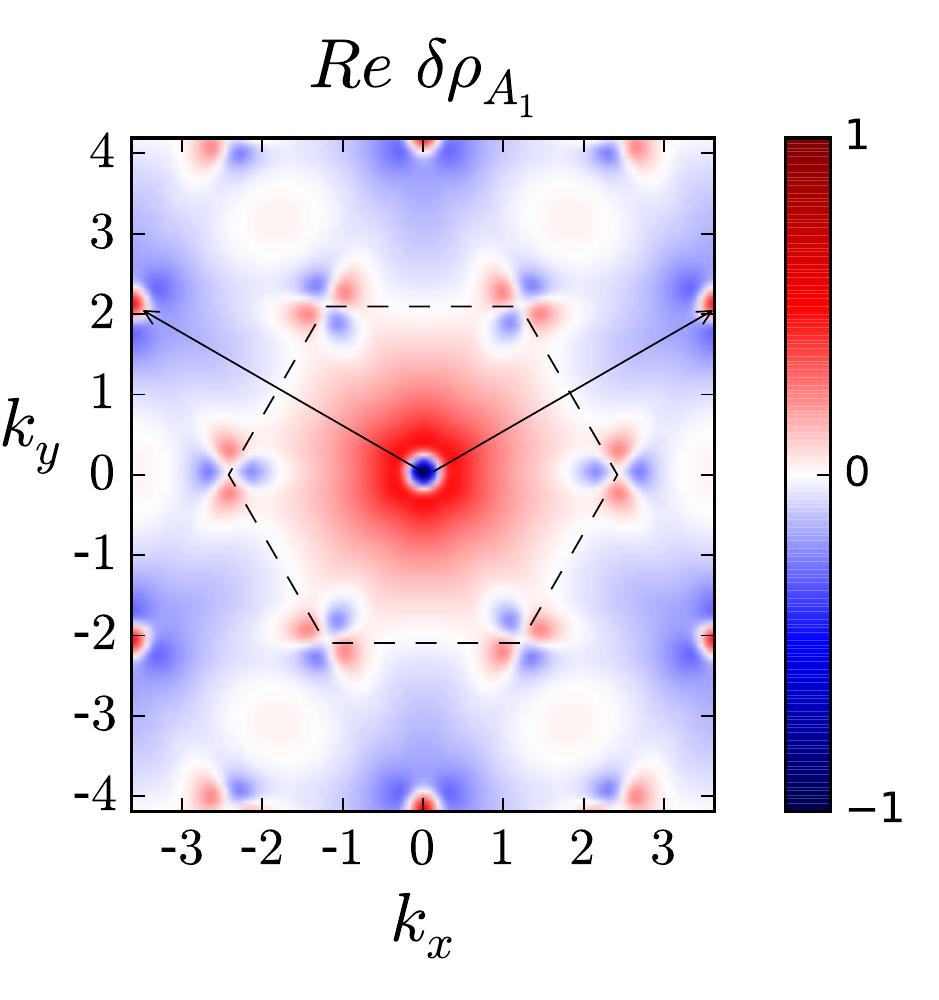}&
\includegraphics[trim = 47mm 0mm 10mm 0mm, clip, width=4.3cm]{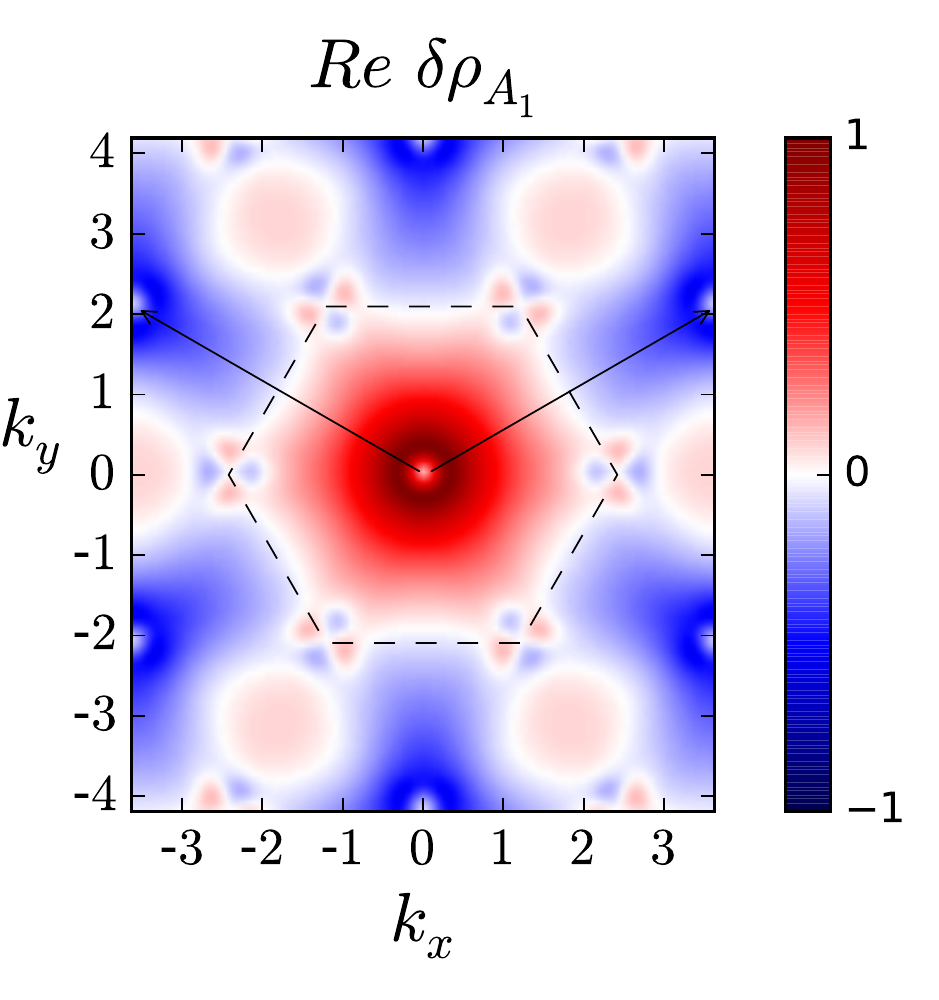}\\
\end{array}$
\caption{\small (Color online) Momentum space pattern of the LDOS on the impurity surface for $N=1$, 2, 3 and 4 (from left to right). Only the real part of the LDOS Fourier transform is depicted when the impurity is located on sublattice A$_{1}$ (first row) and B$_{1}$ (second row). The potential magnitude is $V_{0}= t$. Note that $\delta\rho=\delta\rho_{A_{1}}+\delta\rho_{B_{1}}$ in the case of monolayer graphene, whereas the LDOS modulations on the impurity surface are given by $\delta\rho_{A_{1}}$ otherwise. The dashed-line-made hexagons outline the Brillouin zone and can be used as guides for the eyes. The two vectors that span the reciprocal space are depicted by black arrows. The scattering which occurs between equivalent valleys yields the spots that can be connected to the origin by a linear combination of these basis vectors. They mainly have a circular symmetry. The spots at the hexagon corners are induced by scattering between non-equivalent valleys ($\xi=-\xi'$). They have a twofold rotational symmetry when the impurity is localized on sublattice B$_{1}$ (second row).}
\label{LDOS FT Impurity Surface IR Representation}
\end{figure*}

Besides, the expression (\ref{LDOS Impurity in bulk}) is actually general in the sense that it describes all the cases that we have distinguished so far. Indeed, when $N_{0}=1$ it describes an impurity on the surface sublattice A$_{1}$ and one gets the same expression as the one in Eq. (\ref{LDOS Impurity on A1}). Moreover, $N_{0}=N$ corresponds to an impurity on the surface sublattice A$_{N}$ and leads to the same expression as the one in (\ref{LDOS Impurity on B1}) when exchanging A$_{1}$ and B$_{N}$. Therefore, Eq. (\ref{LDOS Impurity in bulk}) generically describes the low-energy Friedel oscillations induced by a single impurity in rhombohedral multilayer graphene.

\subsubsection{Real space behavior of Friedel oscillations}
Now, let us briefly rephrase the results obtained so far at low energy. A localized impurity, located either on the surface or in the bulk of rhombohedral multilayer graphene, essentially induces Friedel oscillations in the LDOS of the two opposite surfaces. Regardless of the layer on which the impurity is located and regardless of its potential magnitude, the induced long-range oscillations mainly involve one sublattice on each surface, where they have the same energy dependence and exhibit the same algebraic decay.
As a consequence, the Friedel oscillations induced at the surfaces of rhombohedral multilayer graphene behave accordingly to Eq. (\ref{LDOS Impurity in bulk}). They always decay as $1/r$ with the distance to the impurity, and their reduction to a $1/r^{2}$ power law can only occur in monolayer graphene, where the two sublattices involved at low energy belong to the same surface.

\section{Fourier transform analysis}
\label{Fourier transform analysis}

\subsection{Modulations induced on the impurity surface}

The previous section has focused on the real-space description of Friedel oscillations. We now analyze the interference pattern in momentum space, and first discuss the interferences induced on the surface on which the impurity is located.

\subsubsection{Localized impurity on sublattice A$_{1}$}

The momentum space signature of the $1/r$-decaying Friedel oscillations is obtained from Eq. (\ref{LDOS Impurity in bulk}), where we disregard the momentum independent factor $t(\omega)/4^{2}N^{2}\omega^{2-(2N_{0}+1)/N}$. It is given by
\begin{align}\label{LDOS FT Sublattice A1 Impurity A1}
\delta\rho_{A_{1}}({\bf \Delta K} + {\bf q}, \omega) &\simeq -~ \frac{\Theta(q-2q_{F})}{\sqrt{q^{2}-(2q_{F})^{2}}}~,
\end{align}
where $\Theta$ denotes the Heaviside step function and $\bf{q_{F}}$ is the Fermi momentum defined from the nodal point at the center of every valley. It fixes the Fermi contours involved in the elastic scattering according to the low-energy dispersion relation, i.e. $\omega \simeq q_{F}^{N}$. The reader may refer to the case $\nu=0$ in Appendix \ref{Fourier transform of the Friedel oscillations} for more details about the derivation which arises from a first-order expansion in the limit $q \rightarrow 2q_{F}$.

As a result, the Fourier transform of the $1/r$-decaying Friedel oscillations is a real function of the momentum and outlines a $2q_{F}$-radius ring, regardless of the scattering wave vector ${\bf \Delta K}$. Thus this signature reveals the circular symmetry of the Fermi contours involved in the elastic scattering. This is in agreement with the numerical results of the first row in Fig. \ref{LDOS FT Impurity Surface IR Representation} and with the momentum space analysis realized in the case of monolayer graphene in [\onlinecite{PhysRevLett.100.076601}].

Note that, for all the numerical results shown in this paper, the modulus of the Fermi momentum is fixed at $q_{F}=0.1a_{0}^{-1}$. This fixes the radius of the LDOS modulations independently of the value of $N$. From the dispersion relation in Eq. (\ref {Dispersion relation}), the energies are then given by $\omega \simeq 0.150t$, $0.075t$, $0.038t$ and $0.019t$ for $N=1$, 2, 3 and 4, respectively. Thus, the system mainly behaves accordingly to the low-energy descriptions discussed above.

\subsubsection{Localized impurity on sublattice B$_{1}$}

If the impurity lies on sublattice B$_{1}$, then the Fourier transform of the $1/r$-decaying Friedel oscillations is obtained from Eq. (\ref{LDOS Impurity on B1}) when considering the case $\nu=-1$ in Appendix \ref{Fourier transform of the Friedel oscillations}. This results in
\begin{align}\label{LDOS FT Sublattice A1 Impurity B1}
\delta\rho_{A_{1}}({\bf \Delta K} + {\bf q}, \omega) &\simeq 
-~ \frac{\Theta(q-2q_{F})}{\sqrt{q^{2}-(2q_{F})^{2}}} \\
&\times (-\xi \xi')~ e^{i {\bf \Delta K}\cdot{\bf d_{3}}}~ e^{i (\theta^{\xi}({\bf q})-\theta^{\xi'}({\bf q}) )} ~, \notag
\end{align}
where $\theta^{\xi}({\bf q})$ is phase involved in the low-energy description of the Bloch spinors in Eq. (\ref{Bloch eigenstates}). As already emphasized, it also characterizes the orientation of the scattering wave vector ${\bf q}$.

The scattering between equivalent valleys ($\xi=\xi'$) yields a $2q_{F}$-radius ring in momentum space. The real part of the LDOS Fourier transform associated to intravalley scattering (${\bf \Delta K} = 0$) is opposite to the one induced by the impurity on sublattice A$_{1}$ and described in Eq. (\ref{LDOS FT Sublattice A1 Impurity A1}). This is in agreement with the numerical results shown in the second row in Fig. \ref{LDOS FT Impurity Surface IR Representation}.

The scattering between non-equivalent valleys ($\xi=-\xi'$) involves the phase $2\theta^{\xi}({\bf q})$. It breaks the circular symmetry which is reduced to a twofold rotational symmetry, as illustrated in Fig. \ref{LDOS FT Impurity Surface IR Representation}.

\begin{figure}[t]
\centering
$\begin{array}{cc}
\includegraphics[trim = 40mm 0mm 05mm 0mm, clip, width=4.2cm]{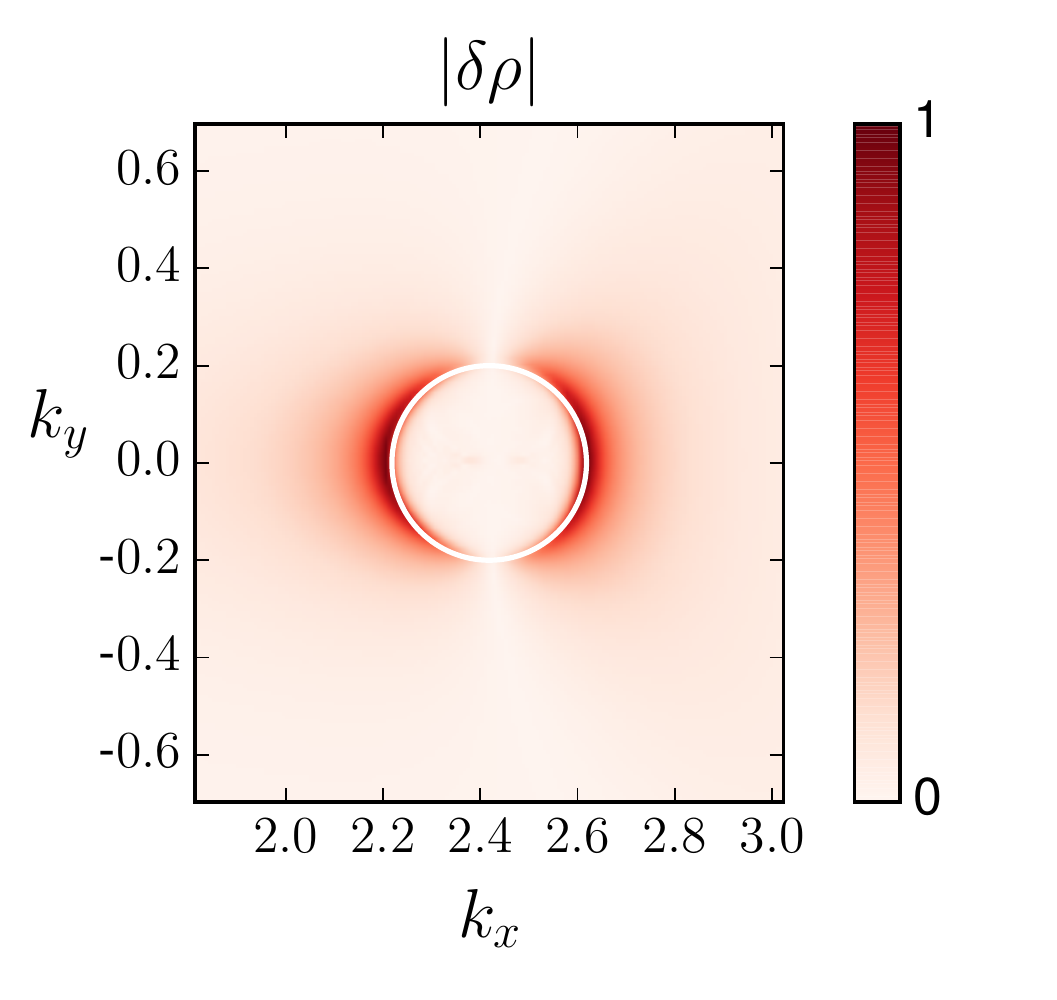}&
\includegraphics[trim = 40mm 0mm 05mm 0mm, clip, width=4.2cm]{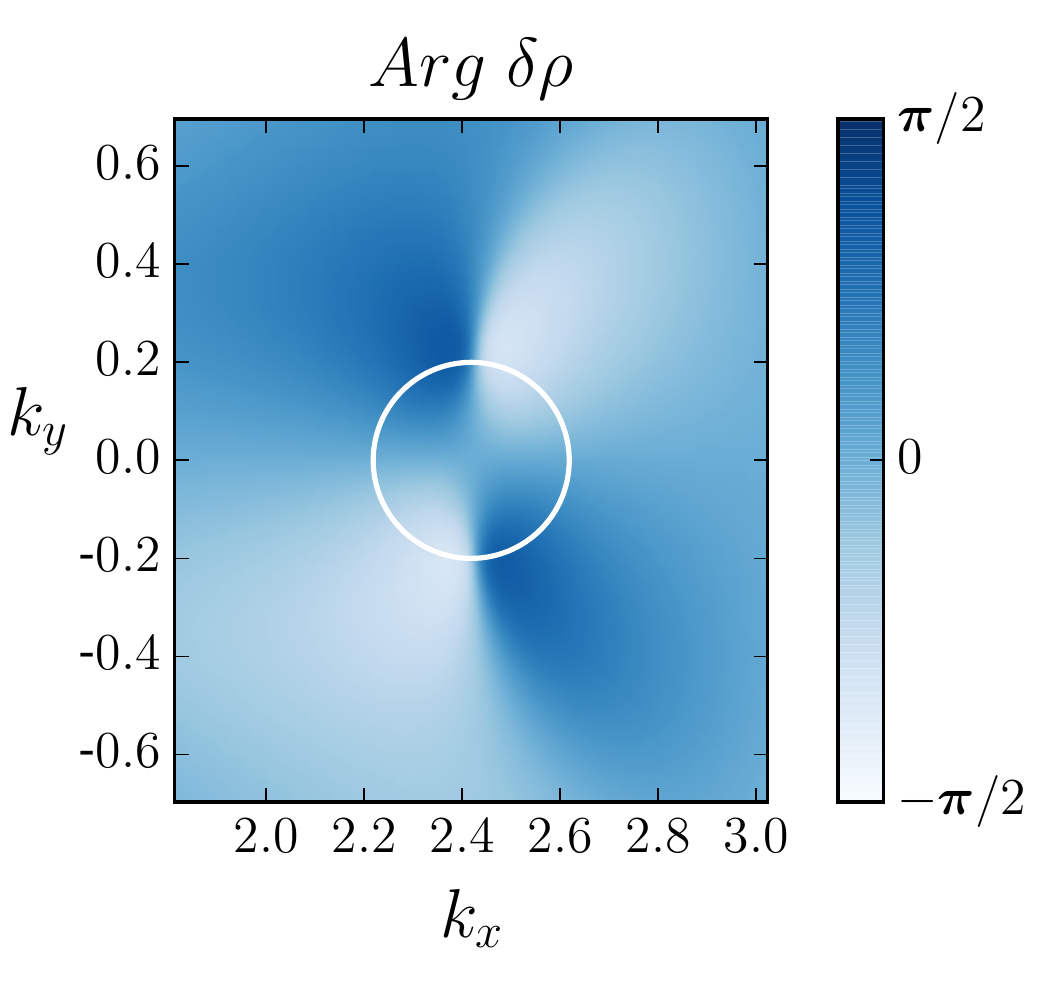}\\
\includegraphics[trim = 40mm 0mm 05mm 0mm, clip, width=4.2cm]{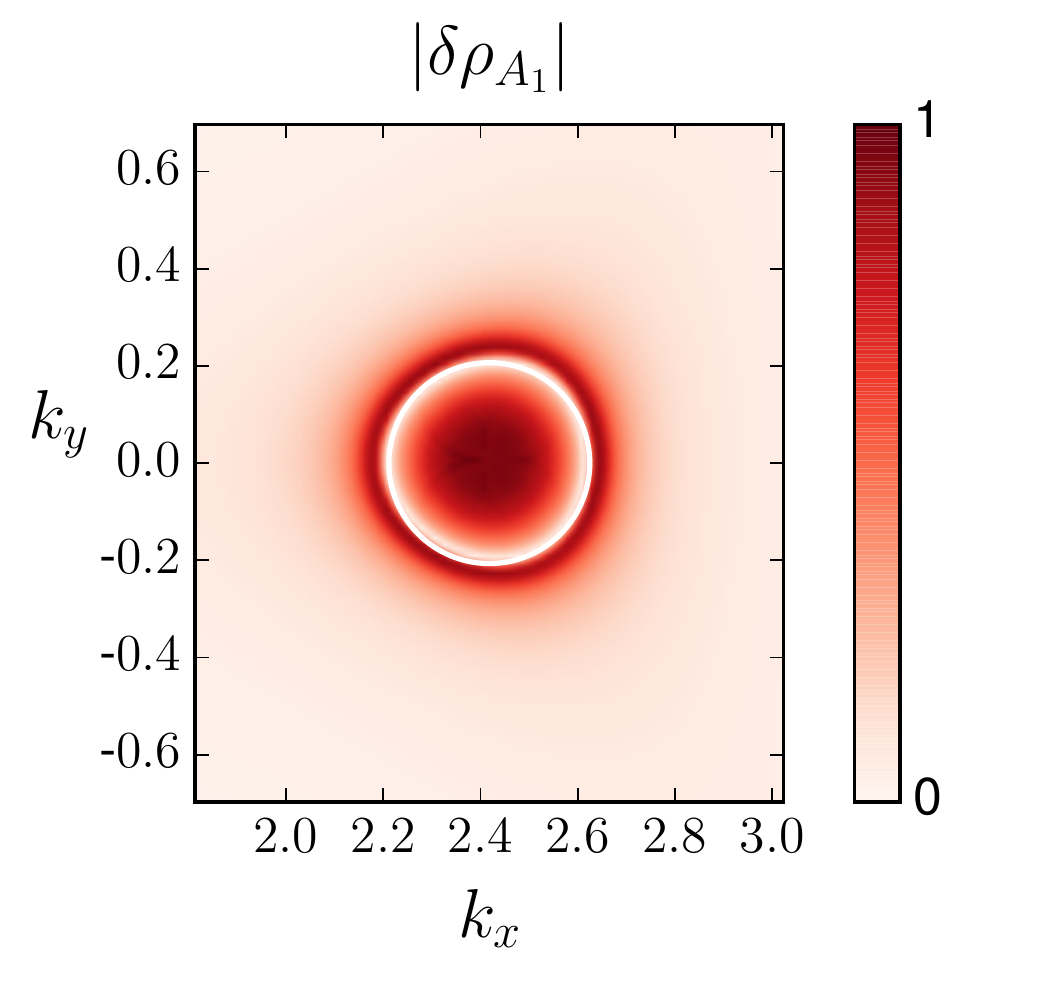}&
\includegraphics[trim = 40mm 0mm 05mm 0mm, clip, width=4.2cm]{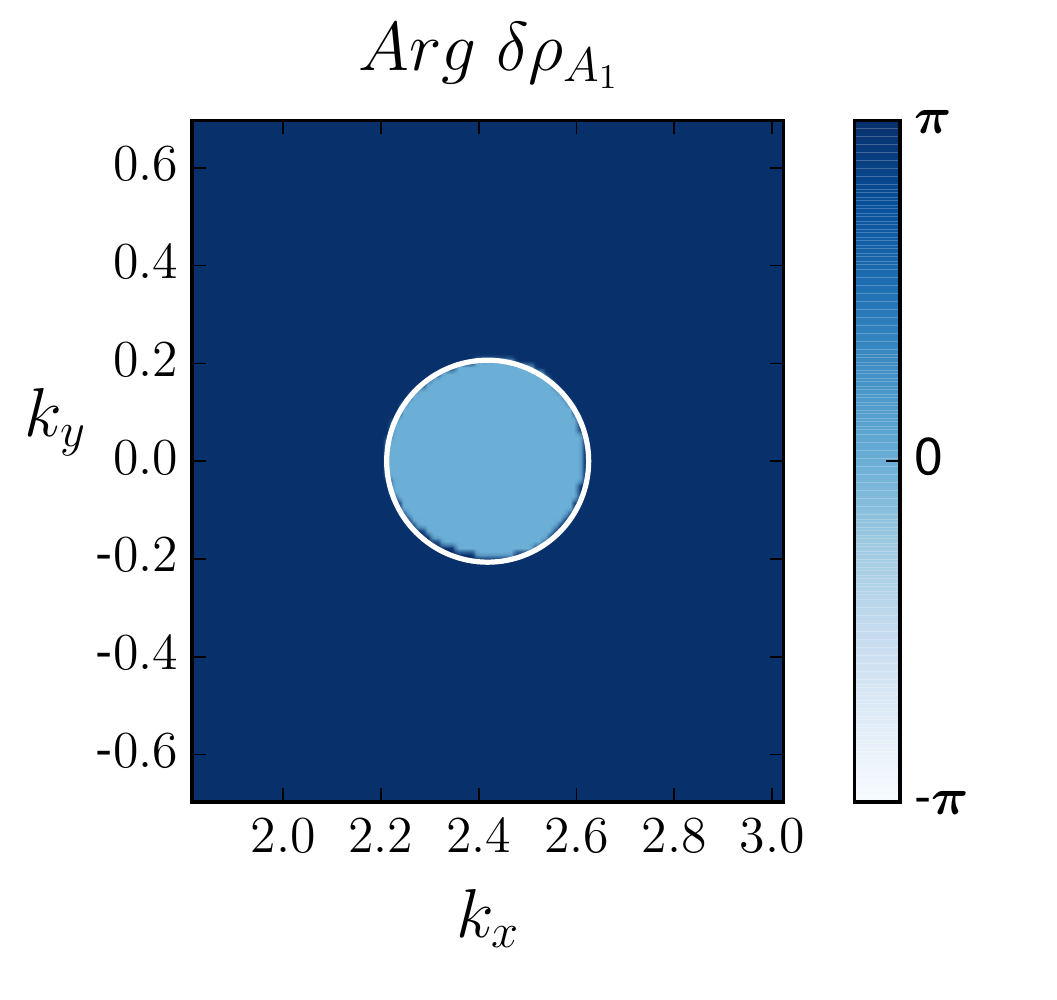}\\
\includegraphics[trim = 40mm 0mm 05mm 0mm, clip, width=4.2cm]{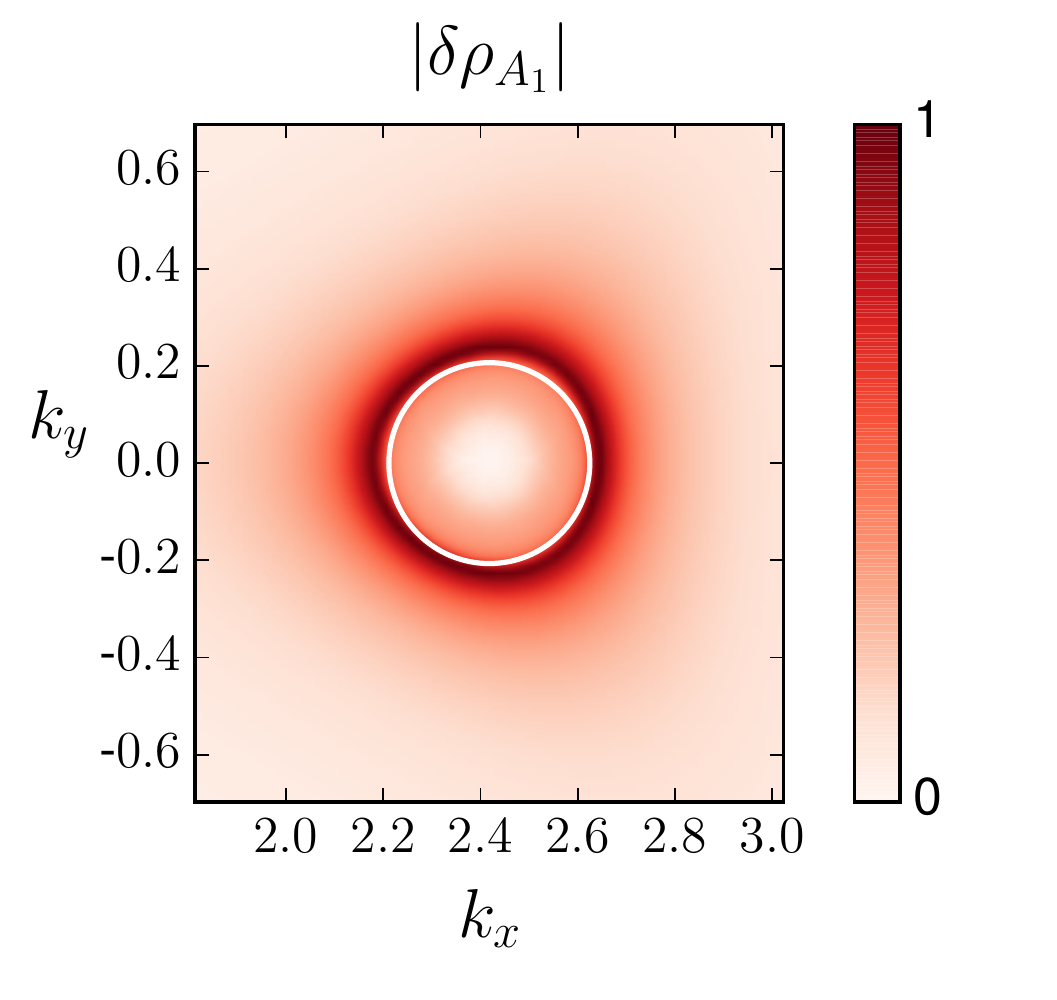}&
\includegraphics[trim = 40mm 0mm 05mm 0mm, clip, width=4.2cm]{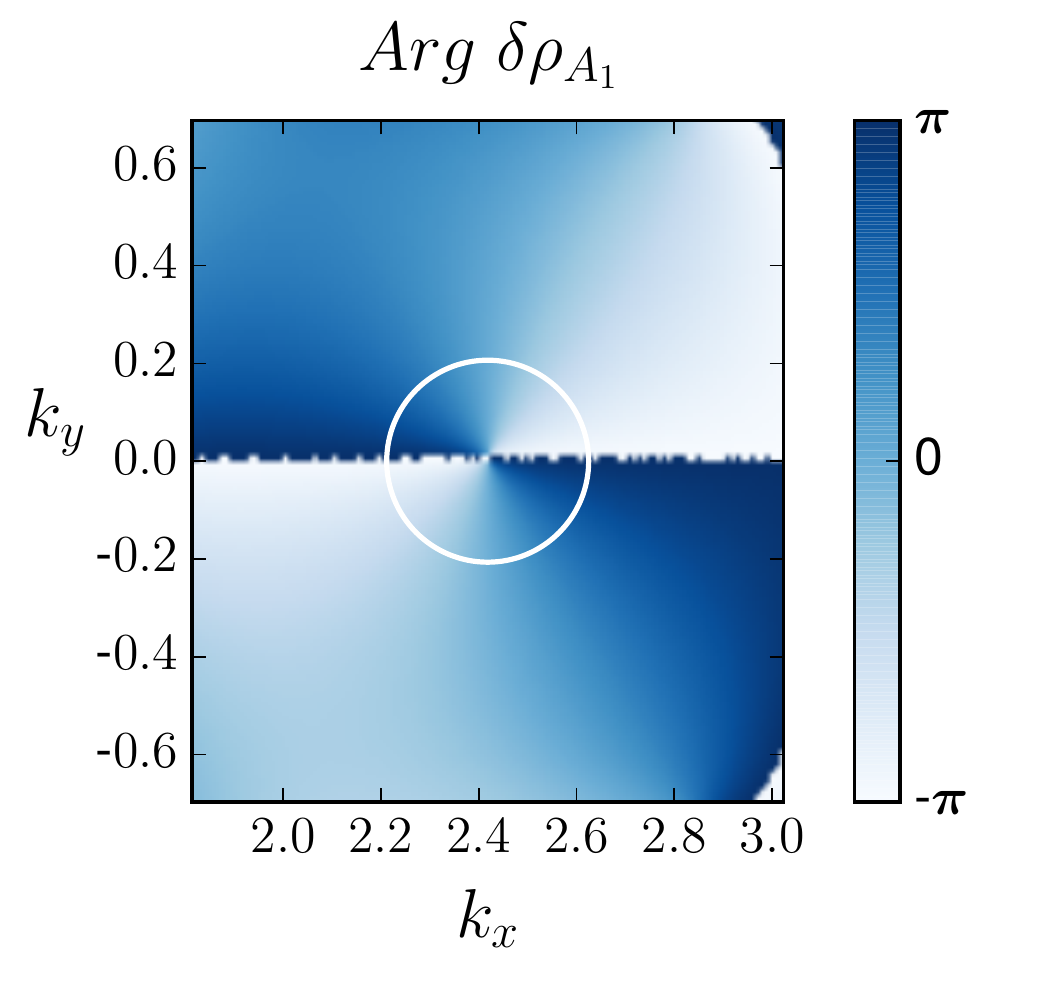}\\
\end{array}$
\caption{\small (Color online) Polar representation of the LDOS Fourier transform induced by scattering between non-equivalent valleys on the surface where there is impurity. It is illustrated for the valleys that are related to one another by $\delta m = 1$, $\delta n = -1$ and $\xi=-\xi'=-1$. It corresponds to the spots located at the right-hand corner of the hexagons in Fig. \ref{LDOS FT Impurity Surface IR Representation}. The $2q_{F}$-radius circle is mentioned in white as a guide for the eyes. The first row refers to the impurity surface of monolayer graphene ($N=1$) and thus $\delta \rho=\delta\rho_{A_{1}}+\delta\rho_{B_{1}}$. The second and third rows are both obtained for bilayer graphene when the impurity lies on the sublattice A$_{1}$ and B$_{1}$, respectively. In both cases, the LDOS modulations at the surface mainly involve sublattice A$_{1}$ at low energy, so that only $\delta\rho_{A_{1}}$ is mentioned.}
\label{LDOS FT Impurity Surface Polar Representation Zoom}
\end{figure}

\begin{figure*}[t]
\centering
$\begin{array}{cccc}
\includegraphics[trim = 40mm 0mm 10mm 0mm, clip, width=4.2cm]{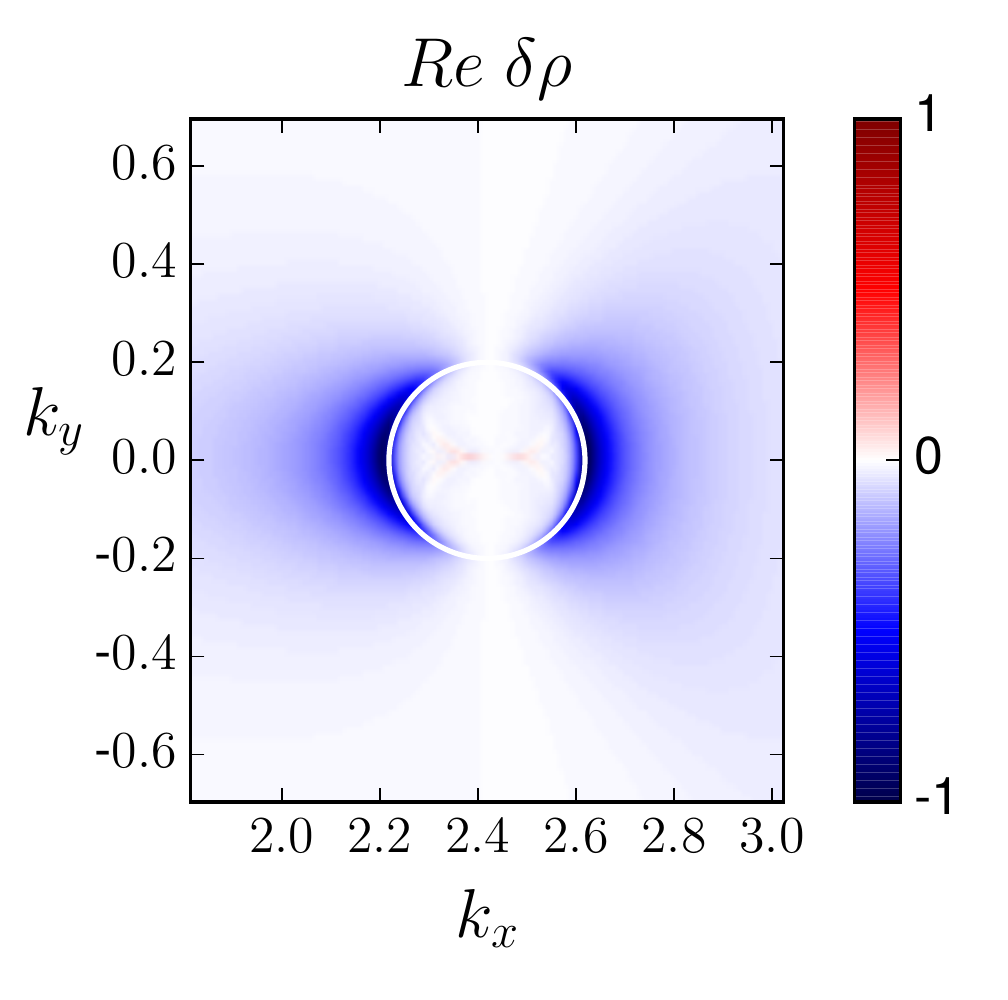}&
\includegraphics[trim = 40mm 0mm 10mm 0mm, clip, width=4.2cm]{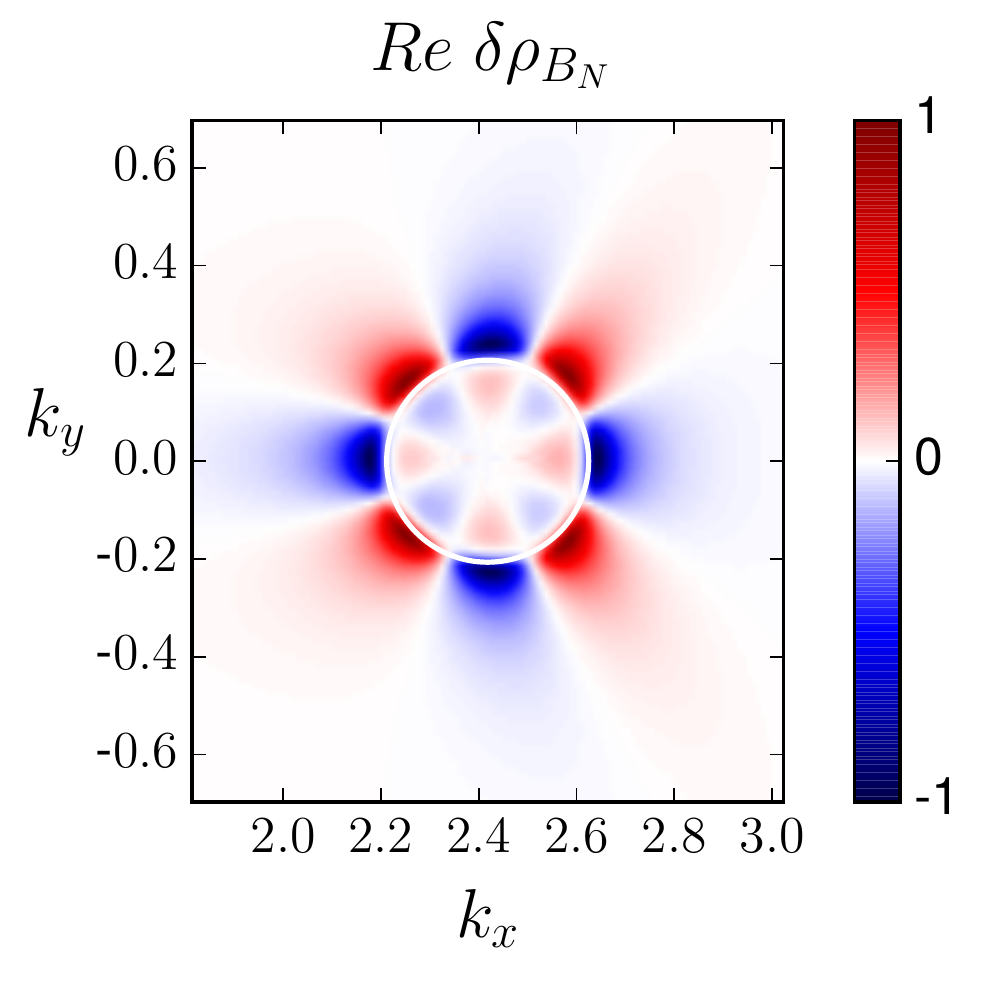}&
\includegraphics[trim = 40mm 0mm 10mm 0mm, clip, width=4.2cm]{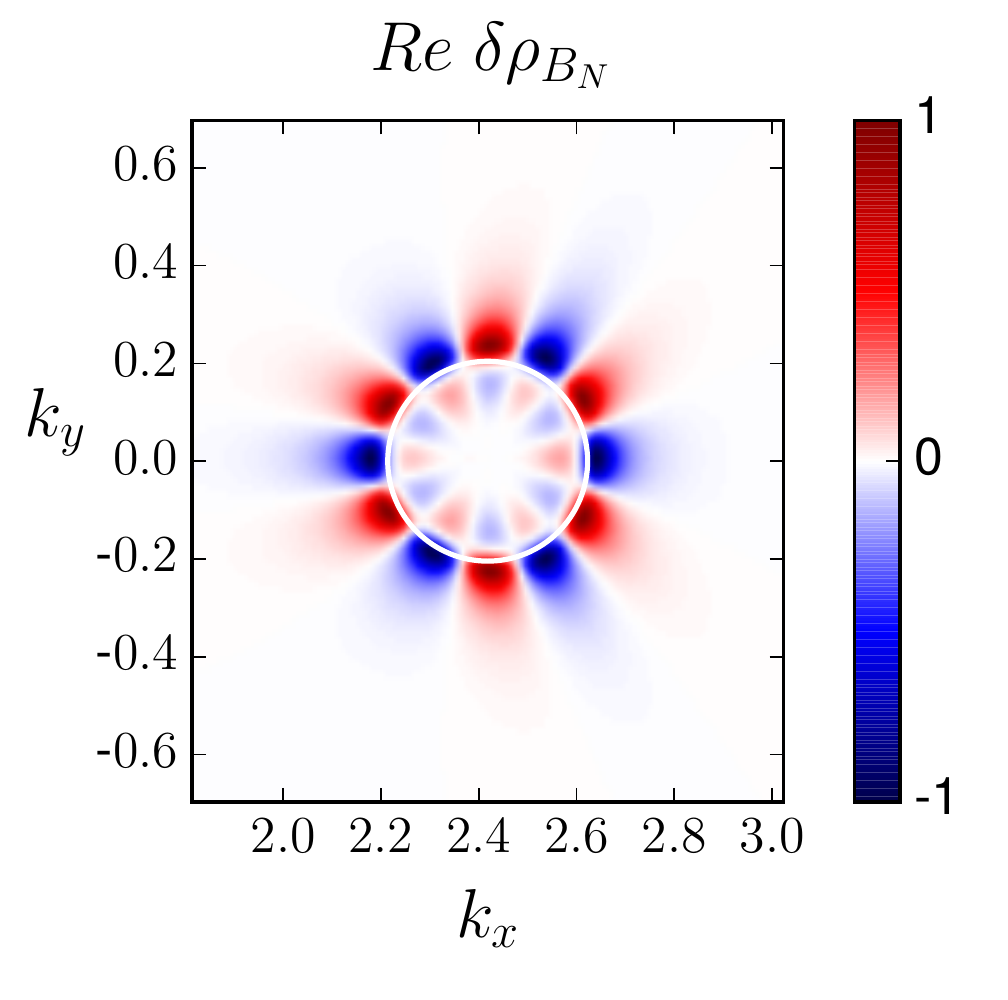}&
\includegraphics[trim = 40mm 0mm 10mm 0mm, clip, width=4.2cm]{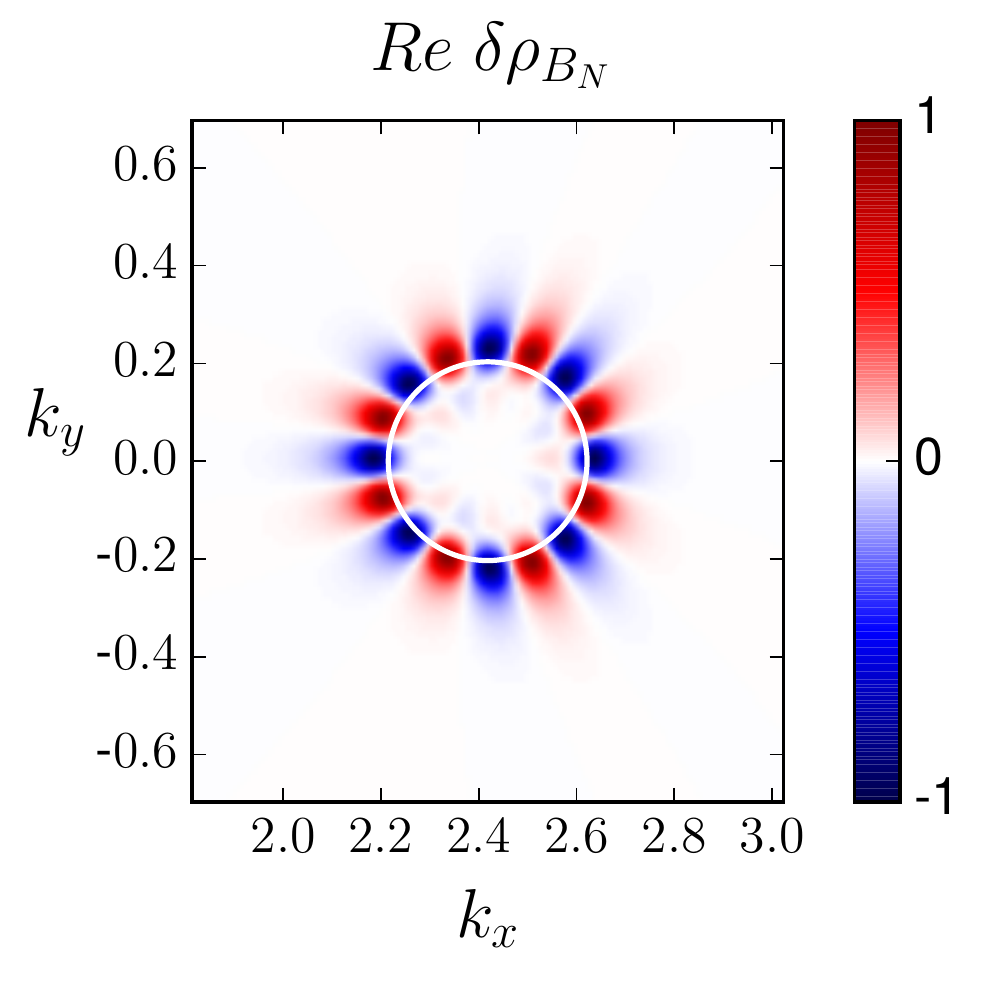}\\
\includegraphics[trim = 40mm 0mm 10mm 0mm, clip, width=4.2cm]{kDOS1zoomRealw150t100tp30N01A1.pdf}&
\includegraphics[trim = 40mm 0mm 10mm 0mm, clip, width=4.2cm]{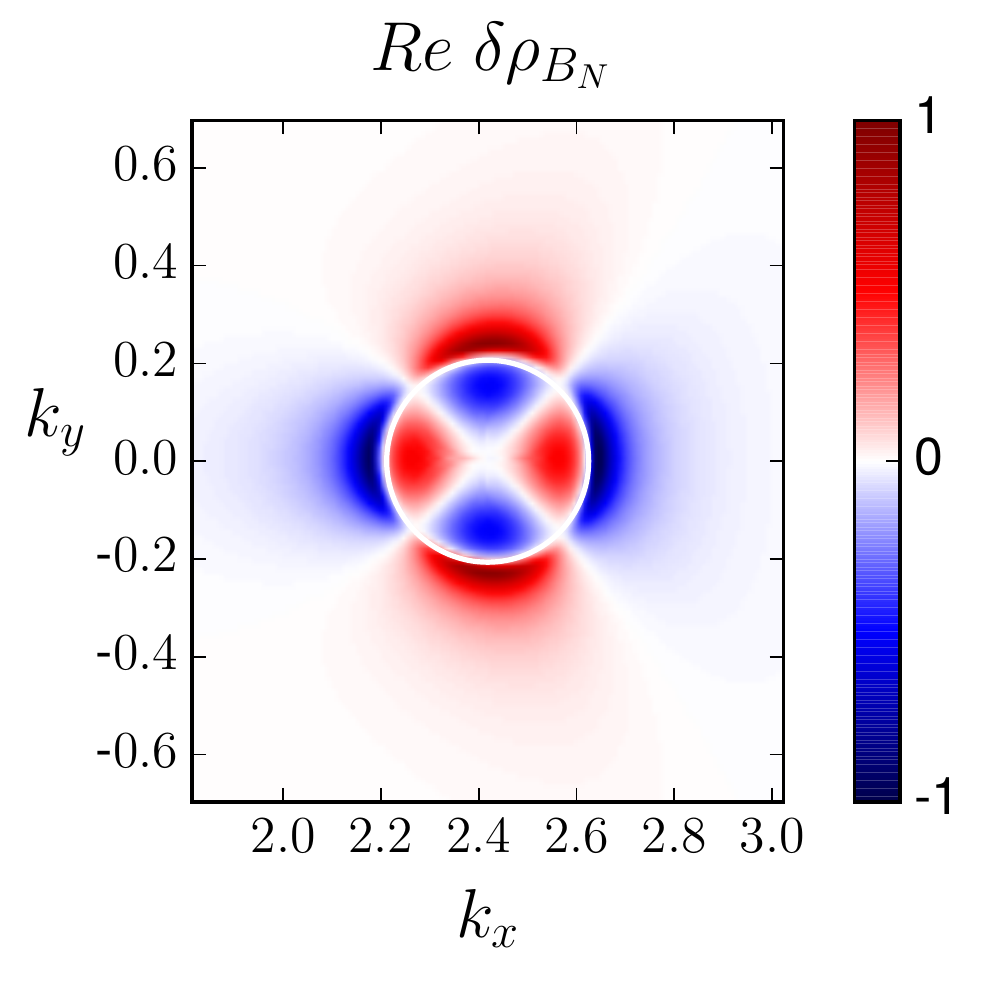}&
\includegraphics[trim = 40mm 0mm 10mm 0mm, clip, width=4.2cm]{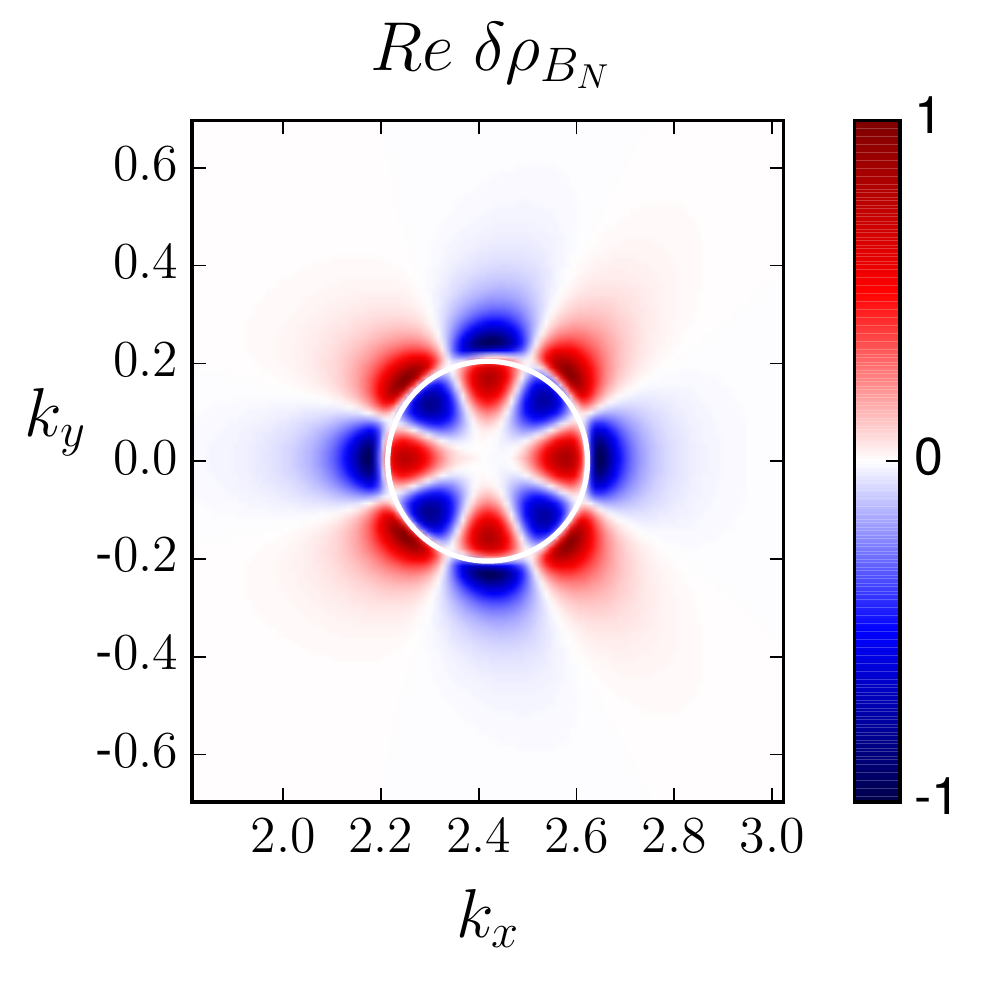}&
\includegraphics[trim = 40mm 0mm 10mm 0mm, clip, width=4.2cm]{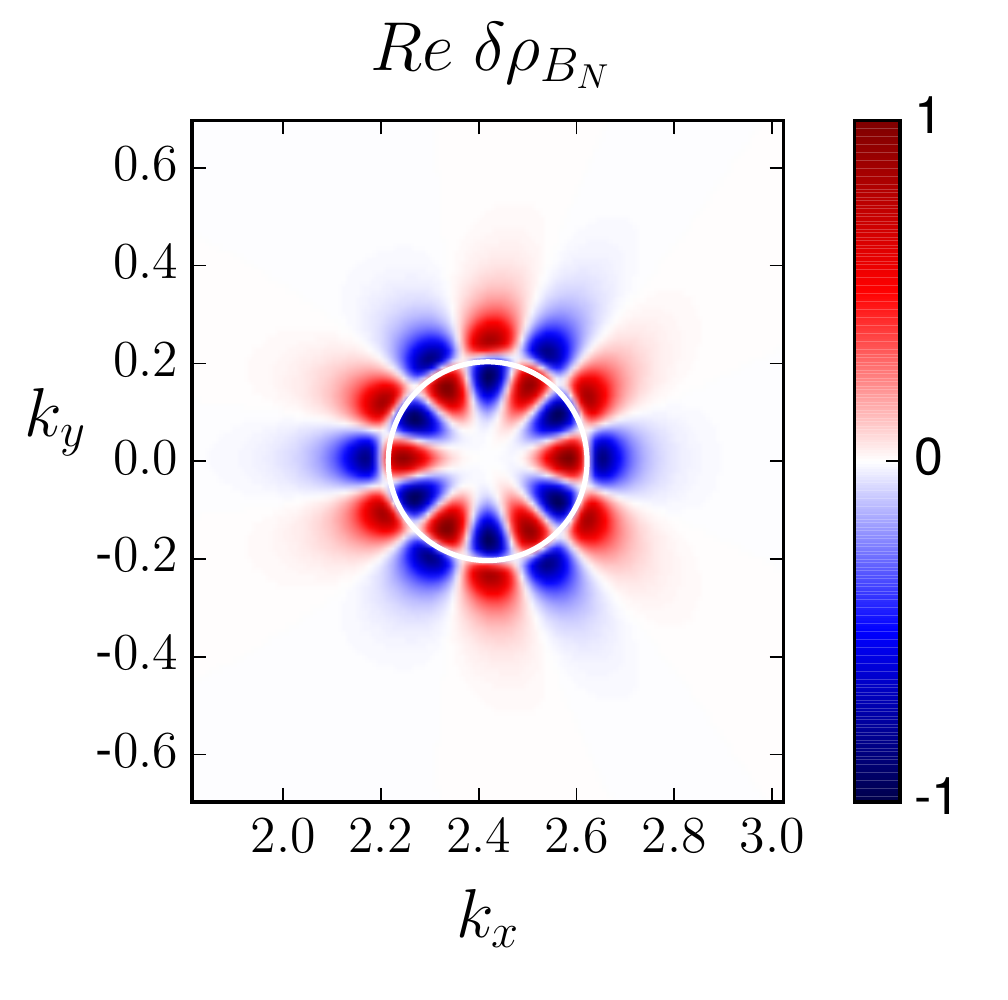}\\
\end{array}$
\caption{\small (Color online) Momentum space pattern of the LDOS modulations induced by scattering between non-equivalent valleys on the pristine surface.  The potential, which is simulated by $V_{0}= t$, lies on the opposite surface, either on sublattice A$_{1}$ (first row), or on sublattice B$_{1}$ (second row). The columns, from left to right, refer to $N=1$, 2, 3 and 4. Only the real part of the LDOS Fourier transform is shown. The circular symmetry is reduced to a $2N$- or 2($N-$1)-fold rotational symmetry, depending on the sublattice on which the impurity is located at the opposite surface.}
\label{LDOS FT Pristine Surface RI Representation Zoom}
\end{figure*}

\subsubsection{General remarks}
In monolayer graphene, the case of an impurity located on sublattice A$_{1}$ is of course equivalent to the one of an impurity on sublattice B$_{1}$, hence the two identical patterns in the first column in Fig. \ref{LDOS FT Impurity Surface IR Representation}. When imaging the surface, one takes the contributions of the two sublattices into account. The consequence is that the scattering wave vectors ${\bf \Delta K}$ which connect two equivalent valleys and satisfy
\begin{align}\label{D3 Criterion Graphene}
{\bf \Delta K}\cdot{\bf d_{3}} &= (\delta m+\delta n) \frac{2\pi}{3} \notag \\
&= 0~[2\pi] ~,
\end{align}
are responsible for a reduction of the Friedel oscillations to a $1/r^{2}$ power law. This is the reason why intravalley scattering, which is characterized by $\delta m= \delta n = 0$, yields a $2q_{F}$-radius disk instead of a $2q_{F}$-radius ring [\onlinecite{PhysRevLett.100.076601}]. We insist that intravalley scattering is not the only one process to satisfy the criterion (\ref{D3 Criterion Graphene}) and to yield a $2q_{F}$-radius disk. Anyway, the Fourier analysis of intravalley scattering is sufficient to identify monolayer graphene in the class of rhombohedral $N$-layer graphene materials.

Actually, the scattering which takes place between non-equivalent valleys turns out to be even more informative. Its Fourier transform is shown for monolayer and bilayer graphene in the polar representation in Fig. \ref{LDOS FT Impurity Surface Polar Representation Zoom}. In the case of monolayer graphene, the Fourier transform of the $1/r$-decaying Friedel oscillations is
\begin{align}
\delta\rho~({\bf \Delta K} + {\bf q}, \omega) &= \delta\rho_{A_{1}}({\bf \Delta K} + {\bf q}, \omega) + \delta\rho_{B_{1}}({\bf \Delta K} + {\bf q}, \omega) \\
&\simeq -~ \frac{\Theta(q-2q_{F})}{\sqrt{q^{2}-(2q_{F})^{2}}}~ \Big( 1 + e^{-i{\bf \Delta K}\cdot{\bf d_{3}}}~ e^{-i 2\theta^{\xi}({\bf q})} \Big) \notag ~.
\end{align}
The modulus vanishes twice along the $2q_{F}$-radius ring, so that there are two discontinuities in the argument. These behaviors are in agreement with the ones depicted in the first row in Fig. \ref{LDOS FT Impurity Surface Polar Representation Zoom}. On the contrary, Eqs. (\ref{LDOS FT Sublattice A1 Impurity A1}) and (\ref{LDOS FT Sublattice A1 Impurity B1}) show that the modulus always outlines a full $2q_{F}$-radius ring in rhombohedral $N$-layer graphene when $N\geq 2$. Therefore, the scattering between non-equivalent valleys is also sufficient to distinguish monolayer graphene from rhombohedral multilayer graphene.

Finally, if the modulus of the LDOS Fourier transform associated to the scattering between non-equivalent valleys show that $N\geq 2$ by outlining a full $2q_{F}$-radius ring, then the argument reveals on which sublattice the impurity is located. From Eqs. (\ref{LDOS FT Sublattice A1 Impurity A1}) and (\ref{LDOS FT Sublattice A1 Impurity B1}), the argument is fixed to $\pi~ [2\pi]$ if the impurity lies on sublattice A$_{1}$, whereas it is given by $2\theta^{\xi}({\bf q})$ if the impurity is on sublattice B$_{1}$. So if the impurity is on sublattice B$_{1}$, the argument of the LDOS Fourier transform winds twice when the wave vector ${\bf q}$ runs once along the $2q_{F}$-radius ring. This explains the two discontinuity lines that cross the $2q_{F}$-radius ring in the last plot in Fig. \ref{LDOS FT Impurity Surface Polar Representation Zoom}.

\subsection{Modulations induced on the pristine surface}

The localized impurity which lies on a surface of rhombohedral multilayer graphene also induces Friedel oscillations on the opposite pristine surface at low energy. Their Fourier transform analysis is the purpose of the subsequent lines.

\subsubsection{Localized impurity on sublattice A$_{1}$}

\begin{figure}[t]
\centering
$\begin{array}{cc}
\includegraphics[trim = 40mm 0mm 05mm 0mm, clip, width=4.2cm]{kDOS1zoomModuw150t100tp30N01A1.pdf}&
\includegraphics[trim = 40mm 0mm 05mm 0mm, clip, width=4.2cm]{kDOS1zoomPhasw150t100tp30N01A1.pdf}\\
\includegraphics[trim = 40mm 0mm 05mm 0mm, clip, width=4.2cm]{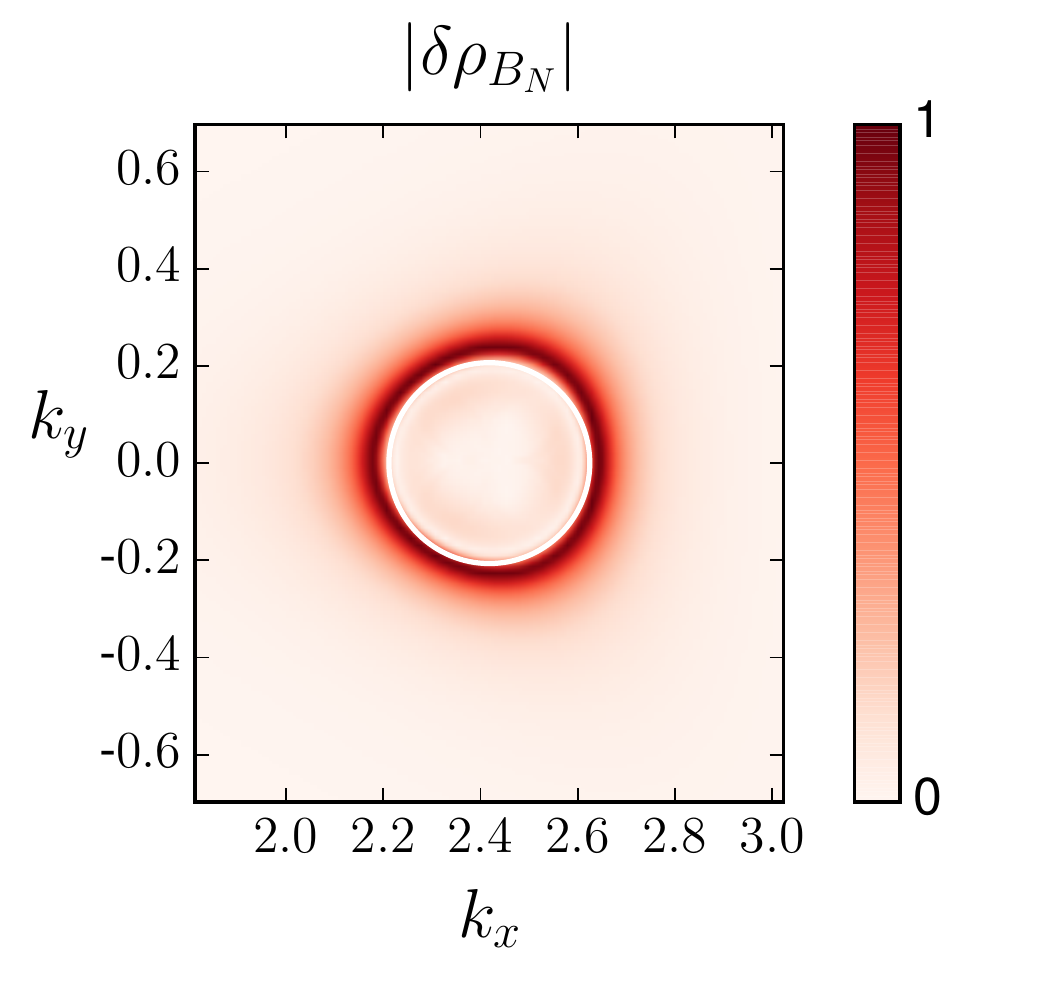}&
\includegraphics[trim = 40mm 0mm 05mm 0mm, clip, width=4.2cm]{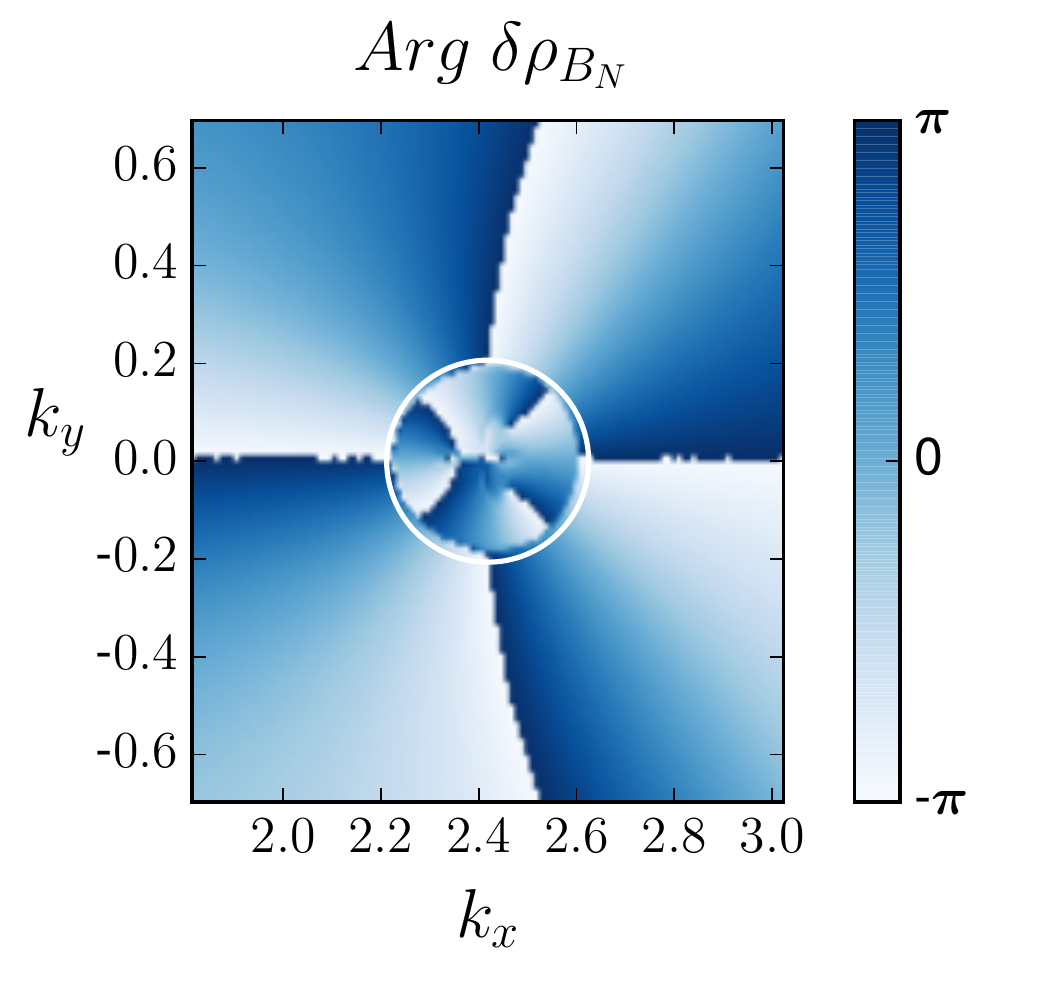}\\
\includegraphics[trim = 40mm 0mm 05mm 0mm, clip, width=4.2cm]{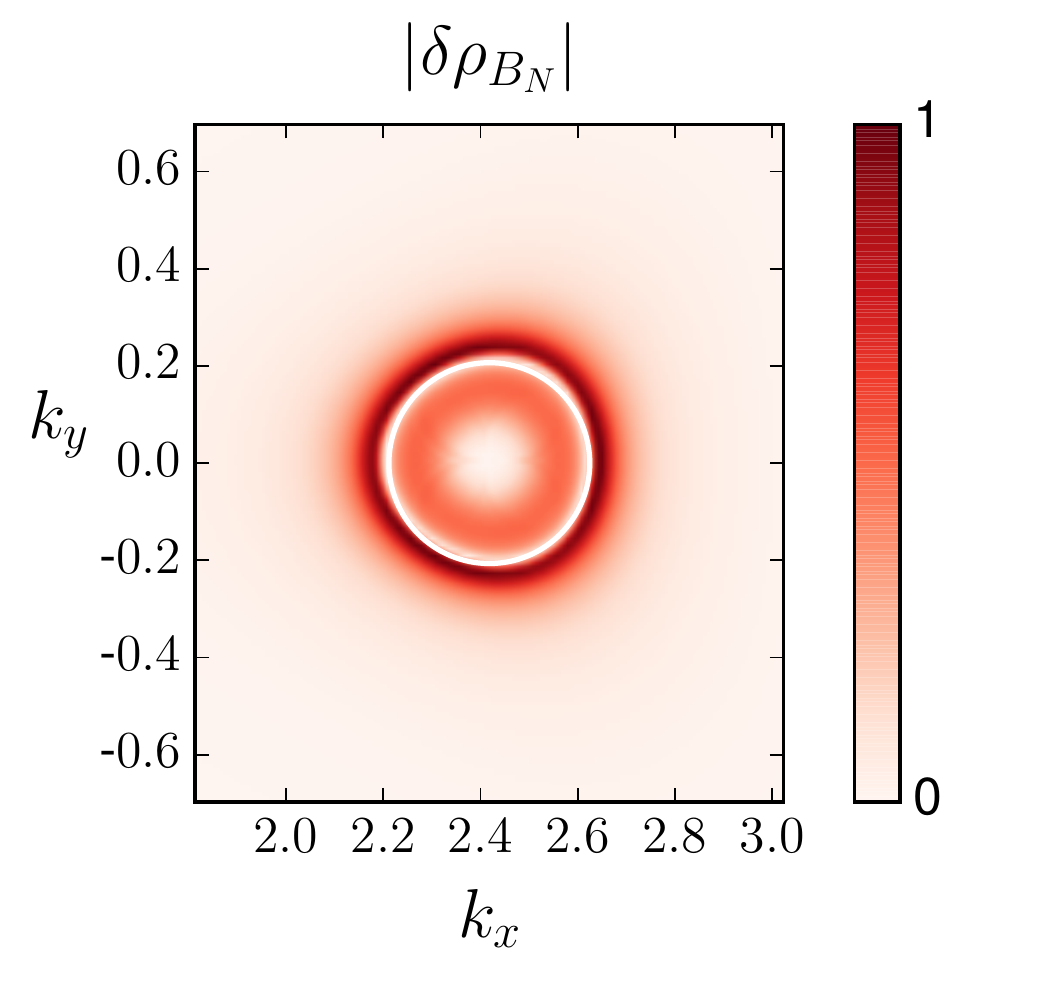}&
\includegraphics[trim = 40mm 0mm 05mm 0mm, clip, width=4.2cm]{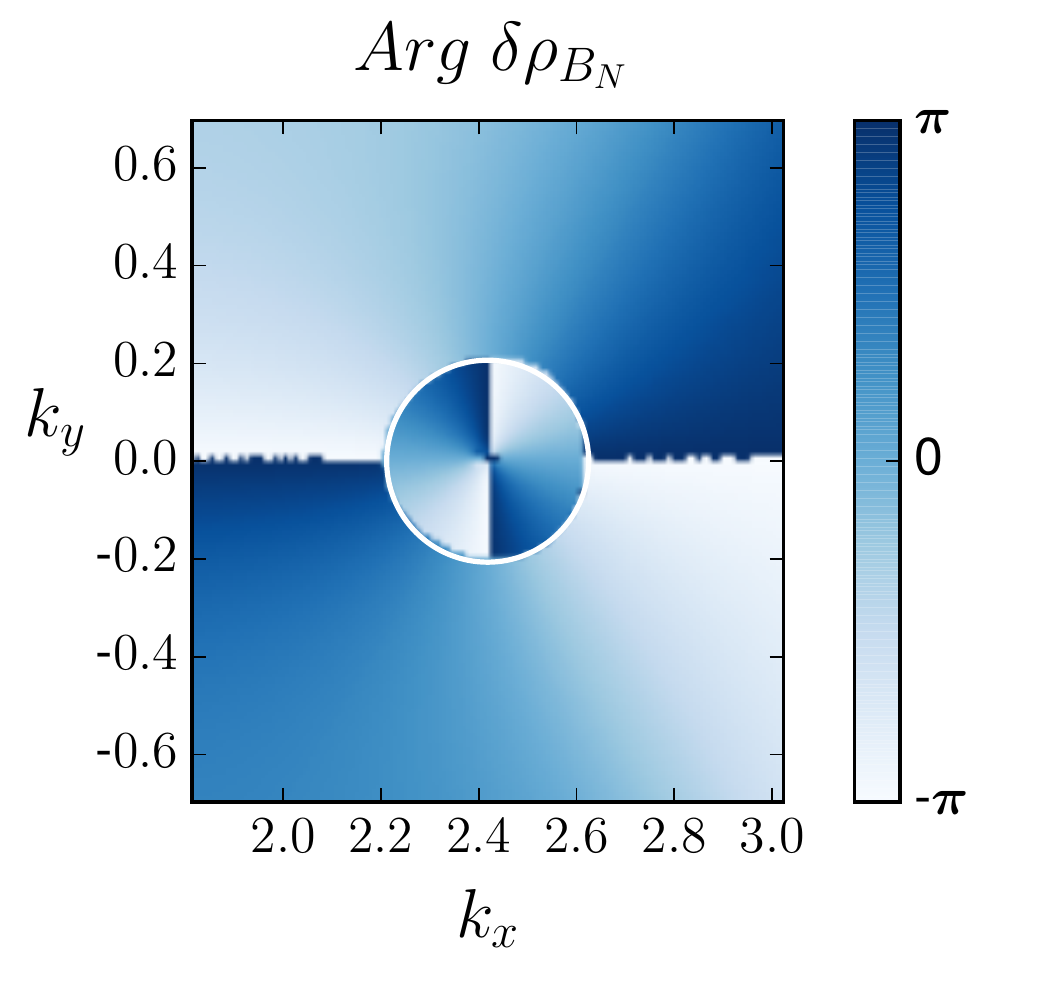}\\
\end{array}$
\caption{\small (Color online) Polar representation of the LDOS Fourier transform induced by the scattering between non-equivalent valleys on the pristine surface . It is illustrated for the valleys that are related to one another by $\delta m = 1$, $\delta n = -1$ and $\xi=-\xi'=-1$. The $2q_{F}$-radius circle is mentioned in white as a guide for the eyes. The first row refers to the impurity surface of monolayer graphene ($N=1$) and thus $\delta \rho=\delta\rho_{A_{1}}+\delta\rho_{B_{1}}$. The second and third rows are both obtained for bilayer graphene when the impurity lies on sublattices A$_{1}$ and B$_{1}$, respectively. In both cases the LDOS modulations on the pristine surface mainly involve sublattice B$_{N}$ at low energy, so that only $\delta\rho_{B_{N}}$ is mentioned.}
\label{LDOS FT Pristine Surface Polar Representation Zoom}
\end{figure}

\begin{figure}[b]
\centering
$\begin{array}{cc}
\includegraphics[trim = 40mm 0mm 05mm 0mm, clip, width=4.2cm]{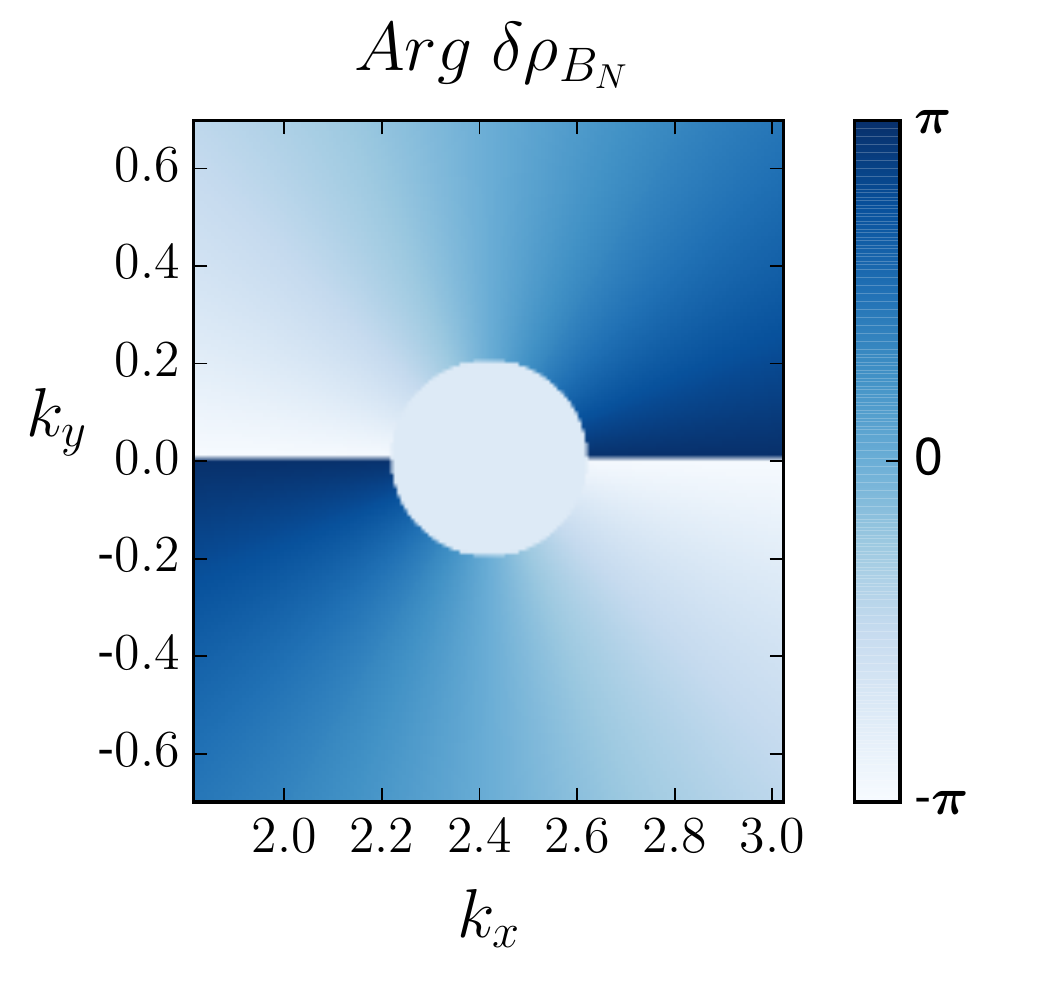}&
\includegraphics[trim = 40mm 0mm 05mm 0mm, clip, width=4.2cm]{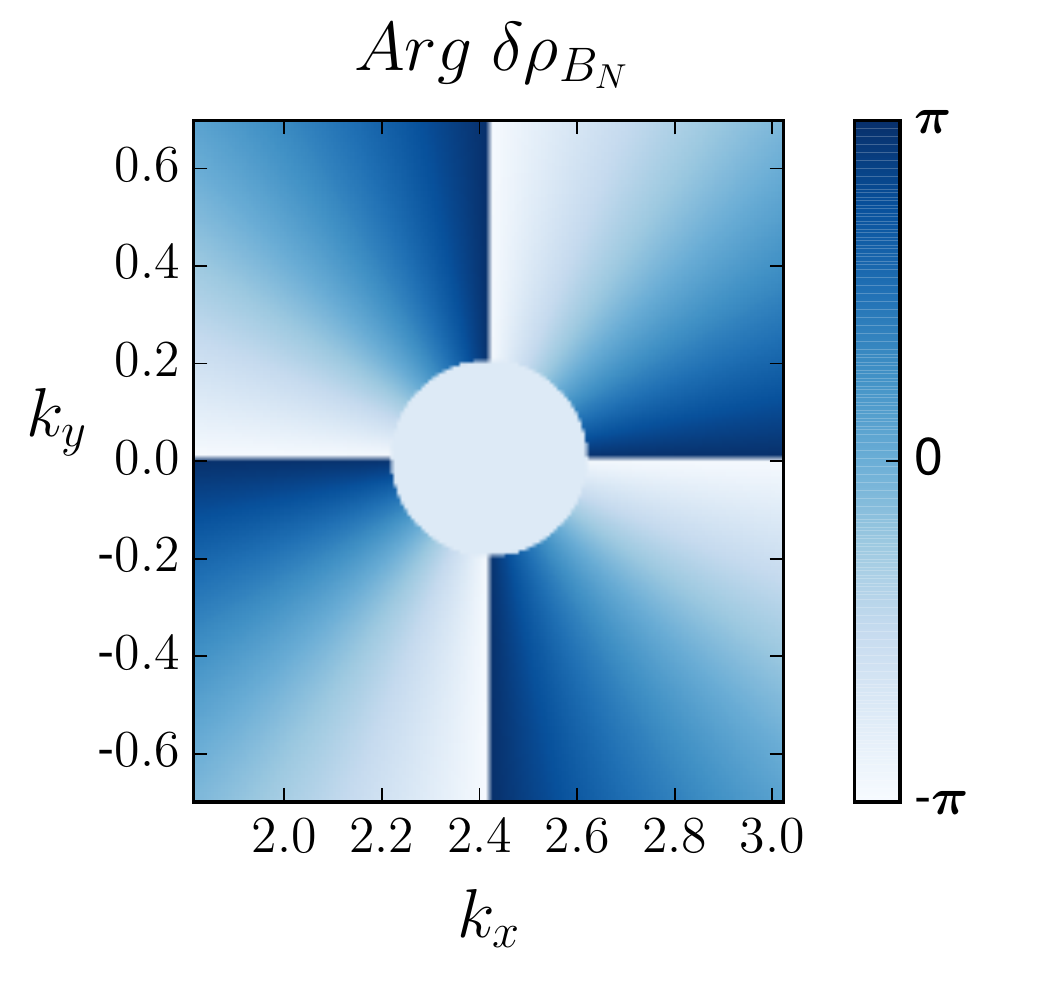}\\
\includegraphics[trim = 40mm 0mm 05mm 0mm, clip, width=4.2cm]{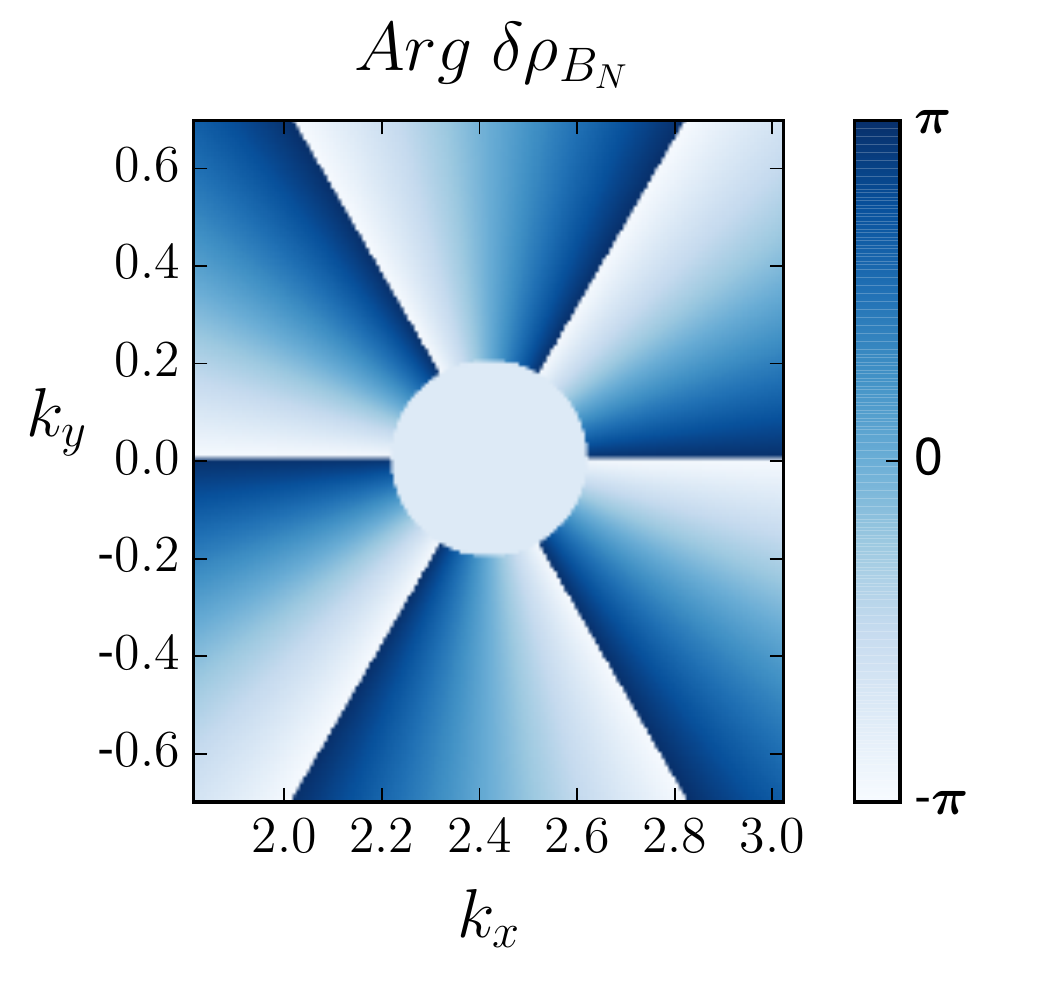}&
\includegraphics[trim = 40mm 0mm 05mm 0mm, clip, width=4.2cm]{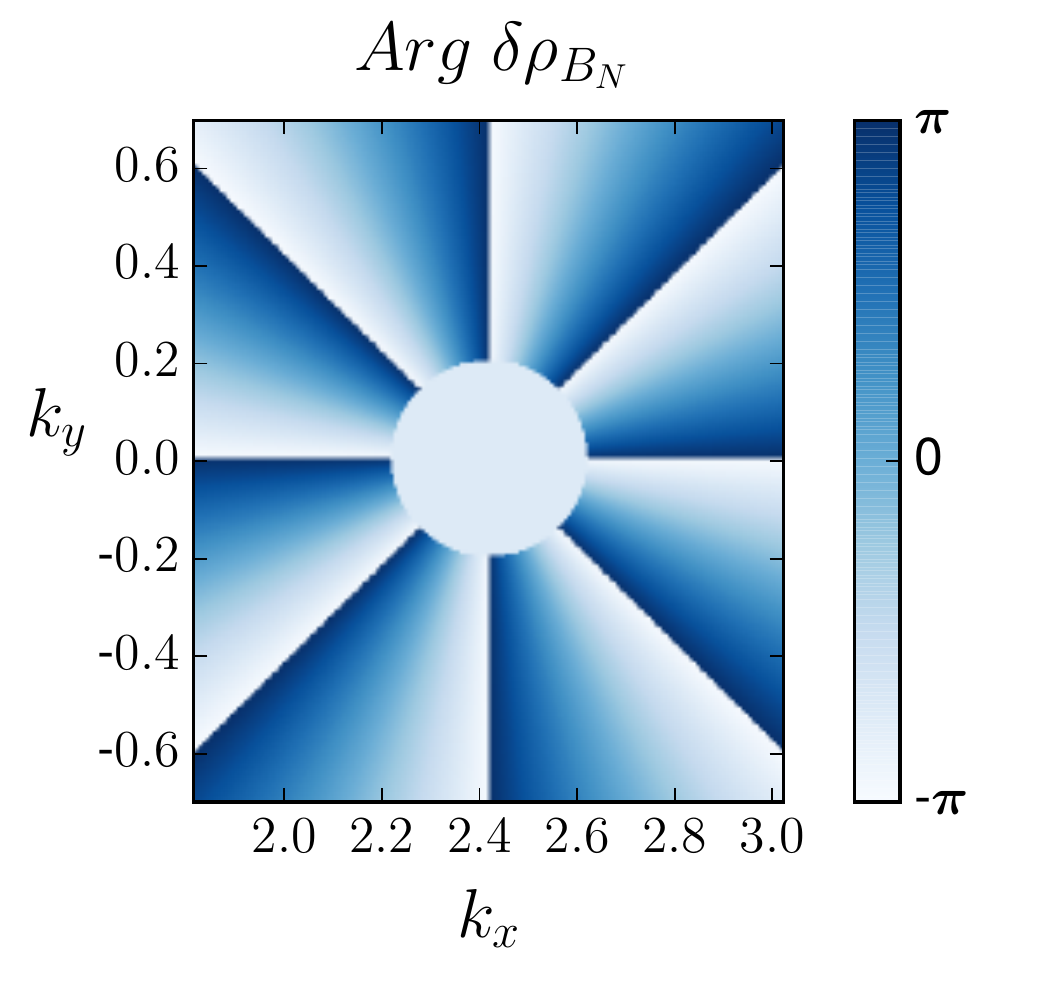}\\
\end{array}$
\caption{\small (Color online) Argument of the LDOS Fourier transform due to scattering between non-equivalent valleys for $N=1$ (top left-hand corner), $N=2$ (top right-hand corner), $N=3$ (bottom left-hand corner) and $N=4$ (bottom right-hand corner). This multivalued function, obtained from the analytic expression (\ref{LDOS FT Sublattice BN Impurity A1}), winds $2N$ times when the wave vector describes a closed path that encloses once the $2q_{F}$-radius ring. This winding number leads to the Berry phase difference between the two non-equivalent valleys involved in the scattering.}
\label{FT STM Zoom Polar Angle}
\end{figure}

If the impurity lies on sublattice A$_{1}$, the $1/r$-decaying Friedel oscillations induced at the opposite pristine surface are described by Eq. (\ref{LDOS Impurity on A1}) and have the following Fourier transform
\begin{align}\label{LDOS FT Sublattice BN Impurity A1}
\delta\rho_{B_{N}}({\bf \Delta K} + {\bf q}, \omega) &\simeq 
-~ \frac{\Theta(q-2q_{F})}{\sqrt{q^{2}-(2q_{F})^{2}}} \\
&\times (-\xi \xi')^{N}~ e^{-iN {\bf \Delta K}\cdot{\bf d_{3}}}~ e^{-iN(\theta^{\xi}({\bf q})-\theta^{\xi'}({\bf q}) )} ~. \notag
\end{align}
This expression is derived in Appendix \ref{Fourier transform of the Friedel oscillations} when considering $\nu=N$. The scattering between equivalent valleys ($\xi=\xi'$) is again responsible for a $2q_{F}$-radius ring regardless of the number of layers. So the Fermi contours involved in this kind of scattering have a circular symmetry whatever the number of layers stacked in the material.

Importantly, the scattering which takes place between non-equivalent valleys ($\xi=-\xi'$) is more instructive about the Bloch band structure at low energy. Zooms of the interference patterns it leads to are depicted in Fig. \ref{LDOS FT Pristine Surface RI Representation Zoom}. It shows that the real part of the LDOS Fourier transform reduces the circular symmetry of the $2q_{F}$-radius ring to a $2N$-fold rotational symmetry, which immediately reveals how many layers are stacked in the material. It is of course more natural to present the LDOS Fourier transform in the polar representation, which is illustrated in Fig. \ref{LDOS FT Pristine Surface Polar Representation Zoom}. Once again, the modulus exhibits a $2q_{F}$-radius ring due to the circular symmetry of the Fermi contours involved in the scattering. However, the argument is given by $2N\theta^{\xi}({\bf q})$ and exactly maps twice the phase that encodes the momentum dependence of the Bloch spinors defined in Eq. (\ref{Bloch eigenstates}). Conceptually, this means that one can image the whole Bloch band structure described by the two-band Hamiltonian matrix $(\ref{Effective hamiltonian})$ at low energy.

Then one can access some features of the band structure such as the Berry phase which characterizes the nodal point in every valley. Indeed, the LDOS Fourier transform winds $2N$ times when the wave vector ${\bf q}$ runs once along the $2q_{F}$-radius circle. So the Berry phase is simply given by $\pi$ times half the number of discontinuity lines that connect the $2q_{F}$-radius ring in the $[-\pi,+\pi]$ representation of Fig. \ref{FT STM Zoom Polar Angle}. It is worth mentioning here that, more generally, the momentum dependent argument of the LDOS Fourier transform is given by the phase difference $\theta^{\xi}({\bf q})-\theta^{\xi'}({\bf q})$, as defined in Eq. (\ref{LDOS FT Sublattice BN Impurity A1}). Then its variation along the $2q_{F}$-radius ring actually leads to the Berry-phase difference between the two valleys ${\bf K^{\xi}_{mn}}$ and ${\bf K^{\xi'}_{m'n'}}$ involved in the scattering. So the fact that the interference pattern leads to twice the Berry phase in the case of rhombohedral multilayer graphene is a consequence of the time-reversal symmetry which requires non-equivalent valleys to have opposite Berry phases.

\subsubsection{Localized impurity on sublattice B$_{1}$}

Finally, when the impurity lies on sublattice B$_{1}$, the $1/r$-decaying Friedel oscillations induced at the opposite pristine surface are described by Eq. (\ref{LDOS Impurity on B1}) and lead to the following Fourier transform
\begin{align}\label{LDOS FT Sublattice BN Impurity B1}
\delta\rho_{B_{N}}({\bf \Delta K} + {\bf q}, \omega) &\simeq 
-~ \frac{\Theta(q-2q_{F})}{\sqrt{q^{2}-(2q_{F})^{2}}} \\
&\times (-\xi \xi')^{N-1}~ e^{-i(N-1){\bf \Delta K}\cdot{\bf d_{3}}}~ e^{-i(N-1)(\theta^{\xi}({\bf q})-\theta^{\xi'}({\bf q}) )} ~. \notag
\end{align}
This expression is derived in Appendix \ref{Fourier transform of the Friedel oscillations} when considering $\nu=N-1$. Again, a $2q_{F}$-radius ring is outlined by the modulus of the LDOS Fourier transform induced by scattering between equivalent valleys ($\xi=\xi'$), as well as by scattering between non-equivalent valleys ($\xi=\xi'$). The argument of the LDOS Fourier transform that refers to scattering between non-equivalent valleys maps 2($N-$1) times the phase $\theta^{\xi}({\bf q})$. This is illustrated in the third row in Fig. \ref{LDOS FT Pristine Surface Polar Representation Zoom} in the case of bilayer graphene.

So the momentum-space interference pattern no longer directly leads to the phase of the Bloch spinors defined in Eq. (\ref{Bloch eigenstates}) nor to the Berry phase, when the impurity lies on sublattice B$_{1}$. Nevertheless, the Fourier transform of the scattering between non-equivalent valleys maps twice the phase $\theta^{\xi}({\bf q})$ on the surface where there is the impurity, and 2($N-$1) times the phase $\theta^{\xi}({\bf q})$ on the opposite pristine surface. This means that the number of discontinuity lines in the argument of the LDOS Fourier transform again highlights the number of layers stacked in the material, if one can access the LDOS of both surfaces.

\subsection{Generic momentum space behavior of Friedel oscillations}
Finally, we present the generic momentum space behavior of Friedel oscillations described by Eq. (\ref{LDOS Impurity in bulk}), which means it also describes bulk impurities and not just surface impurities as discussed above. For more simplicity we again assume that the impurity is localized on sublattice $A_{N_{0}}$. When disregarding the factor $t(\omega)/4^{2}N^{2}\omega^{2-(2N_{0}+1)/N}$ in Eq. (\ref{LDOS Impurity in bulk}), the Fourier transform of the $1/r$-decaying Friedel oscillations is given by
\begin{widetext}
\begin{align}
\left \{
\begin{aligned}
\delta\rho_{A_{1}}({\bf \Delta K} + {\bf q}, \omega) &\simeq -~ \frac{\Theta(q-2q_{F})}{\sqrt{q^{2}-(2q_{F})^{2}}}~
 (-\xi \xi')^{N_{0}-1}~ e^{i(N_{0}-1) {\bf \Delta K}\cdot{\bf d_{3}}}~ e^{i(N_{0}-1)(\xi-\xi')\theta_{\bf q}} ~  \\
\delta\rho_{B_{N}}({\bf \Delta K} + {\bf q}, \omega) &\simeq -~ \frac{\Theta(q-2q_{F})}{\sqrt{q^{2}-(2q_{F})^{2}}}~
 (-\xi \xi')^{N-N_{0}+1}~ e^{-i(N-N_{0}+1) {\bf \Delta K}\cdot{\bf d_{3}}}~ e^{-i(N-N_{0}+1)(\xi-\xi')\theta_{\bf q}} ~
\end{aligned}
\right . ~.
\end{align}
\end{widetext}

In the case of scattering between non-equivalent valleys ($\xi=-\xi'$), the argument of the LDOS Fourier transform on the sublattice A$_{1}$ maps $2(N_{0}-1)$ times the phase $\theta^{\xi}({\bf q})$, while it maps $2(N-N_{0}+1)$ times $-\theta^{\xi}({\bf q})$ on the opposite sublattice B$_{N}$. As a result, $\Arg[\delta\rho_{A_{1}}]-\Arg[\delta\rho_{B_{N}}]$ maps twice the phase that encodes the momentum dependence of the Bloch spinors, namely $2N\theta_{\bf q}$. It follows that the Berry phase of the valley $\xi$ is given by
\begin{align}
\frac{1}{4}\oint_{{\cal C}_{2{\bf q_{F}}}}d{\bf q}\cdot \nabla_{\bf q} [\Arg \delta\rho_{A_{1}}-\Arg \delta\rho_{B_{N}}]=\xi N\pi,
\end{align}
where ${\cal C}_{2{\bf q_{F}}}$ is a path that encloses once the $2q_{F}$ radius circle at the extremity of the scattering wave vector ${\bf \Delta K}$ that connects two non-equivalent valleys. This momentum space signature additionally provides the number of layers that are stacked in the material.

From Appendix \ref{Fourier transform of the Friedel oscillations}, it can be remarked that, even though the modulus of the LDOS Fourier transform consists of the signature of the algebraic decay of Friedel oscillations, the momentum dependent argument appears as an overall factor in the Fourier signature obtained from Eq. (\ref{Appendix LDOS BN 1/r FT}) and, therefore, does not depend on the peculiar algebraic decay involved. This means that the argument of the LDOS Fourier transform should be observable even when imaging the LDOS quite close to the impurity, where higher power laws are dominant.

\section*{Conclusion}
\label{Conclusion}
The low-energy physics of rhombohedral $N$-layer graphene mainly takes place on the two external layers, where the electronic band structure defines a semimetal. In this paper, we have addressed the problem of elastic scattering through a localized impurity located either on the surface, or within the bulk. The scatterer induces Friedel oscillations that always decay as $1/r$ on both surfaces in multilayer graphene. This result depends neither on the layer of the impurity, nor on the magnitude of its potential. Therefore, monolayer graphene is the only material of the rhombohedral class that exhibits $1/r^{2}$-decaying long-range interferences.

The Friedel oscillations have then been analyzed in terms of their Fourier transforms. When imaging the surface on which the impurity is located, the momentum-space interference pattern enables us to clearly distinguish monolayer graphene from rhombohedral multilayer graphene. This distinction can be made from intravalley scattering, as well as from the scattering that takes place between non-equivalent valleys. The latter also reveals the sublattice of the impurity in the case of multilayer graphene.
 
More generally and regardless of the layer on which the impurity is located, the interferences induced on the two opposite surfaces by scattering between non-equivalent valleys highlight the number of layers stacked in the material. We have shown that they also reveal the whole Bloch band structure of rhombohedral multilayer graphene at low energy, i.e. the Fermi contours as well as the momentum dependence of the Bloch spinors along them. Interestingly, this subsequently leads to the $\pi$-quantized Berry phases that characterize the nodal points in the dispersion relation of this time-reversal invariant two-band semimetal. Such an observation by STM would mean that this experimental technique would be able to probe the whole Bloch band structure of a two-dimensional electronic system.

\section*{Acknowledgements}
\label{Acknowledgements}
The authors would like to thank A. N. Rudenko for useful comments. This work was supported by NWO via a Spinoza Prize and by European Union Seventh Framework Programme under Grant Agreement No. 604391 Graphene Flagship.

\appendix

\section{Low-energy band structure}\label{Appendix Low-energy Band Structure}
This appendix reminds of the effective two-band description of rhombohedral multilayer graphene at low energy. The electronic band structure of the material relies on the recursive system introduced in Eq. (\ref{Recursive System}), namely
\begin{align}
\left \{
\begin{aligned}
f({\bf k})~ B_{1} &= E~ A_{1} \\
f^{*}({\bf k})~ A_{n-1}+t_{\perp}~ A_{n} &= E~ B_{n-1} \\
t_{\perp}~ B_{n-1}+f({\bf k})~ B_{n} &= E~ A_{n} \\
f^{*}({\bf k})~ A_{N} &= E~ B_{N}
\end{aligned}
\right . ~,
\end{align}
where $E$ denotes the eigenenergy, $A_{n}$ (respectively $B_{n}$) refers to the electronic orbitals of sublattice A$_{\text{n}}$ (respectively B$_{\text{n}}$), and $n$ runs from $2$ up to $N$. The above system can be rewritten as
\begin{widetext}
\begin{align}
\left \{
\begin{aligned}
f({\bf k})~ B_{1} &= E~ A_{1} \\
f({\bf k})~ B_{n} + t_{\perp}~ A_{n} &= -t_{\perp}\Big(1-\frac{E}{t_{\perp}}-\frac{E^{2}}{t^{2}_{\perp}}\Big)~B_{n-1}-f^{*}({\bf k})\Big(1+\frac{E}{t_{\perp}}\Big)~A_{n-1} \\
f({\bf k})~ B_{n} - t_{\perp}~ A_{n} &= -t_{\perp}\Big(1+\frac{E}{t_{\perp}}-\frac{E^{2}}{t^{2}_{\perp}}\Big)~B_{n-1}+f^{*}({\bf k})\Big(1+\frac{E}{t_{\perp}}\Big)~A_{n-1} \\
f^{*}({\bf k})~ A_{N} &= E~ B_{N}
\end{aligned}
\right . ~
\end{align}
\end{widetext}
and, in the limit $E \ll t_{\perp}$, it reduces to
\begin{align}
\left \{
\begin{aligned}
-t_{\perp} \Big( \frac{f({\bf k})}{-t_{\perp}} \Big)^{N}~ B_{N} &\simeq E~ A_{1} \\
-t_{\perp} \Big( \frac{f^{*}({\bf k})}{-t_{\perp}} \Big)^{N}~ A_{N} &\simeq E~ B_{N}
\end{aligned}
\right . ~
\end{align}

The low-energy band structure is finally given by
\begin{align}
\mathcal{H}_{N}({\bf k})= \Big( \frac{-1}{t_{\perp}} \Big)^{N-1}
\left( \begin{array}{cc} 
0 & f^{N}({\bf k}) \\
f^{*^{N}}({\bf k}) & 0
\end{array} \right) ~
\end{align}
and only involves the outer sublattices A$_{1}$ and B$_{\text{N}}$.

\section{Some integrals}\label{Appendix Some Integrals}

In this appendix, we evaluate some integrals that describe the algebraic decay of the bare Green functions at large distances.

\subsection{Integral $I^{(1)}_{M,N}({\bf r}, \omega)$}
\begin{figure}[t]
\centering
\includegraphics[trim = 00mm 00mm 00mm 00mm, clip, width=6.0cm]{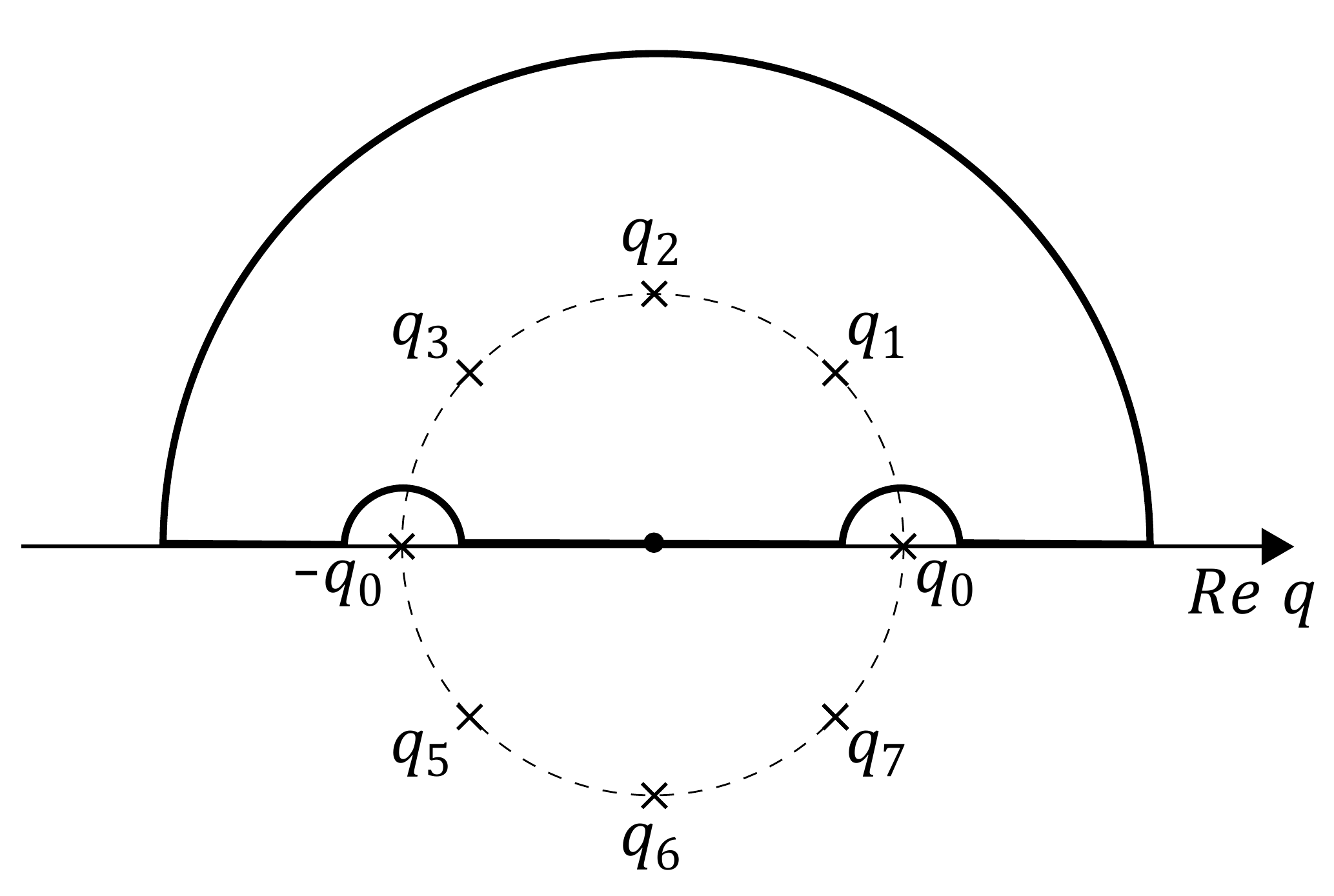}
\caption{\small Integration contour and poles involved in the integral (\ref{Appendix Integral 2}) for $N=4$.}
\label{PolesIllustration}
\end{figure}

For the integers $M$ such as $M<N$, we introduce the following integral:
\begin{align}\label{Appendix Intgral I1 Definition}
I^{(1)}_{M,N}({\bf r}, \omega) &= \int_{{\mathbb{R}}^{2}}\frac{d^{2}q}{(2\pi)^{2}}~ \frac{q^{2M}}{\omega^{2}-q^{2N}}~e^{i{\bf q}\cdot{\bf r}} \notag \\
&= \int_{0}^{+\infty}\frac{dq}{2\pi}~ \frac{q^{2M+1}~J_{0}(qr)}{\omega^{2}-q^{2N}} \notag \\
&= \int_{0}^{\infty}\frac{dq}{\pi}~ \frac{q^{2M+1}}{\omega^{2}-q^{2N}} \int_{1}^{+\infty}\frac{du}{\pi}~\frac{\sin(qru)}{\sqrt{u^{2}-1}} \notag \\
&= \frac{1}{2\pi^{2}} \int_{1}^{+\infty}\frac{du}{\sqrt{u^{2}-1}}~ \int_{\mathbb{R}} dq~ \frac{q^{2M+1}~\sin(qru)}{\omega^{2}-q^{2N}}~.
\end{align}
The fraction that appears in the integral over the momentum has $2N$ simple poles, namely $q_{n}=\omega^{\frac{1}{N}}e^{in\frac{\pi}{N}}$, where the integer $n$ runs from 0 up to $2N-1$. They are illustrated on Fig. \ref{PolesIllustration} when $N=4$. Via its residues, this fraction can be decomposed into a sum of elementary fractions that have a first-order polynomial as denominator
\begin{align}\label{Appendix Integral 2}
\int_{{\mathbb R}} dq~ \frac{q^{2M+1}~\sin(qru)}{\omega^{2}-q^{2N}} &= \sum_{n=0}^{2N-1} \frac{1}{2Nq_{n}^{2N-2M-1}} \int_{{\mathbb R}} dq~ \frac{q~\sin(qru)}{q_{n}-q}
\end{align}
Given that $ru >0$, two cases have to be distinguished:
\begin{itemize}
\item $\Imag q_{n} \neq 0$: if $\Imag q_{n} > 0$, then the residue theorem implies that
\begin{align}
\int_{{\mathbb R}} dq~ \frac{q~\sin(qru)}{q_{n}-q}=-\pi~ q_{n}~e^{iq_{n}ru}~.
\end{align}
All these poles lead to terms that decay exponentially with the distance to the impurity. Here, we disregard such terms, since we are interested in the algebraic decay of the Friedel oscillations. Of course, the same conclusion holds for poles with negative imaginary parts.
\item $\Imag q_{n} = 0$: whatever the number of layers is, there are always two opposite poles on the real axis, namely $q_{0}=\omega^{\frac{1}{N}}$ and $-q_{0}$. The Cauchy principal value of the integral (\ref{Appendix Integral 2}) is obtained from the residue theorem
\begin{align}
\pv \int_{{\mathbb R}} dq~ \frac{q~\sin(qru)}{q_{0}-q} &= \Imag \int_{{\mathbb R}} dq~ \frac{q~e^{iqru}}{q_{0}-q} \notag \\
&= -\pi~q_{0}\cos (q_{0}ru)~,
\end{align}
while its kernel is expressed in terms of the Dirac $\delta$ distribution
\begin{align}
-i\pi~\langle \delta(q_{0}-q) | q \sin (qru)\rangle &= -i\pi~ q_{0}\sin(q_{0}ru)~.
\end{align}
\end{itemize}
When considering only the two real poles, the integral (\ref{Appendix Integral 2}) is estimated by
\begin{align}
\int_{{\mathbb R}} dq~ \frac{q^{2M+1}~\sin(qru)}{\omega^{2}-q^{2N}} &\simeq -\frac{\pi}{Nq_{0}^{2(N-M-1)}}~e^{iq_{0}ru}~.
\end{align}
Finally, the integral introduced in Eq. (\ref{Appendix Intgral I1 Definition}) is rewritten as
\begin{align}\label{Appendix Intgral I1 Result}
I^{(1)}_{M,N}({\bf r}, \omega) &\simeq - \frac{i}{4N\omega^{2(1-\frac{M+1}{N})}}~H_{0}\big(\omega^{\frac{1}{N}}r\big)~,
\end{align}
where $H_{0}$ is the zeroth-order Hankel function of the first kind.

\subsection{Integral $I^{(2)}_{L,M,N}({\bf r}, \omega)$}
At present, let us estimate the following integral:
\begin{align}\label{Appendix Intgral I2 Definition}
I^{(2)}_{L,M,N}({\bf r}, \omega) &= - \int_{{\mathbb{R}}^{2}}\frac{d^{2}q}{(2\pi)^{2}}~ \frac{q^{2M}}{\omega^{2}-q^{2N}}~\Big( q e^{i\xi\theta_{\bf q}}\Big)^{L} ~e^{i{\bf q}\cdot{\bf r}} \notag \\
&=  - \xi^{L}~e^{iL\xi(\theta_{\bf r}+\frac{\pi}{2})}~ \int_{0}^{+\infty}\frac{dq}{2\pi}~ \frac{q^{2M+1}}{\omega^{2}-q^{2N}}~q^{L}J_{L}(qr) ~,
\end{align}
where $\theta_{\bf q}$ and $\theta_{\bf r}$ denote the polar angles of the momentum ${\bf q}$ and the coordinate vector ${\bf r}$, respectively. In the above equation, we have also used the fact that $\xi = \pm1$, which implies that $J_{\xi L}(qr) = \xi^{L} J_{L}(qr)$. Because Bessel functions satisfy the recurrence relation
\begin{align}\label{Bessel recurrence relation}
J_{L}(x)=(-1)^{L} x^{L} \Big( \frac{1}{x} \frac{d}{dx} \Big)^{L}J_{0}(x) ~,
\end{align}
the integral  $I^{(2)}_{L,M,N}$ can be related to the spatial derivatives of $I^{(1)}_{M,N}$. This results in
\begin{align}
I^{(2)}_{L,M,N}({\bf r}, \omega) &\simeq \frac{i~\xi^{L}~e^{iL\xi(\theta_{\bf r}+\frac{\pi}{2})}}{4N\omega^{2(1-\frac{M+1}{N})}}~
(-1)^{L} r^{L} \Big( \frac{1}{r} \frac{d}{dr} \Big)^{L}H_{0}\big(\omega^{\frac{1}{N}}r\big) ~.
\end{align}
But Hankel functions also satisfy the recurrence relation (\ref{Bessel recurrence relation}), so the integral finally behaves as
\begin{align}\label{Appendix Intgral I2 Result}
I^{(2)}_{L,M,N}({\bf r}, \omega) &\simeq \frac{i~\omega^{\frac{L}{N}}~e^{iL\xi\theta_{\bf r}}}{4N\omega^{2(1-\frac{M+1}{N})}}~
~ i^{L} H_{L}\big(\omega^{\frac{1}{N}}r\big) ~,
\end{align}
where $H_{L}$ is the $L$th order Hankel function of the first kind.

\section{Bare Green functions at large distances}\label{Appendix Bare Green Functions}
This appendix focuses on the expressions of the bare Green functions at large distances. Only some of their components are presented here, but they turn out to be sufficient in order to evaluate the LDOS at the surfaces of rhombohedral multilayer graphene. Their expressions are based on the integrals approximated in Appendix \ref{Appendix Some Integrals}.

\subsection{Within the two-band description}
When the impurity is localized sublattice A$_{1}$, the bare Green function is a $2\times 2$ matrix which can be written as
\begin{align}
G^{(0)}({\bf K^{\xi}_{mn}}+{\bf q}, \omega) &\simeq \frac{1}{\omega^{2}-q^{2N}} \notag \\
& \times
\left( \begin{array}{cc} 
\omega & -\Big(\xi q e^{i\theta^{\xi}_{mn}(\bf q)} \Big)^{N} \\
-\Big(\xi q e^{-i\theta^{\xi}_{mn}(\bf q)} \Big)^{N} & \omega
\end{array} \right) ~ \notag \\
\end{align}
in the vicinity of any valley ${\bf K^{\xi}_{mn}}$. The real-space representation of the bare Green function is then obtained from
\begin{align}\label{Appendix Real Space Bare GF}
G^{(0)}({\bf r}, \omega) &\simeq \sum_{m,n,\xi} e^{i{\bf K^{\xi}_{mn}}\cdot{\bf r}}~ \int_{{\mathbb R}^{2}}\frac{d^{2}q}{(2\pi)^{2}}~ G^{(0)}({\bf K^{\xi}_{mn}}+{\bf q}, \omega)~e^{i{\bf q}\cdot{\bf r}} ~.
\end{align}
At large distances, it is found that
\begin{align}\label{Appendix Two-band Bare GF A1A1}
G^{(0)}_{A_{1}A_{1}}({\bf r}, \omega) &\simeq \sum_{m,n,\xi} e^{i{\bf K^{\xi}_{mn}}\cdot{\bf r}}~\omega ~I^{(1)}_{0,N}({\bf r}, \omega) \notag \\
&\simeq \frac{-i}{4N\omega^{1-\frac{2}{N}}}~ H_{0}\big(\omega^{\frac{1}{N}}r\big)~\sum_{m,n,\xi} e^{i{\bf K^{\xi}_{mn}}\cdot{\bf r}} ~,
\end{align}
and
\begin{align}\label{Appendix Two-band Bare GF A1BN}
G^{(0)}_{A_{1}B_{N}}({\bf r}, \omega) &\simeq \sum_{m,n,\xi} e^{i{\bf K^{\xi}_{mn}}\cdot{\bf r}}~ \Big(\xi e^{i{\bf K^{\xi}_{mn}}\cdot{\bf d_{3}}} \Big)^{N}~ I^{(2)}_{N,0,N}({\bf r}, \omega) \\
&\simeq \frac{i}{4N\omega^{1-\frac{2}{N}}} ~ i^{N} H_{N}\big(\omega^{\frac{1}{N}}r\big)~ \sum_{m,n,\xi} \xi^{N}~ e^{i{\bf K^{\xi}_{mn}}\cdot{\bf r}}~ e^{iN\theta^{\xi}_{mn}(\bf r)} \notag ~.
\end{align}
The two other components can be obtained in a similar way.

\subsection{Within the four-band description}
When the impurity is localized on sublattice B$_{1}$, the bare Green function is a $4\times 4$ matrix which can be written as
\begin{align}
G^{(0)}({\bf K^{\xi}_{mn}}+{\bf q}, \omega) &\simeq \frac{1}{D({\bf q},\omega)} \notag \\
& \times
\left( \begin{array}{cccc} 
.... & \xi q e^{i\theta^{\xi}_{mn}(\bf q)}~(\omega^{2}-q^{2(N-1)}) & .... & ....  \\
.... & -\omega~(\omega^{2}-q^{2(N-1)}) & .... & .... \\
.... & .... & .... & .... \\
.... & \Big( \xi q e^{-i\theta^{\xi}_{mn}(\bf q)} \Big)^{N-1}\omega & .... & ....
\end{array} \right)
\end{align}
in the vicinity of any valley ${\bf K^{\xi}_{mn}}$ and where
\begin{align}
D({\bf q},\omega) &= \omega^{4} - \omega^{2}~\Big(1-q^{2}-q^{2(N-1)}\Big)+q^{2N} \notag \\
&\simeq -\omega^{2}+q^{2N}
\end{align}
since the energy we consider, given in units of $t_{\perp}$, satisfies $\omega \ll 1$. At large distances, it is found that
\begin{align}\label{Appendix Four-band Bare GF A1B1}
G^{(0)}_{A_{1}B_{1}}({\bf r}, \omega) &\simeq \sum_{m,n,\xi} e^{i{\bf K^{\xi}_{mn}}\cdot{\bf r}}~ \xi e^{i{\bf K^{\xi}_{mn}}\cdot{\bf d_{3}}}~ 
\Big( I^{(2)}_{1,N-1,N}({\bf r}, \omega) - \omega^{2}I^{(2)}_{1,0,N}({\bf r},\omega) \Big) \notag \\
&\simeq \sum_{m,n,\xi} e^{i{\bf K^{\xi}_{mn}}\cdot{\bf r}}~ \xi e^{i{\bf K^{\xi}_{mn}}\cdot{\bf d_{3}}}~ 
\Big( \omega^{2(1-\frac{1}{N})} - \omega^{2} \Big)~I^{(2)}_{1,0,N}({\bf r},\omega) \notag \\
&\simeq \frac{i\omega^{\frac{1}{N}}}{4N}~ i H_{1}\big(\omega^{\frac{1}{N}}r\big)~
\sum_{m,n,\xi} e^{i{\bf K^{\xi}_{mn}}\cdot{\bf r}}~ \xi e^{i\theta^{\xi}_{mn}(\bf r)}~,
\end{align}
\begin{align}\label{Appendix Four-band Bare GF B1B1}
G^{(0)}_{B_{1}B_{1}}({\bf r}, \omega) &\simeq \sum_{m,n,\xi} e^{i{\bf K^{\xi}_{mn}}\cdot{\bf r}}~ \omega~
\Big( I^{(1)}_{N-1,N}({\bf r}, \omega) - \omega^{2}I^{(1)}_{0,N}({\bf r},\omega) \Big) \notag \\
&\simeq \sum_{m,n,\xi} e^{i{\bf K^{\xi}_{mn}}\cdot{\bf r}}~ \omega~ 
\Big( \omega^{2(1-\frac{1}{N})} - \omega^{2} \Big)~I^{(1)}_{0,N}({\bf r},\omega) \notag \\
&\simeq \frac{-i\omega}{4N}~H_{0}\big(\omega^{\frac{1}{N}}r\big) \sum_{m,n,\xi} e^{i{\bf K^{\xi}_{mn}}\cdot{\bf r}}~,
\end{align}
and
\begin{align}\label{Appendix Four-band Bare GF BNB1}
G^{(0)}_{B_{N}B_{1}}({\bf r}, \omega) &\simeq \sum_{m,n,\xi} e^{i{\bf K^{\xi}_{mn}}\cdot{\bf r}}~ \omega~ 
\Big( \xi e^{-i{\bf K^{\xi}_{mn}}\cdot{\bf d_{3}}} \Big)^{N-1}~ I^{(2)}_{N-1,0,N}({\bf r}, \omega) \notag \\
&\simeq \frac{i\omega^{\frac{1}{N}}}{4N}~ i^{N-1} H_{N-1}\big(\omega^{\frac{1}{N}}r\big)~  \sum_{m,n,\xi} e^{i{\bf K^{\xi}_{mn}}\cdot{\bf r}}~ \Big( \xi e^{-i\theta^{\xi}_{mn}({\bf r})} \Big)^{N-1} ~.
\end{align}
The real space representation of the components $G^{(0)}_{B_{1}A_{1}}$ and $G^{(0)}_{B_{1}B_{N}}$ can be obtained in a similar way given that
\begin{align}
G^{(0)}_{B_{1}A_{1}}({\bf K^{\xi}_{mn}}+{\bf q}, \omega)&=\Big( G^{(0)}_{A_{1}B_{N}}({\bf K^{\xi}_{mn}}+{\bf q}, \omega) \Big)^{*} \notag \\
G^{(0)}_{B_{1}B_{N}}({\bf K^{\xi}_{mn}}+{\bf q}, \omega)&=\Big( G^{(0)}_{B_{N}B_{1}}({\bf K^{\xi}_{mn}}+{\bf q}, \omega) \Big)^{*} ~.
\end{align}
As far as we are concerned, the other components are irrelevant, since they are not involved in the LDOS at the surfaces of the material.

\subsection{Asymptotic behavior of Hankel functions}
Here we just remind the reader of the asymptotic expansion of the $L$th order Hankel function. For $|z|\gg1$, it can be approximated by
\begin{align}\label{Hankel functions asymptotic}
i^{L}~ H_{L}^{(1)}\big(z\big) &\simeq \frac{e^{i(z-\frac{\pi}{4})}}{\sqrt{z}}\Big[ 1 + i~\Big(\frac{L^{2}}{2}-\frac{1}{8} \Big) \frac{1}{z} + ... \Big]~.
\end{align}

\section{Fourier transform of Friedel oscillations}\label{Fourier transform of the Friedel oscillations}
This appendix is devoted to the momentum space representation of the $1/r$-decaying Friedel oscillations. If we disregard the ${\bf r}$-independent factors in the LDOS expressions given in Eqs. (\ref{LDOS Impurity on A1}) and (\ref{LDOS Impurity on B1}), then the Fourier transform of Friedel oscillations is generically obtained from
\begin{widetext}
\begin{align}\label{Appendix LDOS BN 1/r FT}
\delta\rho({\bf k}, \omega, \nu) \simeq& -\frac{1}{\pi}~ \Big(-\xi \xi' \Big)^{\nu}~\int_{\mathbb{R}^2} d{\bf r}~ \frac{\cos \big( 2\omega^{\frac{1}{N}}r \big)}{r}~ \cos \big({\bf \Delta K}\cdot{\bf r} - \nu{\bf \Delta K}\cdot{\bf d_{3}} - \nu(\xi-\xi')\theta_{\bf r} \big)~ e^{-i{\bf k}\cdot{\bf r}} \notag \\
\simeq& -\frac{1}{2}~ \Big(-\xi \xi' \Big)^{\nu}~e^{-i\nu{\bf \Delta K}\cdot{\bf d_{3}}}~ e^{-i\nu(\xi-\xi')(\theta_{{\bf k}-{\bf \Delta K}}-\frac{\pi}{2})}~\lim\limits_{\epsilon \to 0^{+}} \int_{0}^{+\infty} dr~ \Big( e^{-(\epsilon-i2\omega^{\frac{1}{N}})r} + e^{-(\epsilon+i2\omega^{\frac{1}{N}})r} \Big)~ J_{(\xi-\xi')\nu}\big( |{\bf k} - {\bf \Delta K}|r \big) \notag \\
& -\frac{1}{2}~ \Big(-\xi \xi' \Big)^{\nu}~e^{i\nu{\bf \Delta K}\cdot{\bf d_{3}}}~~~ e^{i\nu(\xi-\xi')(\theta_{{\bf k}+{\bf \Delta K}}-\frac{\pi}{2})}~~~ \lim\limits_{\epsilon \to 0^{+}} \int_{0}^{+\infty} dr~ \Big( e^{-(\epsilon-i2\omega^{\frac{1}{N}})r} + e^{-(\epsilon+i2\omega^{\frac{1}{N}})r} \Big)~ J_{(\xi-\xi')\nu}\big( |{\bf k} + {\bf \Delta K}|r \big)~,
\end{align}
\end{widetext}
where the integrals have been regularized by $\epsilon>0$, so that they define Laplace transforms of Bessel functions. By definition, the Bessel function $J_{n}$ is a solution of the following non-linear differential equation:
\begin{align}
r^{2}J''_{n}(r) + r J'_{n}(r) +(r^{2}-n^{2}) J_{n}(r) = 0 ~.
\end{align}
Then $L_{n}$, which denotes the Laplace transform of $J_{n}$, satisfies
\begin{align}
(1+p^{2})~ L_{n}''(p)+3p~ L_{n}'(p)+(1-n^{2})~L_{n}(p)=0~.
\end{align}
It is possible to make this differential equation linear by the means of the following substitutions
\begin{align}
p=\sinh (s)~,~~~~L=\frac{1}{\cosh (s)} {\cal L}_{n}(s)~.
\end{align}
This leads to
\begin{align}\label{Appendix Linear Diff Eq}
{\cal L}''_{n}(s)-n^{2}{\cal L}_{n}(s)=0~.
\end{align}
Changing variables according to
\begin{align}
s=\ln \Big( \sqrt{1+p^{2}} +p~ \Big)
\end{align}
in the solutions of Eq. (\ref{Appendix Linear Diff Eq}) gives the Laplace transform of $J_{n}$, namely
\begin{align}
L_{n}(p)=A_{n}\frac{\Big( \sqrt{1+p^{2}} +p \Big)^{n}}{\sqrt{1+p^{2}}}+B_{n}\frac{\Big( \sqrt{1+p^{2}} +p \Big)^{-n}}{\sqrt{1+p^{2}}}~.
\end{align}
On the one hand, the condition
\begin{align}
\lim\limits_{p \to +\infty}pL_{n}(p)=J_{n}(0^{+})
\end{align}
implies that $A_{n}=0$. On the other hand, the asymptotic behaviors of a function and its Laplace transform are given by the series expansions
\begin{align}
J_{n}(x\rightarrow0^{+}) =& \sum_{k}a_{k}x^{k} \notag \\
L_{n}(p\rightarrow+\infty) =& \sum_{k} a_{k}\frac{k!}{p^{k+1}}  ~,
\end{align}
with the same coefficient $a_{k}$.
Since
\begin{align}
J_{n}(x\rightarrow0^{+}) \sim& \frac{x^{n}}{n!2^{n}} \notag \\
L_{n}(p\rightarrow+\infty) \sim& \frac{B_{n}}{2^{n}p^{n+1}} ~,
\end{align}
the last constant is $B_{n}=1$ and finally
\begin{align}\label{LT of Bessel}
L_{n}(p)=\frac{\Big( \sqrt{1+p^{2}} +p \Big)^{-n}}{\sqrt{1+p^{2}}}~.
\end{align}

This leads to the explicit expression of the LDOS introduced in Eq.(\ref{Appendix LDOS BN 1/r FT}), when considering the limit $\epsilon \rightarrow 0$
\begin{widetext}
\begin{align}
\delta\rho({\bf k}, \omega, \nu) 
&\simeq -\Big(-\xi \xi' \Big)^{\nu}~ \frac{e^{-i\nu{\bf \Delta K}\cdot{\bf d_{3}}}~ e^{-i\nu(\xi-\xi')(\theta_{{\bf k}-{\bf \Delta K}}-\frac{\pi}{2})}}{\sqrt{|{\bf k} - {\bf \Delta K}|^{2}-(2\omega^{\frac{1}{N}})^{2}}}~
\cos \Big( (\xi-\xi')\nu~\varphi_{{\bf k}-{\bf \Delta K}} \Big)~ \Theta \Big(|{\bf k} - {\bf \Delta K}|-2\omega^{\frac{1}{N}} \Big) \notag \\
&~~~ -\Big(-\xi \xi' \Big)^{\nu}~ \frac{e^{i\nu{\bf \Delta K}\cdot{\bf d_{3}}}~~ e^{i\nu(\xi-\xi')(\theta_{{\bf k}+{\bf \Delta K}}-\frac{\pi}{2})}}{\sqrt{|{\bf k} + {\bf \Delta K}|^{2}-(2\omega^{\frac{1}{N}})^{2}}}~~~ 
 \cos \Big( (\xi-\xi')\nu~\varphi_{{\bf k}+{\bf \Delta K}} \Big)~ \Theta \Big(|{\bf k} + {\bf \Delta K}|-2\omega^{\frac{1}{N}} \Big)
\end{align}
\end{widetext}
where $\Theta$ is the Heaviside step function and
\begin{align}\label{Phase Phi}
\varphi_{\bf k} = \Arg \Big[ \sqrt{k^{2}-(2\omega^{\frac{1}{N}})^{2}} +i2\omega^{\frac{1}{N}} \Big] ~.
\end{align}

In the vicinity of each valley, the energy scales as $\omega^{\frac{1}{N}} \sim q_{F}$, with $q_{F}$ the Fermi momentum. And any  scattering process can be associated to the wave vector ${\bf k}={\bf \Delta K} + {\bf q}$ where $ q \ll K_{00}^{\xi}$. Then a first-order expansion in the limit $q \rightarrow 2q_{F}$ leads to
\begin{align}\label{Appendix LDOS Fourier Transform}
\delta\rho({\bf \Delta K} + {\bf q}, \omega, \nu) &\simeq 
-~ \frac{\Theta(q-2q_{F})}{\sqrt{q^{2}-(2q_{F})^{2}}} \notag \\
&\times (-\xi \xi')^{\nu}~ e^{-i\nu {\bf \Delta K}\cdot{\bf d_{3}}}~ e^{-i \nu(\theta^{\xi}({\bf q})-\theta^{\xi'}({\bf q}) )} ~,
\end{align}
where it has been used that
\begin{align}
\lim\limits_{q \to 2q_{F}} \varphi_{\bf q} = \frac{\pi}{2} ~.
\end{align}

\bibliographystyle{apsrev4-1}
\bibliography{references}

\begin{thebibliography}{60}%
\makeatletter
\providecommand \@ifxundefined [1]{%
 \@ifx{#1\undefined}
}%
\providecommand \@ifnum [1]{%
 \ifnum #1\expandafter \@firstoftwo
 \else \expandafter \@secondoftwo
 \fi
}%
\providecommand \@ifx [1]{%
 \ifx #1\expandafter \@firstoftwo
 \else \expandafter \@secondoftwo
 \fi
}%
\providecommand \natexlab [1]{#1}%
\providecommand \enquote  [1]{``#1''}%
\providecommand \bibnamefont  [1]{#1}%
\providecommand \bibfnamefont [1]{#1}%
\providecommand \citenamefont [1]{#1}%
\providecommand \href@noop [0]{\@secondoftwo}%
\providecommand \href [0]{\begingroup \@sanitize@url \@href}%
\providecommand \@href[1]{\@@startlink{#1}\@@href}%
\providecommand \@@href[1]{\endgroup#1\@@endlink}%
\providecommand \@sanitize@url [0]{\catcode `\\12\catcode `\$12\catcode
  `\&12\catcode `\#12\catcode `\^12\catcode `\_12\catcode `\%12\relax}%
\providecommand \@@startlink[1]{}%
\providecommand \@@endlink[0]{}%
\providecommand \url  [0]{\begingroup\@sanitize@url \@url }%
\providecommand \@url [1]{\endgroup\@href {#1}{\urlprefix }}%
\providecommand \urlprefix  [0]{URL }%
\providecommand \Eprint [0]{\href }%
\providecommand \doibase [0]{http://dx.doi.org/}%
\providecommand \selectlanguage [0]{\@gobble}%
\providecommand \bibinfo  [0]{\@secondoftwo}%
\providecommand \bibfield  [0]{\@secondoftwo}%
\providecommand \translation [1]{[#1]}%
\providecommand \BibitemOpen [0]{}%
\providecommand \bibitemStop [0]{}%
\providecommand \bibitemNoStop [0]{.\EOS\space}%
\providecommand \EOS [0]{\spacefactor3000\relax}%
\providecommand \BibitemShut  [1]{\csname bibitem#1\endcsname}%
\let\auto@bib@innerbib\@empty
\bibitem [{\citenamefont {Friedel}(1952)}]{friedel1952xiv}%
  \BibitemOpen
  \bibfield  {author} {\bibinfo {author} {\bibfnamefont {J.}~\bibnamefont
  {Friedel}},\ }\href@noop {} {\bibfield  {journal} {\bibinfo  {journal} {Phil.
  Mag.}\ }\textbf {\bibinfo {volume} {43}},\ \bibinfo {pages} {153} (\bibinfo
  {year} {1952})}\BibitemShut {NoStop}%
\bibitem [{\citenamefont {Ruderman}\ and\ \citenamefont
  {Kittel}(1954)}]{PhysRev.96.99}%
  \BibitemOpen
  \bibfield  {author} {\bibinfo {author} {\bibfnamefont {M.~A.}\ \bibnamefont
  {Ruderman}}\ and\ \bibinfo {author} {\bibfnamefont {C.}~\bibnamefont
  {Kittel}},\ }\href {\doibase 10.1103/PhysRev.96.99} {\bibfield  {journal}
  {\bibinfo  {journal} {Phys. Rev.}\ }\textbf {\bibinfo {volume} {96}},\
  \bibinfo {pages} {99} (\bibinfo {year} {1954})}\BibitemShut {NoStop}%
\bibitem [{\citenamefont {Vonsovsky}(1974)}]{vonsovsky1974magnetism}%
  \BibitemOpen
  \bibfield  {author} {\bibinfo {author} {\bibfnamefont {S.~V.}\ \bibnamefont
  {Vonsovsky}},\ }\href@noop {} {\emph {\bibinfo {title} {Magnetism, vol. 2}}}\
  (\bibinfo  {publisher} {Johh Wiley \& Sons, New York},\ \bibinfo {year}
  {1974})\BibitemShut {NoStop}%
\bibitem [{\citenamefont {Adhikari}(1986)}]{adhikari1986quantum}%
  \BibitemOpen
  \bibfield  {author} {\bibinfo {author} {\bibfnamefont {S.~K.}\ \bibnamefont
  {Adhikari}},\ }\href@noop {} {\bibfield  {journal} {\bibinfo  {journal}
  {American Journal of Physics}\ }\textbf {\bibinfo {volume} {54}},\ \bibinfo
  {pages} {362} (\bibinfo {year} {1986})}\BibitemShut {NoStop}%
\bibitem [{\citenamefont {Crommie}\ \emph {et~al.}(1993)\citenamefont
  {Crommie}, \citenamefont {Lutz},\ and\ \citenamefont
  {Eigler}}]{crommie1993imaging}%
  \BibitemOpen
  \bibfield  {author} {\bibinfo {author} {\bibfnamefont {M.~F.}\ \bibnamefont
  {Crommie}}, \bibinfo {author} {\bibfnamefont {C.~P.}\ \bibnamefont {Lutz}}, \
  and\ \bibinfo {author} {\bibfnamefont {D.~M.}\ \bibnamefont {Eigler}},\
  }\href@noop {} {\bibfield  {journal} {\bibinfo  {journal} {Nature}\ }\textbf
  {\bibinfo {volume} {363}},\ \bibinfo {pages} {524} (\bibinfo {year}
  {1993})}\BibitemShut {NoStop}%
\bibitem [{\citenamefont {Binnig}\ and\ \citenamefont
  {Rohrer}(1987)}]{RevModPhys.59.615}%
  \BibitemOpen
  \bibfield  {author} {\bibinfo {author} {\bibfnamefont {G.}~\bibnamefont
  {Binnig}}\ and\ \bibinfo {author} {\bibfnamefont {H.}~\bibnamefont
  {Rohrer}},\ }\href {\doibase 10.1103/RevModPhys.59.615} {\bibfield  {journal}
  {\bibinfo  {journal} {Rev. Mod. Phys.}\ }\textbf {\bibinfo {volume} {59}},\
  \bibinfo {pages} {615} (\bibinfo {year} {1987})}\BibitemShut {NoStop}%
\bibitem [{\citenamefont {Sprunger}\ \emph {et~al.}(1997)\citenamefont
  {Sprunger}, \citenamefont {Petersen}, \citenamefont {Plummer}, \citenamefont
  {L{\ae}gsgaard},\ and\ \citenamefont {Besenbacher}}]{Sprunger21031997}%
  \BibitemOpen
  \bibfield  {author} {\bibinfo {author} {\bibfnamefont {P.~T.}\ \bibnamefont
  {Sprunger}}, \bibinfo {author} {\bibfnamefont {L.}~\bibnamefont {Petersen}},
  \bibinfo {author} {\bibfnamefont {E.~W.}\ \bibnamefont {Plummer}}, \bibinfo
  {author} {\bibfnamefont {E.}~\bibnamefont {L{\ae}gsgaard}}, \ and\ \bibinfo
  {author} {\bibfnamefont {F.}~\bibnamefont {Besenbacher}},\ }\href {\doibase
  10.1126/science.275.5307.1764} {\bibfield  {journal} {\bibinfo  {journal}
  {Science}\ }\textbf {\bibinfo {volume} {275}},\ \bibinfo {pages} {1764}
  (\bibinfo {year} {1997})}\BibitemShut {NoStop}%
\bibitem [{\citenamefont {Petersen}\ \emph {et~al.}(1998)\citenamefont
  {Petersen}, \citenamefont {Sprunger}, \citenamefont {Hofmann}, \citenamefont
  {L\ae{}gsgaard}, \citenamefont {Briner}, \citenamefont {Doering},
  \citenamefont {Rust}, \citenamefont {Bradshaw}, \citenamefont {Besenbacher},\
  and\ \citenamefont {Plummer}}]{PhysRevB.57.R6858}%
  \BibitemOpen
  \bibfield  {author} {\bibinfo {author} {\bibfnamefont {L.}~\bibnamefont
  {Petersen}}, \bibinfo {author} {\bibfnamefont {P.~T.}\ \bibnamefont
  {Sprunger}}, \bibinfo {author} {\bibfnamefont {P.}~\bibnamefont {Hofmann}},
  \bibinfo {author} {\bibfnamefont {E.}~\bibnamefont {L\ae{}gsgaard}}, \bibinfo
  {author} {\bibfnamefont {B.~G.}\ \bibnamefont {Briner}}, \bibinfo {author}
  {\bibfnamefont {M.}~\bibnamefont {Doering}}, \bibinfo {author} {\bibfnamefont
  {H.-P.}\ \bibnamefont {Rust}}, \bibinfo {author} {\bibfnamefont {A.~M.}\
  \bibnamefont {Bradshaw}}, \bibinfo {author} {\bibfnamefont {F.}~\bibnamefont
  {Besenbacher}}, \ and\ \bibinfo {author} {\bibfnamefont {E.~W.}\ \bibnamefont
  {Plummer}},\ }\href {\doibase 10.1103/PhysRevB.57.R6858} {\bibfield
  {journal} {\bibinfo  {journal} {Phys. Rev. B}\ }\textbf {\bibinfo {volume}
  {57}},\ \bibinfo {pages} {R6858} (\bibinfo {year} {1998})}\BibitemShut
  {NoStop}%
\bibitem [{\citenamefont {Novoselov}\ \emph {et~al.}(2004)\citenamefont
  {Novoselov}, \citenamefont {Geim}, \citenamefont {Morozov}, \citenamefont
  {Jiang}, \citenamefont {Zhang}, \citenamefont {Dubonos}, \citenamefont
  {Grigorieva},\ and\ \citenamefont {Firsov}}]{Novoselov22102004}%
  \BibitemOpen
  \bibfield  {author} {\bibinfo {author} {\bibfnamefont {K.~S.}\ \bibnamefont
  {Novoselov}}, \bibinfo {author} {\bibfnamefont {A.~K.}\ \bibnamefont {Geim}},
  \bibinfo {author} {\bibfnamefont {S.~V.}\ \bibnamefont {Morozov}}, \bibinfo
  {author} {\bibfnamefont {D.}~\bibnamefont {Jiang}}, \bibinfo {author}
  {\bibfnamefont {Y.}~\bibnamefont {Zhang}}, \bibinfo {author} {\bibfnamefont
  {S.~V.}\ \bibnamefont {Dubonos}}, \bibinfo {author} {\bibfnamefont {I.~V.}\
  \bibnamefont {Grigorieva}}, \ and\ \bibinfo {author} {\bibfnamefont {A.~A.}\
  \bibnamefont {Firsov}},\ }\href {\doibase 10.1126/science.1102896} {\bibfield
   {journal} {\bibinfo  {journal} {Science}\ }\textbf {\bibinfo {volume}
  {306}},\ \bibinfo {pages} {666} (\bibinfo {year} {2004})}\BibitemShut
  {NoStop}%
\bibitem [{\citenamefont {Wehling}\ \emph {et~al.}(2007)\citenamefont
  {Wehling}, \citenamefont {Balatsky}, \citenamefont {Katsnelson},
  \citenamefont {Lichtenstein}, \citenamefont {Scharnberg},\ and\ \citenamefont
  {Wiesendanger}}]{PhysRevB.75.125425}%
  \BibitemOpen
  \bibfield  {author} {\bibinfo {author} {\bibfnamefont {T.~O.}\ \bibnamefont
  {Wehling}}, \bibinfo {author} {\bibfnamefont {A.~V.}\ \bibnamefont
  {Balatsky}}, \bibinfo {author} {\bibfnamefont {M.~I.}\ \bibnamefont
  {Katsnelson}}, \bibinfo {author} {\bibfnamefont {A.~I.}\ \bibnamefont
  {Lichtenstein}}, \bibinfo {author} {\bibfnamefont {K.}~\bibnamefont
  {Scharnberg}}, \ and\ \bibinfo {author} {\bibfnamefont {R.}~\bibnamefont
  {Wiesendanger}},\ }\href {\doibase 10.1103/PhysRevB.75.125425} {\bibfield
  {journal} {\bibinfo  {journal} {Phys. Rev. B}\ }\textbf {\bibinfo {volume}
  {75}},\ \bibinfo {pages} {125425} (\bibinfo {year} {2007})}\BibitemShut
  {NoStop}%
\bibitem [{\citenamefont {Uchoa}\ \emph {et~al.}(2009)\citenamefont {Uchoa},
  \citenamefont {Yang}, \citenamefont {Tsai}, \citenamefont {Peres},\ and\
  \citenamefont {Castro~Neto}}]{PhysRevLett.103.206804}%
  \BibitemOpen
  \bibfield  {author} {\bibinfo {author} {\bibfnamefont {B.}~\bibnamefont
  {Uchoa}}, \bibinfo {author} {\bibfnamefont {L.}~\bibnamefont {Yang}},
  \bibinfo {author} {\bibfnamefont {S.-W.}\ \bibnamefont {Tsai}}, \bibinfo
  {author} {\bibfnamefont {N.~M.~R.}\ \bibnamefont {Peres}}, \ and\ \bibinfo
  {author} {\bibfnamefont {A.~H.}\ \bibnamefont {Castro~Neto}},\ }\href
  {\doibase 10.1103/PhysRevLett.103.206804} {\bibfield  {journal} {\bibinfo
  {journal} {Phys. Rev. Lett.}\ }\textbf {\bibinfo {volume} {103}},\ \bibinfo
  {pages} {206804} (\bibinfo {year} {2009})}\BibitemShut {NoStop}%
\bibitem [{\citenamefont {Wehling}\ \emph {et~al.}(2010)\citenamefont
  {Wehling}, \citenamefont {Dahal}, \citenamefont {Lichtenstein}, \citenamefont
  {Katsnelson}, \citenamefont {Manoharan},\ and\ \citenamefont
  {Balatsky}}]{PhysRevB.81.085413}%
  \BibitemOpen
  \bibfield  {author} {\bibinfo {author} {\bibfnamefont {T.~O.}\ \bibnamefont
  {Wehling}}, \bibinfo {author} {\bibfnamefont {H.~P.}\ \bibnamefont {Dahal}},
  \bibinfo {author} {\bibfnamefont {A.~I.}\ \bibnamefont {Lichtenstein}},
  \bibinfo {author} {\bibfnamefont {M.~I.}\ \bibnamefont {Katsnelson}},
  \bibinfo {author} {\bibfnamefont {H.~C.}\ \bibnamefont {Manoharan}}, \ and\
  \bibinfo {author} {\bibfnamefont {A.~V.}\ \bibnamefont {Balatsky}},\ }\href
  {\doibase 10.1103/PhysRevB.81.085413} {\bibfield  {journal} {\bibinfo
  {journal} {Phys. Rev. B}\ }\textbf {\bibinfo {volume} {81}},\ \bibinfo
  {pages} {085413} (\bibinfo {year} {2010})}\BibitemShut {NoStop}%
\bibitem [{\citenamefont {Ugeda}\ \emph {et~al.}(2010)\citenamefont {Ugeda},
  \citenamefont {Brihuega}, \citenamefont {Guinea},\ and\ \citenamefont
  {G\'omez-Rodr\'{i}guez}}]{PhysRevLett.104.096804}%
  \BibitemOpen
  \bibfield  {author} {\bibinfo {author} {\bibfnamefont {M.~M.}\ \bibnamefont
  {Ugeda}}, \bibinfo {author} {\bibfnamefont {I.}~\bibnamefont {Brihuega}},
  \bibinfo {author} {\bibfnamefont {F.}~\bibnamefont {Guinea}}, \ and\ \bibinfo
  {author} {\bibfnamefont {J.~M.}\ \bibnamefont {G\'omez-Rodr\'{i}guez}},\
  }\href {\doibase 10.1103/PhysRevLett.104.096804} {\bibfield  {journal}
  {\bibinfo  {journal} {Phys. Rev. Lett.}\ }\textbf {\bibinfo {volume} {104}},\
  \bibinfo {pages} {096804} (\bibinfo {year} {2010})}\BibitemShut {NoStop}%
\bibitem [{\citenamefont {{Brar}}\ \emph {et~al.}(2011)\citenamefont {{Brar}},
  \citenamefont {{Decker}}, \citenamefont {{Solowan}}, \citenamefont {{Wang}},
  \citenamefont {{Maserati}}, \citenamefont {{Chan}}, \citenamefont {{Lee}},
  \citenamefont {{Girit}}, \citenamefont {{Zettl}}, \citenamefont {{Louie}},
  \citenamefont {{Cohen}},\ and\ \citenamefont
  {{Crommie}}}]{2011NatPh...7...43B}%
  \BibitemOpen
  \bibfield  {author} {\bibinfo {author} {\bibfnamefont {V.~W.}\ \bibnamefont
  {{Brar}}}, \bibinfo {author} {\bibfnamefont {R.}~\bibnamefont {{Decker}}},
  \bibinfo {author} {\bibfnamefont {H.-M.}\ \bibnamefont {{Solowan}}}, \bibinfo
  {author} {\bibfnamefont {Y.}~\bibnamefont {{Wang}}}, \bibinfo {author}
  {\bibfnamefont {L.}~\bibnamefont {{Maserati}}}, \bibinfo {author}
  {\bibfnamefont {K.~T.}\ \bibnamefont {{Chan}}}, \bibinfo {author}
  {\bibfnamefont {H.}~\bibnamefont {{Lee}}}, \bibinfo {author} {\bibfnamefont
  {{\c C}.~O.}\ \bibnamefont {{Girit}}}, \bibinfo {author} {\bibfnamefont
  {A.}~\bibnamefont {{Zettl}}}, \bibinfo {author} {\bibfnamefont {S.~G.}\
  \bibnamefont {{Louie}}}, \bibinfo {author} {\bibfnamefont {M.~L.}\
  \bibnamefont {{Cohen}}}, \ and\ \bibinfo {author} {\bibfnamefont {M.~F.}\
  \bibnamefont {{Crommie}}},\ }\href {\doibase 10.1038/nphys1807} {\bibfield
  {journal} {\bibinfo  {journal} {Nature Physics}\ }\textbf {\bibinfo {volume}
  {7}},\ \bibinfo {pages} {43} (\bibinfo {year} {2011})}\BibitemShut {NoStop}%
\bibitem [{\citenamefont {Wallace}(1947)}]{PhysRev.71.622}%
  \BibitemOpen
  \bibfield  {author} {\bibinfo {author} {\bibfnamefont {P.~R.}\ \bibnamefont
  {Wallace}},\ }\href {\doibase 10.1103/PhysRev.71.622} {\bibfield  {journal}
  {\bibinfo  {journal} {Phys. Rev.}\ }\textbf {\bibinfo {volume} {71}},\
  \bibinfo {pages} {622} (\bibinfo {year} {1947})}\BibitemShut {NoStop}%
\bibitem [{\citenamefont {Klein}(1929)}]{klein1929reflexion}%
  \BibitemOpen
  \bibfield  {author} {\bibinfo {author} {\bibfnamefont {O.}~\bibnamefont
  {Klein}},\ }\href@noop {} {\bibfield  {journal} {\bibinfo  {journal}
  {Zeitschrift f{\"u}r Physik}\ }\textbf {\bibinfo {volume} {53}},\ \bibinfo
  {pages} {157} (\bibinfo {year} {1929})}\BibitemShut {NoStop}%
\bibitem [{\citenamefont {Ando}\ \emph {et~al.}(1998)\citenamefont {Ando},
  \citenamefont {Nakanishi},\ and\ \citenamefont {Saito}}]{ando1998berry}%
  \BibitemOpen
  \bibfield  {author} {\bibinfo {author} {\bibfnamefont {T.}~\bibnamefont
  {Ando}}, \bibinfo {author} {\bibfnamefont {T.}~\bibnamefont {Nakanishi}}, \
  and\ \bibinfo {author} {\bibfnamefont {R.}~\bibnamefont {Saito}},\
  }\href@noop {} {\bibfield  {journal} {\bibinfo  {journal} {Journal of the
  Physical Society of Japan}\ }\textbf {\bibinfo {volume} {67}},\ \bibinfo
  {pages} {2857} (\bibinfo {year} {1998})}\BibitemShut {NoStop}%
\bibitem [{\citenamefont {Katsnelson}\ \emph {et~al.}(2006)\citenamefont
  {Katsnelson}, \citenamefont {Novoselov},\ and\ \citenamefont
  {Geim}}]{Katsnelson:2006qf}%
  \BibitemOpen
  \bibfield  {author} {\bibinfo {author} {\bibfnamefont {M.~I.}\ \bibnamefont
  {Katsnelson}}, \bibinfo {author} {\bibfnamefont {K.~S.}\ \bibnamefont
  {Novoselov}}, \ and\ \bibinfo {author} {\bibfnamefont {A.~K.}\ \bibnamefont
  {Geim}},\ }\href {http://dx.doi.org/10.1038/nphys384} {\bibfield  {journal}
  {\bibinfo  {journal} {Nature Physics}\ }\textbf {\bibinfo {volume} {2}},\
  \bibinfo {pages} {620} (\bibinfo {year} {2006})}\BibitemShut {NoStop}%
\bibitem [{\citenamefont {Cheianov}\ \emph {et~al.}(2007)\citenamefont
  {Cheianov}, \citenamefont {Fal'ko},\ and\ \citenamefont
  {Altshuler}}]{Cheianov02032007}%
  \BibitemOpen
  \bibfield  {author} {\bibinfo {author} {\bibfnamefont {V.~V.}\ \bibnamefont
  {Cheianov}}, \bibinfo {author} {\bibfnamefont {V.}~\bibnamefont {Fal'ko}}, \
  and\ \bibinfo {author} {\bibfnamefont {B.~L.}\ \bibnamefont {Altshuler}},\
  }\href {\doibase 10.1126/science.1138020} {\bibfield  {journal} {\bibinfo
  {journal} {Science}\ }\textbf {\bibinfo {volume} {315}},\ \bibinfo {pages}
  {1252} (\bibinfo {year} {2007})}\BibitemShut {NoStop}%
\bibitem [{\citenamefont {Young}\ and\ \citenamefont
  {Kim}(2009)}]{young2009quantum}%
  \BibitemOpen
  \bibfield  {author} {\bibinfo {author} {\bibfnamefont {A.~F.}\ \bibnamefont
  {Young}}\ and\ \bibinfo {author} {\bibfnamefont {P.}~\bibnamefont {Kim}},\
  }\href@noop {} {\bibfield  {journal} {\bibinfo  {journal} {Nature Physics}\
  }\textbf {\bibinfo {volume} {5}},\ \bibinfo {pages} {222} (\bibinfo {year}
  {2009})}\BibitemShut {NoStop}%
\bibitem [{\citenamefont {Tudorovskiy}\ \emph {et~al.}(2012)\citenamefont
  {Tudorovskiy}, \citenamefont {Reijnders},\ and\ \citenamefont
  {Katsnelson}}]{1402-4896-2012-T146-014010}%
  \BibitemOpen
  \bibfield  {author} {\bibinfo {author} {\bibfnamefont {T.}~\bibnamefont
  {Tudorovskiy}}, \bibinfo {author} {\bibfnamefont {K.~J.~A.}\ \bibnamefont
  {Reijnders}}, \ and\ \bibinfo {author} {\bibfnamefont {M.~I.}\ \bibnamefont
  {Katsnelson}},\ }\href {http://stacks.iop.org/1402-4896/2012/i=T146/a=014010}
  {\bibfield  {journal} {\bibinfo  {journal} {Physica Scripta}\ }\textbf
  {\bibinfo {volume} {2012}},\ \bibinfo {pages} {014010} (\bibinfo {year}
  {2012})}\BibitemShut {NoStop}%
\bibitem [{\citenamefont {Reijnders}\ \emph {et~al.}(2013)\citenamefont
  {Reijnders}, \citenamefont {Tudorovskiy},\ and\ \citenamefont
  {Katsnelson}}]{Reijnders2013155}%
  \BibitemOpen
  \bibfield  {author} {\bibinfo {author} {\bibfnamefont {K.~J.~A.}\
  \bibnamefont {Reijnders}}, \bibinfo {author} {\bibfnamefont {T.}~\bibnamefont
  {Tudorovskiy}}, \ and\ \bibinfo {author} {\bibfnamefont {M.~I.}\ \bibnamefont
  {Katsnelson}},\ }\href {\doibase http://dx.doi.org/10.1016/j.aop.2013.03.001}
  {\bibfield  {journal} {\bibinfo  {journal} {Annals of Physics}\ }\textbf
  {\bibinfo {volume} {333}},\ \bibinfo {pages} {155 } (\bibinfo {year}
  {2013})}\BibitemShut {NoStop}%
\bibitem [{\citenamefont {Katsnelson}(2012)}]{katsnelson2012graphene}%
  \BibitemOpen
  \bibfield  {author} {\bibinfo {author} {\bibfnamefont {M.~I.}\ \bibnamefont
  {Katsnelson}},\ }\href@noop {} {\emph {\bibinfo {title} {Graphene: Carbon in
  Two Dimensions}}}\ (\bibinfo  {publisher} {Cambridge University Press},\
  \bibinfo {year} {2012})\BibitemShut {NoStop}%
\bibitem [{\citenamefont {Cheianov}\ and\ \citenamefont
  {Fal'ko}(2006)}]{PhysRevLett.97.226801}%
  \BibitemOpen
  \bibfield  {author} {\bibinfo {author} {\bibfnamefont {V.~V.}\ \bibnamefont
  {Cheianov}}\ and\ \bibinfo {author} {\bibfnamefont {V.~I.}\ \bibnamefont
  {Fal'ko}},\ }\href {\doibase 10.1103/PhysRevLett.97.226801} {\bibfield
  {journal} {\bibinfo  {journal} {Phys. Rev. Lett.}\ }\textbf {\bibinfo
  {volume} {97}},\ \bibinfo {pages} {226801} (\bibinfo {year}
  {2006})}\BibitemShut {NoStop}%
\bibitem [{\citenamefont {Bena}(2008)}]{PhysRevLett.100.076601}%
  \BibitemOpen
  \bibfield  {author} {\bibinfo {author} {\bibfnamefont {C.}~\bibnamefont
  {Bena}},\ }\href {\doibase 10.1103/PhysRevLett.100.076601} {\bibfield
  {journal} {\bibinfo  {journal} {Phys. Rev. Lett.}\ }\textbf {\bibinfo
  {volume} {100}},\ \bibinfo {pages} {076601} (\bibinfo {year}
  {2008})}\BibitemShut {NoStop}%
\bibitem [{\citenamefont {Brihuega}\ \emph {et~al.}(2008)\citenamefont
  {Brihuega}, \citenamefont {Mallet}, \citenamefont {Bena}, \citenamefont
  {Bose}, \citenamefont {Michaelis}, \citenamefont {Vitali}, \citenamefont
  {Varchon}, \citenamefont {Magaud}, \citenamefont {Kern},\ and\ \citenamefont
  {Veuillen}}]{PhysRevLett.101.206802}%
  \BibitemOpen
  \bibfield  {author} {\bibinfo {author} {\bibfnamefont {I.}~\bibnamefont
  {Brihuega}}, \bibinfo {author} {\bibfnamefont {P.}~\bibnamefont {Mallet}},
  \bibinfo {author} {\bibfnamefont {C.}~\bibnamefont {Bena}}, \bibinfo {author}
  {\bibfnamefont {S.}~\bibnamefont {Bose}}, \bibinfo {author} {\bibfnamefont
  {C.}~\bibnamefont {Michaelis}}, \bibinfo {author} {\bibfnamefont
  {L.}~\bibnamefont {Vitali}}, \bibinfo {author} {\bibfnamefont
  {F.}~\bibnamefont {Varchon}}, \bibinfo {author} {\bibfnamefont
  {L.}~\bibnamefont {Magaud}}, \bibinfo {author} {\bibfnamefont
  {K.}~\bibnamefont {Kern}}, \ and\ \bibinfo {author} {\bibfnamefont {J.~Y.}\
  \bibnamefont {Veuillen}},\ }\href {\doibase 10.1103/PhysRevLett.101.206802}
  {\bibfield  {journal} {\bibinfo  {journal} {Phys. Rev. Lett.}\ }\textbf
  {\bibinfo {volume} {101}},\ \bibinfo {pages} {206802} (\bibinfo {year}
  {2008})}\BibitemShut {NoStop}%
\bibitem [{\citenamefont {Simon}\ \emph {et~al.}(2009)\citenamefont {Simon},
  \citenamefont {Bena}, \citenamefont {Vonau}, \citenamefont {Aubel},
  \citenamefont {Nasrallah}, \citenamefont {Habar},\ and\ \citenamefont
  {Peruchetti}}]{Simon:2009eu}%
  \BibitemOpen
  \bibfield  {author} {\bibinfo {author} {\bibfnamefont {L.}~\bibnamefont
  {Simon}}, \bibinfo {author} {\bibfnamefont {C.}~\bibnamefont {Bena}},
  \bibinfo {author} {\bibfnamefont {F.}~\bibnamefont {Vonau}}, \bibinfo
  {author} {\bibfnamefont {D.}~\bibnamefont {Aubel}}, \bibinfo {author}
  {\bibfnamefont {H.}~\bibnamefont {Nasrallah}}, \bibinfo {author}
  {\bibfnamefont {M.}~\bibnamefont {Habar}}, \ and\ \bibinfo {author}
  {\bibfnamefont {J.~C.}\ \bibnamefont {Peruchetti}},\ }\href {\doibase
  10.1140/epjb/e2009-00142-3} {\bibfield  {journal} {\bibinfo  {journal} {Eur.
  Phys. J. B}\ }\textbf {\bibinfo {volume} {69}},\ \bibinfo {pages} {351}
  (\bibinfo {year} {2009})}\BibitemShut {NoStop}%
\bibitem [{\citenamefont {Xia}\ \emph {et~al.}(2009)\citenamefont {Xia},
  \citenamefont {Qian}, \citenamefont {Hsieh}, \citenamefont {Wray},
  \citenamefont {Pal}, \citenamefont {Lin}, \citenamefont {Bansil},
  \citenamefont {Grauer}, \citenamefont {Hor}, \citenamefont {Cava} \emph
  {et~al.}}]{xia2009observation}%
  \BibitemOpen
  \bibfield  {author} {\bibinfo {author} {\bibfnamefont {Y.}~\bibnamefont
  {Xia}}, \bibinfo {author} {\bibfnamefont {D.}~\bibnamefont {Qian}}, \bibinfo
  {author} {\bibfnamefont {D.}~\bibnamefont {Hsieh}}, \bibinfo {author}
  {\bibfnamefont {L.}~\bibnamefont {Wray}}, \bibinfo {author} {\bibfnamefont
  {A.}~\bibnamefont {Pal}}, \bibinfo {author} {\bibfnamefont {H.}~\bibnamefont
  {Lin}}, \bibinfo {author} {\bibfnamefont {A.}~\bibnamefont {Bansil}},
  \bibinfo {author} {\bibfnamefont {D.}~\bibnamefont {Grauer}}, \bibinfo
  {author} {\bibfnamefont {Y.}~\bibnamefont {Hor}}, \bibinfo {author}
  {\bibfnamefont {R.}~\bibnamefont {Cava}},  \emph {et~al.},\ }\href@noop {}
  {\bibfield  {journal} {\bibinfo  {journal} {Nature Physics}\ }\textbf
  {\bibinfo {volume} {5}},\ \bibinfo {pages} {398} (\bibinfo {year}
  {2009})}\BibitemShut {NoStop}%
\bibitem [{\citenamefont {Zhang}\ \emph {et~al.}(2009)\citenamefont {Zhang},
  \citenamefont {Cheng}, \citenamefont {Chen}, \citenamefont {Jia},
  \citenamefont {Ma}, \citenamefont {He}, \citenamefont {Wang}, \citenamefont
  {Zhang}, \citenamefont {Dai}, \citenamefont {Fang}, \citenamefont {Xie},\
  and\ \citenamefont {Xue}}]{PhysRevLett.103.266803}%
  \BibitemOpen
  \bibfield  {author} {\bibinfo {author} {\bibfnamefont {T.}~\bibnamefont
  {Zhang}}, \bibinfo {author} {\bibfnamefont {P.}~\bibnamefont {Cheng}},
  \bibinfo {author} {\bibfnamefont {X.}~\bibnamefont {Chen}}, \bibinfo {author}
  {\bibfnamefont {J.-F.}\ \bibnamefont {Jia}}, \bibinfo {author} {\bibfnamefont
  {X.}~\bibnamefont {Ma}}, \bibinfo {author} {\bibfnamefont {K.}~\bibnamefont
  {He}}, \bibinfo {author} {\bibfnamefont {L.}~\bibnamefont {Wang}}, \bibinfo
  {author} {\bibfnamefont {H.}~\bibnamefont {Zhang}}, \bibinfo {author}
  {\bibfnamefont {X.}~\bibnamefont {Dai}}, \bibinfo {author} {\bibfnamefont
  {Z.}~\bibnamefont {Fang}}, \bibinfo {author} {\bibfnamefont {X.}~\bibnamefont
  {Xie}}, \ and\ \bibinfo {author} {\bibfnamefont {Q.-K.}\ \bibnamefont
  {Xue}},\ }\href {\doibase 10.1103/PhysRevLett.103.266803} {\bibfield
  {journal} {\bibinfo  {journal} {Phys. Rev. Lett.}\ }\textbf {\bibinfo
  {volume} {103}},\ \bibinfo {pages} {266803} (\bibinfo {year}
  {2009})}\BibitemShut {NoStop}%
\bibitem [{\citenamefont {\ifmmode~\check{S}\else \v{S}\fi{}op\'{i}k}\ \emph
  {et~al.}(2014)\citenamefont {\ifmmode~\check{S}\else \v{S}\fi{}op\'{i}k},
  \citenamefont {Kailasvuori},\ and\ \citenamefont
  {Trushin}}]{PhysRevB.89.165308}%
  \BibitemOpen
  \bibfield  {author} {\bibinfo {author} {\bibfnamefont {B.~c.~v.}\
  \bibnamefont {\ifmmode~\check{S}\else \v{S}\fi{}op\'{i}k}}, \bibinfo {author}
  {\bibfnamefont {J.}~\bibnamefont {Kailasvuori}}, \ and\ \bibinfo {author}
  {\bibfnamefont {M.}~\bibnamefont {Trushin}},\ }\href {\doibase
  10.1103/PhysRevB.89.165308} {\bibfield  {journal} {\bibinfo  {journal} {Phys.
  Rev. B}\ }\textbf {\bibinfo {volume} {89}},\ \bibinfo {pages} {165308}
  (\bibinfo {year} {2014})}\BibitemShut {NoStop}%
\bibitem [{\citenamefont {Das~Sarma}\ and\ \citenamefont
  {Hwang}(2015)}]{PhysRevB.91.195104}%
  \BibitemOpen
  \bibfield  {author} {\bibinfo {author} {\bibfnamefont {S.}~\bibnamefont
  {Das~Sarma}}\ and\ \bibinfo {author} {\bibfnamefont {E.~H.}\ \bibnamefont
  {Hwang}},\ }\href {\doibase 10.1103/PhysRevB.91.195104} {\bibfield  {journal}
  {\bibinfo  {journal} {Phys. Rev. B}\ }\textbf {\bibinfo {volume} {91}},\
  \bibinfo {pages} {195104} (\bibinfo {year} {2015})}\BibitemShut {NoStop}%
\bibitem [{\citenamefont {Haering}(1958)}]{haering1958band}%
  \BibitemOpen
  \bibfield  {author} {\bibinfo {author} {\bibfnamefont {R.}~\bibnamefont
  {Haering}},\ }\href@noop {} {\bibfield  {journal} {\bibinfo  {journal}
  {Canadian Journal of Physics}\ }\textbf {\bibinfo {volume} {36}},\ \bibinfo
  {pages} {352} (\bibinfo {year} {1958})}\BibitemShut {NoStop}%
\bibitem [{\citenamefont {Guinea}\ \emph {et~al.}(2006)\citenamefont {Guinea},
  \citenamefont {Castro~Neto},\ and\ \citenamefont
  {Peres}}]{PhysRevB.73.245426}%
  \BibitemOpen
  \bibfield  {author} {\bibinfo {author} {\bibfnamefont {F.}~\bibnamefont
  {Guinea}}, \bibinfo {author} {\bibfnamefont {A.~H.}\ \bibnamefont
  {Castro~Neto}}, \ and\ \bibinfo {author} {\bibfnamefont {N.~M.~R.}\
  \bibnamefont {Peres}},\ }\href {\doibase 10.1103/PhysRevB.73.245426}
  {\bibfield  {journal} {\bibinfo  {journal} {Phys. Rev. B}\ }\textbf {\bibinfo
  {volume} {73}},\ \bibinfo {pages} {245426} (\bibinfo {year}
  {2006})}\BibitemShut {NoStop}%
\bibitem [{\citenamefont {Zan}\ \emph {et~al.}(2012)\citenamefont {Zan},
  \citenamefont {Muryn}, \citenamefont {Bangert}, \citenamefont {Mattocks},
  \citenamefont {Wincott}, \citenamefont {Vaughan}, \citenamefont {Li},
  \citenamefont {Colombo}, \citenamefont {Ruoff}, \citenamefont {Hamilton},\
  and\ \citenamefont {Novoselov}}]{C2NR30162H}%
  \BibitemOpen
  \bibfield  {author} {\bibinfo {author} {\bibfnamefont {R.}~\bibnamefont
  {Zan}}, \bibinfo {author} {\bibfnamefont {C.}~\bibnamefont {Muryn}}, \bibinfo
  {author} {\bibfnamefont {U.}~\bibnamefont {Bangert}}, \bibinfo {author}
  {\bibfnamefont {P.}~\bibnamefont {Mattocks}}, \bibinfo {author}
  {\bibfnamefont {P.}~\bibnamefont {Wincott}}, \bibinfo {author} {\bibfnamefont
  {D.}~\bibnamefont {Vaughan}}, \bibinfo {author} {\bibfnamefont
  {X.}~\bibnamefont {Li}}, \bibinfo {author} {\bibfnamefont {L.}~\bibnamefont
  {Colombo}}, \bibinfo {author} {\bibfnamefont {R.~S.}\ \bibnamefont {Ruoff}},
  \bibinfo {author} {\bibfnamefont {B.}~\bibnamefont {Hamilton}}, \ and\
  \bibinfo {author} {\bibfnamefont {K.~S.}\ \bibnamefont {Novoselov}},\ }\href
  {\doibase 10.1039/C2NR30162H} {\bibfield  {journal} {\bibinfo  {journal}
  {Nanoscale}\ }\textbf {\bibinfo {volume} {4}},\ \bibinfo {pages} {3065}
  (\bibinfo {year} {2012})}\BibitemShut {NoStop}%
\bibitem [{\citenamefont {Reich}\ \emph {et~al.}(2002)\citenamefont {Reich},
  \citenamefont {Maultzsch}, \citenamefont {Thomsen},\ and\ \citenamefont
  {Ordej\'on}}]{PhysRevB.66.035412}%
  \BibitemOpen
  \bibfield  {author} {\bibinfo {author} {\bibfnamefont {S.}~\bibnamefont
  {Reich}}, \bibinfo {author} {\bibfnamefont {J.}~\bibnamefont {Maultzsch}},
  \bibinfo {author} {\bibfnamefont {C.}~\bibnamefont {Thomsen}}, \ and\
  \bibinfo {author} {\bibfnamefont {P.}~\bibnamefont {Ordej\'on}},\ }\href
  {\doibase 10.1103/PhysRevB.66.035412} {\bibfield  {journal} {\bibinfo
  {journal} {Phys. Rev. B}\ }\textbf {\bibinfo {volume} {66}},\ \bibinfo
  {pages} {035412} (\bibinfo {year} {2002})}\BibitemShut {NoStop}%
\bibitem [{\citenamefont {Hasegawa}\ \emph {et~al.}(2006)\citenamefont
  {Hasegawa}, \citenamefont {Konno}, \citenamefont {Nakano},\ and\
  \citenamefont {Kohmoto}}]{PhysRevB.74.033413}%
  \BibitemOpen
  \bibfield  {author} {\bibinfo {author} {\bibfnamefont {Y.}~\bibnamefont
  {Hasegawa}}, \bibinfo {author} {\bibfnamefont {R.}~\bibnamefont {Konno}},
  \bibinfo {author} {\bibfnamefont {H.}~\bibnamefont {Nakano}}, \ and\ \bibinfo
  {author} {\bibfnamefont {M.}~\bibnamefont {Kohmoto}},\ }\href {\doibase
  10.1103/PhysRevB.74.033413} {\bibfield  {journal} {\bibinfo  {journal} {Phys.
  Rev. B}\ }\textbf {\bibinfo {volume} {74}},\ \bibinfo {pages} {033413}
  (\bibinfo {year} {2006})}\BibitemShut {NoStop}%
\bibitem [{\citenamefont {Dietl}\ \emph {et~al.}(2008)\citenamefont {Dietl},
  \citenamefont {Pi\'echon},\ and\ \citenamefont
  {Montambaux}}]{PhysRevLett.100.236405}%
  \BibitemOpen
  \bibfield  {author} {\bibinfo {author} {\bibfnamefont {P.}~\bibnamefont
  {Dietl}}, \bibinfo {author} {\bibfnamefont {F.}~\bibnamefont {Pi\'echon}}, \
  and\ \bibinfo {author} {\bibfnamefont {G.}~\bibnamefont {Montambaux}},\
  }\href {\doibase 10.1103/PhysRevLett.100.236405} {\bibfield  {journal}
  {\bibinfo  {journal} {Phys. Rev. Lett.}\ }\textbf {\bibinfo {volume} {100}},\
  \bibinfo {pages} {236405} (\bibinfo {year} {2008})}\BibitemShut {NoStop}%
\bibitem [{\citenamefont {Liu}\ \emph {et~al.}(2007)\citenamefont {Liu},
  \citenamefont {Ming},\ and\ \citenamefont {Li}}]{PhysRevB.76.064120}%
  \BibitemOpen
  \bibfield  {author} {\bibinfo {author} {\bibfnamefont {F.}~\bibnamefont
  {Liu}}, \bibinfo {author} {\bibfnamefont {P.}~\bibnamefont {Ming}}, \ and\
  \bibinfo {author} {\bibfnamefont {J.}~\bibnamefont {Li}},\ }\href {\doibase
  10.1103/PhysRevB.76.064120} {\bibfield  {journal} {\bibinfo  {journal} {Phys.
  Rev. B}\ }\textbf {\bibinfo {volume} {76}},\ \bibinfo {pages} {064120}
  (\bibinfo {year} {2007})}\BibitemShut {NoStop}%
\bibitem [{\citenamefont {Ma\~nes}\ \emph {et~al.}(2007)\citenamefont
  {Ma\~nes}, \citenamefont {Guinea},\ and\ \citenamefont
  {Vozmediano}}]{PhysRevB.75.155424}%
  \BibitemOpen
  \bibfield  {author} {\bibinfo {author} {\bibfnamefont {J.~L.}\ \bibnamefont
  {Ma\~nes}}, \bibinfo {author} {\bibfnamefont {F.}~\bibnamefont {Guinea}}, \
  and\ \bibinfo {author} {\bibfnamefont {M.~A.~H.}\ \bibnamefont
  {Vozmediano}},\ }\href {\doibase 10.1103/PhysRevB.75.155424} {\bibfield
  {journal} {\bibinfo  {journal} {Phys. Rev. B}\ }\textbf {\bibinfo {volume}
  {75}},\ \bibinfo {pages} {155424} (\bibinfo {year} {2007})}\BibitemShut
  {NoStop}%
\bibitem [{\citenamefont {Allain}\ and\ \citenamefont
  {Fuchs}(2011)}]{Allain:2011sf}%
  \BibitemOpen
  \bibfield  {author} {\bibinfo {author} {\bibfnamefont {P.~E.}\ \bibnamefont
  {Allain}}\ and\ \bibinfo {author} {\bibfnamefont {J.~N.}\ \bibnamefont
  {Fuchs}},\ }\href {\doibase 10.1140/epjb/e2011-20351-3} {\bibfield  {journal}
  {\bibinfo  {journal} {Eur. Phys. J. B}\ }\textbf {\bibinfo {volume} {83}},\
  \bibinfo {pages} {301} (\bibinfo {year} {2011})}\BibitemShut {NoStop}%
\bibitem [{\citenamefont {Mallet}\ \emph {et~al.}(2012)\citenamefont {Mallet},
  \citenamefont {Brihuega}, \citenamefont {Bose}, \citenamefont {Ugeda},
  \citenamefont {G\'omez-Rodr\'{i}guez}, \citenamefont {Kern},\ and\
  \citenamefont {Veuillen}}]{PhysRevB.86.045444}%
  \BibitemOpen
  \bibfield  {author} {\bibinfo {author} {\bibfnamefont {P.}~\bibnamefont
  {Mallet}}, \bibinfo {author} {\bibfnamefont {I.}~\bibnamefont {Brihuega}},
  \bibinfo {author} {\bibfnamefont {S.}~\bibnamefont {Bose}}, \bibinfo {author}
  {\bibfnamefont {M.~M.}\ \bibnamefont {Ugeda}}, \bibinfo {author}
  {\bibfnamefont {J.~M.}\ \bibnamefont {G\'omez-Rodr\'{i}guez}}, \bibinfo
  {author} {\bibfnamefont {K.}~\bibnamefont {Kern}}, \ and\ \bibinfo {author}
  {\bibfnamefont {J.~Y.}\ \bibnamefont {Veuillen}},\ }\href {\doibase
  10.1103/PhysRevB.86.045444} {\bibfield  {journal} {\bibinfo  {journal} {Phys.
  Rev. B}\ }\textbf {\bibinfo {volume} {86}},\ \bibinfo {pages} {045444}
  (\bibinfo {year} {2012})}\BibitemShut {NoStop}%
\bibitem [{\citenamefont {Pereg-Barnea}\ and\ \citenamefont
  {MacDonald}(2008)}]{PhysRevB.78.014201}%
  \BibitemOpen
  \bibfield  {author} {\bibinfo {author} {\bibfnamefont {T.}~\bibnamefont
  {Pereg-Barnea}}\ and\ \bibinfo {author} {\bibfnamefont {A.~H.}\ \bibnamefont
  {MacDonald}},\ }\href {\doibase 10.1103/PhysRevB.78.014201} {\bibfield
  {journal} {\bibinfo  {journal} {Phys. Rev. B}\ }\textbf {\bibinfo {volume}
  {78}},\ \bibinfo {pages} {014201} (\bibinfo {year} {2008})}\BibitemShut
  {NoStop}%
\bibitem [{\citenamefont {Pancharatnam}(1956)}]{Pancharatnam}%
  \BibitemOpen
  \bibfield  {author} {\bibinfo {author} {\bibfnamefont {S.}~\bibnamefont
  {Pancharatnam}},\ }\href@noop {} {\bibfield  {journal} {\bibinfo  {journal}
  {Proc. Indian Acad. Sci.}\ }\textbf {\bibinfo {volume} {44}},\ \bibinfo
  {pages} {247} (\bibinfo {year} {1956})}\BibitemShut {NoStop}%
\bibitem [{\citenamefont {Berry}(1984)}]{1984}%
  \BibitemOpen
  \bibfield  {author} {\bibinfo {author} {\bibfnamefont {M.~V.}\ \bibnamefont
  {Berry}},\ }\href {http://www.jstor.org/stable/2397741} {\bibfield  {journal}
  {\bibinfo  {journal} {Proc. R. Soc. London, Ser. A}\ }\textbf {\bibinfo
  {volume} {392}},\ \bibinfo {pages} {45} (\bibinfo {year} {1984})}\BibitemShut
  {NoStop}%
\bibitem [{\citenamefont {Shapere}\ and\ \citenamefont
  {Wilczek}(1989)}]{shapere1989geometric}%
  \BibitemOpen
  \bibfield  {author} {\bibinfo {author} {\bibfnamefont {A.}~\bibnamefont
  {Shapere}}\ and\ \bibinfo {author} {\bibfnamefont {F.}~\bibnamefont
  {Wilczek}},\ }\href@noop {} {\emph {\bibinfo {title} {Geometric Phases in
  Physics, volume 5 of Advanced Series in Mathematical Physics}}}\ (\bibinfo
  {publisher} {World Scientific, Singapore},\ \bibinfo {year}
  {1989})\BibitemShut {NoStop}%
\bibitem [{\citenamefont {King-Smith}\ and\ \citenamefont
  {Vanderbilt}(1993)}]{PhysRevB.47.1651}%
  \BibitemOpen
  \bibfield  {author} {\bibinfo {author} {\bibfnamefont {R.~D.}\ \bibnamefont
  {King-Smith}}\ and\ \bibinfo {author} {\bibfnamefont {D.}~\bibnamefont
  {Vanderbilt}},\ }\href {\doibase 10.1103/PhysRevB.47.1651} {\bibfield
  {journal} {\bibinfo  {journal} {Phys. Rev. B}\ }\textbf {\bibinfo {volume}
  {47}},\ \bibinfo {pages} {1651} (\bibinfo {year} {1993})}\BibitemShut
  {NoStop}%
\bibitem [{\citenamefont {Resta}(1994)}]{RevModPhys.66.899}%
  \BibitemOpen
  \bibfield  {author} {\bibinfo {author} {\bibfnamefont {R.}~\bibnamefont
  {Resta}},\ }\href {\doibase 10.1103/RevModPhys.66.899} {\bibfield  {journal}
  {\bibinfo  {journal} {Rev. Mod. Phys.}\ }\textbf {\bibinfo {volume} {66}},\
  \bibinfo {pages} {899} (\bibinfo {year} {1994})}\BibitemShut {NoStop}%
\bibitem [{\citenamefont {Thonhauser}\ \emph {et~al.}(2005)\citenamefont
  {Thonhauser}, \citenamefont {Ceresoli}, \citenamefont {Vanderbilt},\ and\
  \citenamefont {Resta}}]{PhysRevLett.95.137205}%
  \BibitemOpen
  \bibfield  {author} {\bibinfo {author} {\bibfnamefont {T.}~\bibnamefont
  {Thonhauser}}, \bibinfo {author} {\bibfnamefont {D.}~\bibnamefont
  {Ceresoli}}, \bibinfo {author} {\bibfnamefont {D.}~\bibnamefont
  {Vanderbilt}}, \ and\ \bibinfo {author} {\bibfnamefont {R.}~\bibnamefont
  {Resta}},\ }\href {\doibase 10.1103/PhysRevLett.95.137205} {\bibfield
  {journal} {\bibinfo  {journal} {Phys. Rev. Lett.}\ }\textbf {\bibinfo
  {volume} {95}},\ \bibinfo {pages} {137205} (\bibinfo {year}
  {2005})}\BibitemShut {NoStop}%
\bibitem [{\citenamefont {Raoux}\ \emph {et~al.}(2014)\citenamefont {Raoux},
  \citenamefont {Morigi}, \citenamefont {Fuchs}, \citenamefont {Pi\'echon},\
  and\ \citenamefont {Montambaux}}]{PhysRevLett.112.026402}%
  \BibitemOpen
  \bibfield  {author} {\bibinfo {author} {\bibfnamefont {A.}~\bibnamefont
  {Raoux}}, \bibinfo {author} {\bibfnamefont {M.}~\bibnamefont {Morigi}},
  \bibinfo {author} {\bibfnamefont {J.-N.}\ \bibnamefont {Fuchs}}, \bibinfo
  {author} {\bibfnamefont {F.}~\bibnamefont {Pi\'echon}}, \ and\ \bibinfo
  {author} {\bibfnamefont {G.}~\bibnamefont {Montambaux}},\ }\href {\doibase
  10.1103/PhysRevLett.112.026402} {\bibfield  {journal} {\bibinfo  {journal}
  {Phys. Rev. Lett.}\ }\textbf {\bibinfo {volume} {112}},\ \bibinfo {pages}
  {026402} (\bibinfo {year} {2014})}\BibitemShut {NoStop}%
\bibitem [{\citenamefont {Shockley}(1939)}]{PhysRev.56.317}%
  \BibitemOpen
  \bibfield  {author} {\bibinfo {author} {\bibfnamefont {W.}~\bibnamefont
  {Shockley}},\ }\href {\doibase 10.1103/PhysRev.56.317} {\bibfield  {journal}
  {\bibinfo  {journal} {Phys. Rev.}\ }\textbf {\bibinfo {volume} {56}},\
  \bibinfo {pages} {317} (\bibinfo {year} {1939})}\BibitemShut {NoStop}%
\bibitem [{\citenamefont {Thouless}\ \emph {et~al.}(1982)\citenamefont
  {Thouless}, \citenamefont {Kohmoto}, \citenamefont {Nightingale},\ and\
  \citenamefont {den Nijs}}]{PhysRevLett.49.405}%
  \BibitemOpen
  \bibfield  {author} {\bibinfo {author} {\bibfnamefont {D.~J.}\ \bibnamefont
  {Thouless}}, \bibinfo {author} {\bibfnamefont {M.}~\bibnamefont {Kohmoto}},
  \bibinfo {author} {\bibfnamefont {M.~P.}\ \bibnamefont {Nightingale}}, \ and\
  \bibinfo {author} {\bibfnamefont {M.}~\bibnamefont {den Nijs}},\ }\href
  {\doibase 10.1103/PhysRevLett.49.405} {\bibfield  {journal} {\bibinfo
  {journal} {Phys. Rev. Lett.}\ }\textbf {\bibinfo {volume} {49}},\ \bibinfo
  {pages} {405} (\bibinfo {year} {1982})}\BibitemShut {NoStop}%
\bibitem [{\citenamefont {Kane}\ and\ \citenamefont
  {Mele}(2005)}]{PhysRevLett.95.146802}%
  \BibitemOpen
  \bibfield  {author} {\bibinfo {author} {\bibfnamefont {C.~L.}\ \bibnamefont
  {Kane}}\ and\ \bibinfo {author} {\bibfnamefont {E.~J.}\ \bibnamefont
  {Mele}},\ }\href {\doibase 10.1103/PhysRevLett.95.146802} {\bibfield
  {journal} {\bibinfo  {journal} {Phys. Rev. Lett.}\ }\textbf {\bibinfo
  {volume} {95}},\ \bibinfo {pages} {146802} (\bibinfo {year}
  {2005})}\BibitemShut {NoStop}%
\bibitem [{\citenamefont {Schnyder}\ \emph {et~al.}(2008)\citenamefont
  {Schnyder}, \citenamefont {Ryu}, \citenamefont {Furusaki},\ and\
  \citenamefont {Ludwig}}]{schnyder2008classification}%
  \BibitemOpen
  \bibfield  {author} {\bibinfo {author} {\bibfnamefont {A.~P.}\ \bibnamefont
  {Schnyder}}, \bibinfo {author} {\bibfnamefont {S.}~\bibnamefont {Ryu}},
  \bibinfo {author} {\bibfnamefont {A.}~\bibnamefont {Furusaki}}, \ and\
  \bibinfo {author} {\bibfnamefont {A.~W.}\ \bibnamefont {Ludwig}},\
  }\href@noop {} {\bibfield  {journal} {\bibinfo  {journal} {Physical Review
  B}\ }\textbf {\bibinfo {volume} {78}},\ \bibinfo {pages} {195125} (\bibinfo
  {year} {2008})}\BibitemShut {NoStop}%
\bibitem [{\citenamefont {Xiao}\ \emph {et~al.}(2010)\citenamefont {Xiao},
  \citenamefont {Chang},\ and\ \citenamefont {Niu}}]{RevModPhys.82.1959}%
  \BibitemOpen
  \bibfield  {author} {\bibinfo {author} {\bibfnamefont {D.}~\bibnamefont
  {Xiao}}, \bibinfo {author} {\bibfnamefont {M.-C.}\ \bibnamefont {Chang}}, \
  and\ \bibinfo {author} {\bibfnamefont {Q.}~\bibnamefont {Niu}},\ }\href
  {\doibase 10.1103/RevModPhys.82.1959} {\bibfield  {journal} {\bibinfo
  {journal} {Rev. Mod. Phys.}\ }\textbf {\bibinfo {volume} {82}},\ \bibinfo
  {pages} {1959} (\bibinfo {year} {2010})}\BibitemShut {NoStop}%
\bibitem [{\citenamefont {Mikitik}\ and\ \citenamefont
  {Sharlai}(1999)}]{PhysRevLett.82.2147}%
  \BibitemOpen
  \bibfield  {author} {\bibinfo {author} {\bibfnamefont {G.~P.}\ \bibnamefont
  {Mikitik}}\ and\ \bibinfo {author} {\bibfnamefont {Y.~V.}\ \bibnamefont
  {Sharlai}},\ }\href {\doibase 10.1103/PhysRevLett.82.2147} {\bibfield
  {journal} {\bibinfo  {journal} {Phys. Rev. Lett.}\ }\textbf {\bibinfo
  {volume} {82}},\ \bibinfo {pages} {2147} (\bibinfo {year}
  {1999})}\BibitemShut {NoStop}%
\bibitem [{\citenamefont {Fuchs}\ \emph {et~al.}(2010)\citenamefont {Fuchs},
  \citenamefont {Pi{\'e}chon}, \citenamefont {Goerbig},\ and\ \citenamefont
  {Montambaux}}]{10.1140/epjb/e2010}%
  \BibitemOpen
  \bibfield  {author} {\bibinfo {author} {\bibfnamefont {J.~N.}\ \bibnamefont
  {Fuchs}}, \bibinfo {author} {\bibfnamefont {F.}~\bibnamefont {Pi{\'e}chon}},
  \bibinfo {author} {\bibfnamefont {M.~O.}\ \bibnamefont {Goerbig}}, \ and\
  \bibinfo {author} {\bibfnamefont {G.}~\bibnamefont {Montambaux}},\ }\href
  {\doibase 10.1140/epjb/e2010-00259-2} {\bibfield  {journal} {\bibinfo
  {journal} {The European Physical Journal B}\ }\textbf {\bibinfo {volume}
  {77}},\ \bibinfo {pages} {351} (\bibinfo {year} {2010})}\BibitemShut
  {NoStop}%
\bibitem [{\citenamefont {Novoselov}\ \emph {et~al.}(2005)\citenamefont
  {Novoselov}, \citenamefont {Geim}, \citenamefont {Morozov}, \citenamefont
  {Jiang}, \citenamefont {Katsnelson}, \citenamefont {Grigorieva},
  \citenamefont {Dubonos},\ and\ \citenamefont {Firsov}}]{novoselov2005two}%
  \BibitemOpen
  \bibfield  {author} {\bibinfo {author} {\bibfnamefont {K.}~\bibnamefont
  {Novoselov}}, \bibinfo {author} {\bibfnamefont {A.~K.}\ \bibnamefont {Geim}},
  \bibinfo {author} {\bibfnamefont {S.}~\bibnamefont {Morozov}}, \bibinfo
  {author} {\bibfnamefont {D.}~\bibnamefont {Jiang}}, \bibinfo {author}
  {\bibfnamefont {M.~I.}\ \bibnamefont {Katsnelson}}, \bibinfo {author}
  {\bibfnamefont {I.}~\bibnamefont {Grigorieva}}, \bibinfo {author}
  {\bibfnamefont {S.}~\bibnamefont {Dubonos}}, \ and\ \bibinfo {author}
  {\bibfnamefont {A.}~\bibnamefont {Firsov}},\ }\href@noop {} {\bibfield
  {journal} {\bibinfo  {journal} {Nature}\ }\textbf {\bibinfo {volume} {438}},\
  \bibinfo {pages} {197} (\bibinfo {year} {2005})}\BibitemShut {NoStop}%
\bibitem [{\citenamefont {Zhang}\ \emph {et~al.}(2005)\citenamefont {Zhang},
  \citenamefont {Tan}, \citenamefont {Stormer},\ and\ \citenamefont
  {Kim}}]{zhang2005experimental}%
  \BibitemOpen
  \bibfield  {author} {\bibinfo {author} {\bibfnamefont {Y.}~\bibnamefont
  {Zhang}}, \bibinfo {author} {\bibfnamefont {Y.-W.}\ \bibnamefont {Tan}},
  \bibinfo {author} {\bibfnamefont {H.~L.}\ \bibnamefont {Stormer}}, \ and\
  \bibinfo {author} {\bibfnamefont {P.}~\bibnamefont {Kim}},\ }\href@noop {}
  {\bibfield  {journal} {\bibinfo  {journal} {Nature}\ }\textbf {\bibinfo
  {volume} {438}},\ \bibinfo {pages} {201} (\bibinfo {year}
  {2005})}\BibitemShut {NoStop}%
\bibitem [{\citenamefont {Tamm}(1932)}]{tamm1932possible}%
  \BibitemOpen
  \bibfield  {author} {\bibinfo {author} {\bibfnamefont {I.}~\bibnamefont
  {Tamm}},\ }\href@noop {} {\bibfield  {journal} {\bibinfo  {journal} {Phys. Z.
  Sowjetunion}\ }\textbf {\bibinfo {volume} {1}},\ \bibinfo {pages} {733}
  (\bibinfo {year} {1932})}\BibitemShut {NoStop}%
\bibitem [{\citenamefont {Dutreix}\ \emph {et~al.}(2013)\citenamefont
  {Dutreix}, \citenamefont {Bilteanu}, \citenamefont {Jagannathan},\ and\
  \citenamefont {Bena}}]{PhysRevB.87.245413}%
  \BibitemOpen
  \bibfield  {author} {\bibinfo {author} {\bibfnamefont {C.}~\bibnamefont
  {Dutreix}}, \bibinfo {author} {\bibfnamefont {L.}~\bibnamefont {Bilteanu}},
  \bibinfo {author} {\bibfnamefont {A.}~\bibnamefont {Jagannathan}}, \ and\
  \bibinfo {author} {\bibfnamefont {C.}~\bibnamefont {Bena}},\ }\href {\doibase
  10.1103/PhysRevB.87.245413} {\bibfield  {journal} {\bibinfo  {journal} {Phys.
  Rev. B}\ }\textbf {\bibinfo {volume} {87}},\ \bibinfo {pages} {245413}
  (\bibinfo {year} {2013})}\BibitemShut {NoStop}%
\end{thebibliography}%


%

\end{document}